\documentclass[journal=mamobx,manuscript=article]{achemso}
\usepackage{achemso}

\setkeys{acs}{usetitle = true}
\usepackage[version=3]{mhchem}
\usepackage[T1]{fontenc}
\usepackage{times,mathptmx}
\usepackage{ltablex}
\usepackage{type1cm}
\usepackage{graphicx}
\usepackage{dcolumn}

\usepackage{bm}
\usepackage{units}
\usepackage{color}
\usepackage{xcolor}
\usepackage{latexsym}
\usepackage{subfig}
\usepackage{amsmath,amssymb}
\SectionNumbersOn

\setcitestyle{super}

\newcommand*{\citen}[1]{%
  \begingroup
    \romannumeral-`\x 
    \setcitestyle{numbers}%
    \cite{#1}%
  \endgroup   
}
\title{On the Validity of Effective Potentials in Crowded Solutions of  Linear and Ring Polymers with Reversible Bonds}

\author{Mariarita Paciolla}
\affiliation{Centro de F\'isica de Materiales (CSIC, UPV/EHU) and Materials Physics Center MPC, Paseo Manuel de Lardizabal 5, 20018 San Sebasti\'an, Spain}

\author{Christos N. Likos}
\affiliation{Faculty of Physics, University of Vienna, Boltzmanngasse 5, A-1090 Vienna, Austria}

\author{Angel J. Moreno}
\affiliation{Centro de F\'isica de Materiales (CSIC, UPV/EHU) and Materials Physics Center MPC, Paseo Manuel de Lardizabal 5, 20018 San Sebasti\'an, Spain} 
\alsoaffiliation{Donostia International Physics Center, Paseo Manuel de Lardizabal 4, 20018 San Sebasti\'an, Spain}
\email{angeljose.moreno@ehu.es}

\begin{document}

\newpage

\begin{center}
\LARGE
{\bf Graphical TOC Entry}
\end{center}
\vspace{2 cm}

\begin{figure}[ht]
\centering
  \includegraphics[width=12.8 cm,height=7 cm]{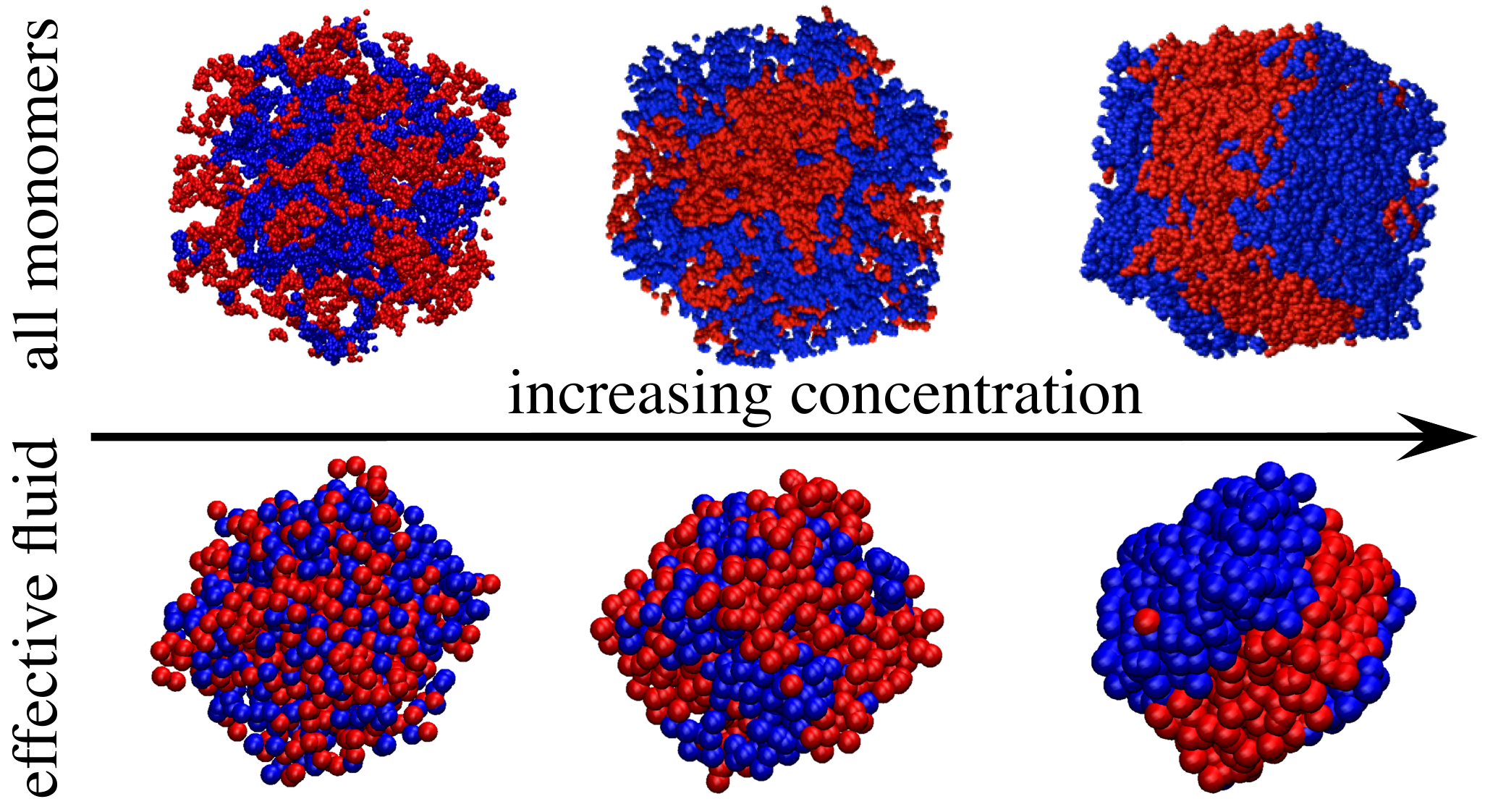}
\end{figure}

\newpage

\begin{abstract}

We perform simulations to compute the effective potential between the centers-of-mass of two polymers with reversible bonds. We investigate the influence of the topology on the potential by employing linear and ring backbones for the precursor (unbonded) polymer, finding that it  leads to qualitatively different effective potentials. In the linear and ring cases the potentials can be described by Gaussians and generalized exponentials, respectively. The interactions are more repulsive for the ring topology, in analogy with known results in the absence of bonding. 
We also investigate  the effect of the specific sequence of the reactive groups along the backbone
(periodic or with different degrees of randomness), establishing that it has a significant impact on the 
effective potentials. When the reactive sites of both polymers are chemically orthogonal so that only intramolecular bonds are possible, the interactions become more repulsive the closer to periodic the sequence is. The opposite effect is found if both polymers have the same type of reactive sites and intermolecular bonds can be formed.
We test the validity of the effective potentials in solution, in a broad range of concentrations from high dilution to far above the overlap concentration. For this purpose, we compare simulations of the effective fluid and test particle route calculations with simulations of the real all-monomer system. Very good agreement is found for the reversible linear polymers, indicating that unlike in their non-bonding counterparts many-body effects are minor even far above the overlap concentration.  The agreement for the reversible rings is less satisfactory, and at high concentration the real system does not show the clustering behavior predicted by the effective potential. 
Results similar to the former ones are found for the partial self-correlations in ring/linear mixtures. Finally, we investigate the possibility of creating, at high concentrations, a gel of two interpenetrated reversible networks. For this purpose we simulate a 50/50 two-component mixture of reversible polymers with orthogonal chemistry for the reactive sites, so that intermolecular bonds are only formed between polymers of the same component. As predicted by both the theoretical phase diagram and the simulations of the effective fluid, the two networks in the all-monomer mixture do not interpenetrate and phase separation (demixing) is observed instead.

\end{abstract}

\maketitle

\section{Introduction}

Single-chain nanoparticles (SCNPs) are soft nano-objects, of size in the range of 3 to 30 nm, which are  synthesized through purely intramolecular cross-linking of functionalized polymers (precursors) \cite{Pomposo2017}.  
Both for their size and for their internal malleability that allows for quick response to environmental changes, 
SCNPs are promising macromolecules for applications as catalytic nanoreactors, 
drug delivery nanocarriers, and biosensing probes, to name a few \cite{Lyon2015,MGonzalez2015,hanlon2016,Rothfuss2018,Kroger2019,Verde2020,Deng2021}. 
Depending on several factors implemented on the precursor, as the solvent conditions, its  molecular topology, chain stiffness or the presence 
of  crowders, the resulting SCNPs present a broad range of structural conformations, 
from very sparse objects \cite{Moreno2013,Rabbel2017} to more compact and even nanogel-like SCNPs \cite{LoVerso2014,LoVerso2015,Formanek2017}.
In the usual route (linear precursors in good solvent), the resulting SCNPs are sparse objects where short-range loops dominate the distribution
of cross-links. This is a direct consequence of the self-avoiding conformations of the linear precursor in good solvent.
In such conformations contacts between monomers separated by long contour distances and formation of long-range loops ---which are efficient for folding into globular shapes--- are unfrequent. The SCNP conformations are dominated by short loops and have scaling 
exponents of $\nu \sim 0.5$ for the dependence of the size on the number of monomers ($R \sim N^\nu$), 
far from the globular state $\nu \sim 1/3$. 

Synthesis of SCNPs has been traditionally dominated by irreversible intrachain cross-linking of the precursor.
In recent years, growing efforts have been dedicated to broaden the functionalities and areas of applicability of SCNPs through the implementation of  reversible bonds in their backbone via non-covalent and dynamically covalent interactions.
The current SCNP chemistry toolbox of reversible bonds includes, among others, 
hydrogen bonds, metal complex formation, hydrazone, enamine, anthrazene, etc \cite{Fulton2011,SanchezChemCom2014,Liu2015,Chen2018}.
Since breaking and formation of these bonds can be activated and tuned through factors as temperature, pH or light, the single-chain character of these objects in solution is lost when their concentration is high enough, leading to the formation of  aggregates and eventually a percolating network. Due to the reversibility of the bonds the bonding pattern of the network is dynamic, allowing for viscous flow of the material, and for physical gelation if the external stimuli are switched off (e.g., by decreasing temperature). The possibility of designing smart polymers that can reversibly transform from solutions of SCNPs to hydrogels has been demonstrated experimentally \cite{Fulton2012,Hebel2021}. These findings pave the way to use polypeptide-based SCNPs as building blocks
for biocompatible and biodegradable materials with self-healing properties and applications in tissue engineering \cite{Hebel2021}.

Recent simulations have investigated the transition from a solution of sparse SCNPs at high dilution to a dynamic network in semidilute and concentrated conditions for a system of linear chains with reversibly bonding sites in their backbones \cite{Formanek2021}. Some remarkable results have been reported: i) the intramolecular bonds still form the majority of the overall bonding 
and the connectivity of the network is mediated by a few intermolecular bonds per chain; 
ii) the bonding pattern of the network is dynamic and the polymers are able to diffuse long distances through breaking and formation of bonds at different sites whithout losing their connection to the percolating cluster; iii) the size and shape of the SCNP conformations at high dilution
are essentially unaffected by crowding  and remain in the network even at densities far above the overlap concentration. The latter is a rather unusual result, in clear contrast with the shrinkage found for other sparse objects as simple (unreactive) linear chains and rings, which by increasing the concentration change their conformations from self-avoiding (Flory exponent $\nu \approx 0.59$) to random walks ($\nu = 1/2$) in the case of linear chains \cite{RubinsteinColby}, and to fractal (`crumpled') globules ($\nu =1/3$) in the case of rings \cite{Halverson2011,Halverson2014}. The weak effect of crowding on the molecular conformations of the reversible SCNPs is inherently related to the formation of intermolecular bonds. Indeed, when the SCNPs are prepared at high dilution through irreversible intramolecular cross-linking and are transferred to high concentration, with no intermolecular bonding, they show a collapse similar to that of rings to crumpled globular conformations \cite{Moreno2016JPCL,gonzalezburgos2018}.  

The effective potential between two macromolecules separated by a given distance is the free energy needed to bring them from the infinity to that distance. Unlike in hard-core colloids, the free energy cost for full-interpenetration of the macromolecules (zero distance) is finite because their centers-of-mass can coincide in space. The cost of full interpenetration strongly depends on the topology and internal deformability of the two macromolecules, typically varying between a few and tenths of times the thermal energy \cite{Likos2001,Likos2006}. Averaging out the molecular internal degrees of freedom and keeping one or a few relevant coordinates (usually the centers-of-mass) reduces the system to an effective fluid of ultrasoft particles interacting through the effective potential \cite{Likos1998,Gotze2004,Narros2013,Bernabei2013,Stiakakis2021}. This methodology allows not only for simulating much larger scales than in the monomer-resolved models, but also for the treatment of the system by methods from liquid state theory, producing a powerful tool for predicting large-scale organization and phase behavior \cite{LIKOS2007,Mladek2008}. A major limitation of this approach is that, because the effective potential is derived for a pair of polymers in the absence of others, it neglects the many-body interactions that are present in a crowded solution or a melt. This approximation works well below and even slightly above the overlap concentration (i.e., the concentration at which the mean intermolecular distance 
is of the order of the unperturbed molecular size). However, it fails dramatically far above the overlap concentration when many-body effects become
a dominant contribution (shrinkage of molecular size being a manifestation of them). A paradigmatic example is the non-emergence of the cluster crystal phase
predicted by the effective potential for flexible ring polymers \cite{Narros2010}, which instead collapse to crumpled globular conformations that hinder
the full interpenetration required to form the cluster nodes.

As mentioned above, when linear polymers with reversible bonds assemble into a dynamic percolating network, they essentially mantain the SCNP conformations
adopted at high dilution \cite{Formanek2021}. This result suggests that many-body effects can be neglibible for this system, 
and the interaction of a tagged pair with their neighboring molecules is effectively given by a flat energy landscape not affecting the effective mutual force
between the two polymers of the tagged pair. In such a case, the validity of the effective potential to describe the static correlations between molecular centers-of-mass could extend to unprecedented densities far above the overlap concentration. With this idea in mind we investigate, by means of simulations, 
the validity of the effective potential for a system of generic bead-spring polymers that switches from a solution of SCNPs at high dilution to a reversibly cross-linked polymer network at high concentrations. We explore a broad concentration range between both limits, as well as the effect of the molecular topology of the unbonded polymer (linear or ring) and the sequence of reactive sites (with different degrees of randomness) along the molecular backbone. We test the accuracy of the effective potentials by comparing simulations of the real all-monomer systems with their corresponding effective fluids of ultrasoft particles, as well as with predictions 
from the test particle route \cite{FoundInhomoFluids}. We also test the approach for a ring-linear mixture, as well as for a two-component linear/linear mixture with orthogonal bonding chemistry, where intermolecular bonding is only allowed between chains of the same component. We find that both the topology of the unbonded polymer and the specific sequence have a strong impact in the effective potential. As suggested by the weak effect of the concentration on the size and shape of the linear polymers with reversible bonds, 
the simulations confirm that the effective fluid provides a very good description of the real system at densities far above the overlap concentration.
In a similar fashion to the case of ring polymers with no bonding, the effective fluid approach is less satisfactory for the ring-based system, and the predicted
clustering behavior is not found in the real system. The effective potential becomes much more repulsive when intermolecular bonding is switched off. As a consequence, the effective binary fluid representing the mixture with orthogonal bonding chemistry shows demixing. This behavior is confirmed in the all-monomer real mixture, which shows spontaneous demixing within the simulation time scale. This striking result suggests that experimental interpenetrated networks with reversible bonds are kinetically trapped states where demixing is prevented by large barriers arising from long bond lifetimes and entanglements.

The article is organised as follows.
In \ref{sec2} we define the model and interactions implemented in the all-monomer simulations. We also give the simulation details for the computation of the effective potentials and briefly describe the analytical test particle route approach. In \ref{sec3} we report a critical analysis of the obtained effective interactions as a function of the topology of the precursor and the specific sequence of reactive sites. 
In \ref{sec4} we present theory and simulation results for the solutions at different concentrations and for the phase behavior of the mixtures, and discuss the validity of the effective potentials to describe the behavior of the all-monomer real systems. In \ref{sec5} we summarize our conclusions.

\section{Model and simulation details}\label{sec2}

The precursors are modeled as fully flexible linear chains or rings made of 200 beads (monomers). A fraction of these monomers  $f = N_{\rm r}/N_{\rm m} = 20/200 = 0.1$ is reactive, where $N_{\rm r}$ and $N_{\rm m}$ are respectively the number of reactive sites and the total number of monomers. The reactive sites are able to form and break bonds with other reactive sites within the same polymer (intra-bonding) or with reactive sites belonging to other polymers (inter-bonding).       
In all cases, we perform Langevin dynamics simulations. In the first step we use them to obtain the effective potentials (\ref{sec:comput});  subsequently we use them to simulate the effective fluid at different concentrations and we compare results with simulations of the corresponding all-monomer system (\ref{sec:allmon}). Moreover, we compare simulation results with theoretical calculations by the test particle route (\ref{sec:TPR}). 
   
\subsection{Model}\label{sec:numer}

We describe the polymer chains by the bead-spring model of Kremer and Grest \cite{Kremer1990}. Thus, excluded volume interactions among the beads are modeled by the Weeks-Chandler-Andersen (WCA) potential: 

\begin{equation}
V_{\rm WCA}(r) = \left\{
\begin{array}{lll}
4\epsilon\left[\left(\frac{\sigma}{r}\right)^{12} - \left(\frac{\sigma}{r}\right)^{6} +\frac{1}{4} \right]  &\ &\  r\leq 2^{1/6}\sigma \\
0  &\ &\  r > 2^{1/6}\sigma 

\end{array}
\right.
\label{eq:WCA}
\end{equation}
The permanent bonds leading to the connectivity of the precursor are implemented via a Finite Extensible Nonlinear Elastic (FENE) potential between consecutive monomers. 
This is given by
\begin{equation}V_{\rm FENE}(r) = -\epsilon K_{\rm F}R_0^2 \ln\left[1-\left(\frac{r}{R_0\sigma}\right)^2 \right] ,
\label{eq:fene}
\end{equation}
where $K_{\rm F} = 15$  and $R_0\sigma = 1.5\sigma$ is the maximum elongation of the bond.


\begin{figure}[pt!]
	\centering
	\includegraphics[width=0.55\textwidth]{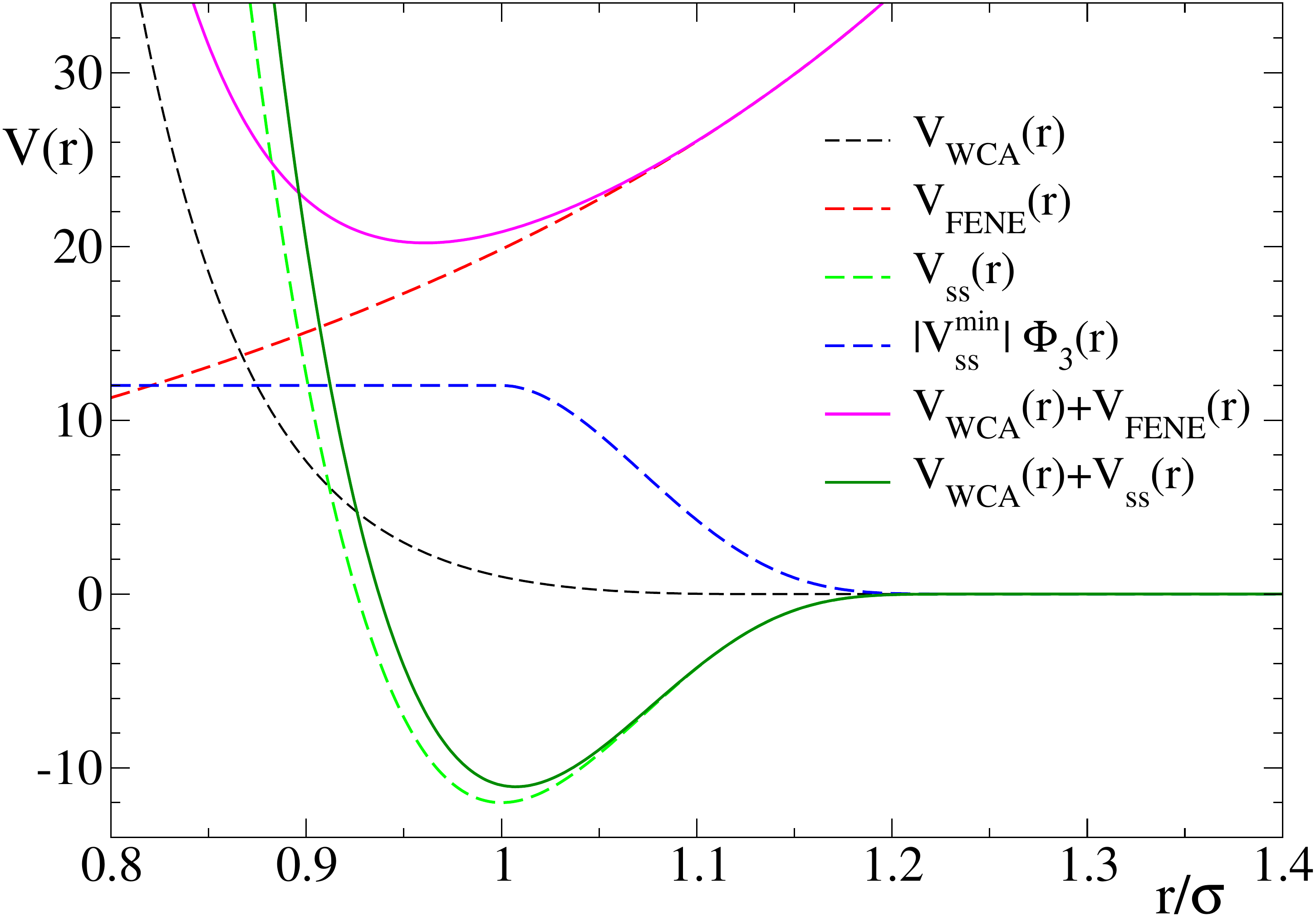}\hspace{0.001cm}
	\vspace{0.001cm}
	\caption{Representation of the different interaction potentials used in this study. The combination of the WCA and FENE potentials results in a deep potential well that sets the mean length of the permanent bonds at $r_{\rm min,irrev} \approx  0.96\sigma$. The combination of the WCA and the reversible bonding potential $V_{\rm ss}(r)$ defines the mean length of the reversible bonds  at $r_{\rm min,rev} \approx  1.0\sigma$. The function 
$|V_{\rm ss}^{\rm min}|\Phi_3(r)$ represents the contribution of a bond belonging 
to a triplet to the 3-body potential (see text).}
	\label{fig:potentials} 
\end{figure}

For implementing the reversible bonds between the reactive sites, we adopt the potential introduced by Rovigatti et al.\cite{Rovigatti2018Vitrimers}: 
%
\begin{equation}
V_{\rm ss}(r) = \left\{
\begin{array}{lll} \epsilon_{\rm ss} e^{\frac{\sigma}{(r_{\rm c}-\sigma)}}  \left[K_{\rm ss}\left(\left(\frac{\sigma}{r}\right)^{4} - 1\right)-1\right]e^{\frac{\sigma}{(r-r_{\rm c})}} &\ &\  r < r_{\rm c} \\
0  &\ &\  r \geq r_{\rm c} 
\end{array}
\right.
\label{eq:SS}
\end{equation}
In our system we set the capture radius $r_{\rm c} = 1.3\sigma$ while $ \epsilon_{\rm ss} = 12\epsilon$ and $K_{\rm ss} = [\sigma/2(\sigma- r_{\rm c})]^{2}$. With these choices the potential and force are continuous and zero at the capture radius. Moreover, the potential is short-ranged and has a deep attractive minimum 
of energy $V_{\rm ss}^{\rm min} = -12 k_{\rm B}T$ (which can be seen as the bond energy) at the distance $r_{\rm min} = 1.0 \sigma$.
When the distance between two reactive sites is smaller than $r_{\rm c}$ the interaction of \ref{eq:SS} switchs on and the sites form a mutual bond. The bond is broken when a fluctuation moves the mutual distance beyond $r_{\rm c}$. Since we wish to limit the valence to a single reversible bond per reactive site, we make use of  the swapping algorithm introduced by Sciortino \cite{SwapSciortino}. Thus we add a repulsive three-body contribution in such a way that it is switched on
when a reactive site $k$ enters the capture radius of a reactive site $i$ that is already bonded to another one $j$. The three-body potential is defined as
\begin{equation}
V_{\rm 3body}(r) = |V_{\rm ss}^{\rm min}| \sum_{i,j,k} \Phi_{3}(r_{ij})\Phi_{3}(r_{jk}) , 
\label{eq:3BODY}
\end{equation}
where the sum includes all $i,j,k$ triplets, and 
\begin{equation}
\Phi_3(r) = \left\{
\begin{array}{lllll} 
1 &\ &\ &\  r \leq \sigma  &\ \\
V_{\rm ss}(r)/V_{\rm ss}^{\rm min}  &\ &\ &\ \sigma < r \leq r_{\rm c} &\ \\
0 &\ &\ &\  r > r_{\rm c}. \ &\ \\ 
\end{array} 
\right.
\label{eq:v3}
\end{equation}
Therefore,
$0 < V_{\rm 3body}(r) \leq |V_{\rm ss}^{\rm min}|$ for each triplet, and when a triplet is formed the energy decrease resulting from the new bond is compensated by the three-body repulsive term without changing the potential energy of the system. This three-body term makes triplets short-lived and spontaneously leads to bond swapping, which speeds up the exploration of different patterns of the bond network.

The units of energy, length, mass and time are respectively $\epsilon$, $\sigma$, $m$ and $\tau=\sqrt{\sigma^2 m/\epsilon}$ where $m$ is the monomer mass. 
The simulations were carried out at temperature $T = \dfrac{\epsilon}{k_{\rm B}} = 1.0$ by using a Langevin
thermostat with a friction coefficient $\gamma = 0.05$ \cite{Smith2009}. Equations of motion were integrated 
within the scheme of Ref.~\cite{Izaguirre2001} by using a time step $\delta t = 0.005$.

\subsection{Computation of the Effective Potential}\label{sec:comput}


The effective potential acting between the two polymers (1,2) can be calculated by integration of the
net force over the axis joining their centers-of-mass (see e.g., Ref.~\citen{Gotze2004}): $\textbf{F}_{\rm eff,12} = -\nabla_{\textbf{R}_{12}} V_{\rm eff}(R_{12})$, where $R_{12}$ is the distance between the two centers-of-mass. The net force experienced by one of the polymers 
is computed as the total force (non-bonded and bonded) exerted on its monomers by the monomers of the other polymer.  
In the expression below we consider the force exerted by polymer 2 on polymer 1: 
\begin{equation}
\textbf{F}_{\rm eff,12} = \left\langle \sum^{N_{\rm m}}_{i,j=1} \textbf{F}_{i(1),j(2)}  \right\rangle_{R_{12}},
\label{eq:Feff}
\end{equation}
where $\textbf{F}_{i(1),j(2)}$ is the force exterted by the $j$th monomer of polymer 2 on the $i$th monomer of polymer 1, the sum
runs over all $i(1),j(2)$-pairs and the subscript at the right-hand side means that the average must be evaluated at the fixed separation $R_{12}$. 
Obviously the expression accounting for the interactions of polymer 1 on polymer 2 just produces the opposite result, and integration leads to the same effective potential. The statistical averages of the components perpendicular 
to the axis joining the centers-for-mass are zero.

We performed Langevin Dynamics simulations where the positions of the centers-of-mass of the two polymers, and therefore their mutual distance $R_{12}$, were kept fixed at every time step. We peformed the simulation runs at the fixed distances ${R}_{12}/\sigma = 0, 1, 2,...,34, 35$. For each distance we performed an equilibration run of $10^7$ time steps, followed
by a production run of $4 \times 10^{8}$ steps. To improve statistics as much as possible, the total force  $\textbf{F}_{\rm eff}$ was obtained by on-the-fly averaging the summation of \ref{eq:Feff} over all the time steps of the production run.
In order to test if there is any dependence of the effective interactions on the specific sequence of reactive sites along the backbone of the precursor,
we consider 3 different cases to simulate for a couple of polymers with reversible bonds: 
i) A random sequence of the 20 reactive sites with the constraint that there is at least $n_{\rm min} = 1$ non-reactive sites between consecutive reactive sites (in order to prevent trivial bonding). This case will be denoted as `gap1'; 
ii) A random sequence with the constraint $n_{\rm min} = 4$. This case will be denoted as `gap4';
iii) A periodic sequence, i.e., there is a constant separation $n_{\rm min} = 9$ between consecutive reactive sites.
This case will be denoted as `periodic'.
In both cases i) and ii) the sequences of the two polymers are different, with the only condition 
that they have the same $n_{\rm min}$. Moreover, to assess whether even by using the same value of $n_{\rm min}$ 
the specific realization of the sequences affects to the effective potential, we simulated two different couples 
(denoted as couple1 and couple2) for each of the cases `gap1' and `gap4'. 
Figures S1 (linear) and S2 (rings) in the Supporting Information (SI) show the specific sequences of the  simulated couples.
Moreover, we investigated a couple formed by a linear chain and a ring. In this case the simulations were limited to the case
$n_{\rm min} = 1$ (gap1) and we used the first polymer of their corresponding couple1. 




\subsection{Simulations of all-monomer and effective fluids}\label{sec:allmon}

We performed Langevin Dynamics simulations of solutions of the real all-monomer polymers and of the corresponding effective fluids. 
We explored the validity of the effective fluid approach in a broad concentration range below and above the overlap concentration \cite{RubinsteinColby}, 
which we define as $\rho^\star = N_{\rm m}/(2R_{\rm g0})^{3}$, where $R_{\rm g0}$ is the radius of gyration of the isolated polymer
(i.e., in the absence of all intermolecular interactions). 
Therefore, if $\rho = N_{\rm p}N_{\rm m}/V$ is the absolute density (number of monomers per volume), with $V$ the volume of the simulation box 
and $N_{\rm p}$ the number of polymers in the box,
the reduced density (normalized by $\rho^\star$) is $\rho/\rho^\star = N_{\rm p}(2R_{\rm g0})^3 /V$. 
In the case of a binary mixture of components (1,2) we define the reduced concentration as 
$\rho/\rho^\star =V^{-1}[ N_{\rm p,1}(2R_{\rm g0, 1})^3 + N_{\rm p,2}(2R_{\rm g0, 2})^3]$.
Therefore the overlap concentration of the binary mixture is 
$\rho^\star =  (N_{\rm p,1}N_{m,1} + N_{\rm p,2}N_{m,2})/[N_{\rm p,1}(2R_{\rm g0, 1})^3 + N_{\rm p,2}(2R_{\rm g0, 2})^3] $.
For the isolated linear and ring polymers we find $R_{\rm g0}/\sigma = 9.93$ and 7.72 respectively. 
Therefore the density of monomers at the overlap concentration is
$\rho^\star\sigma^{-3} = 0.025$, 0.064 for the pure solutions of linear chains and rings with reversible bonds, respectively.
For a 50/50 linear/ring mixture the overlap concentration is $\rho^\star\sigma^{-3} = 0.036$.

In the all-monomer simulations we investigated the pure systems of linear chains and rings with reversible bonds, a 50/50 linear/ring mixture  and  mixture of linear chains with orthogonal bonding. In the latter, bonding (intra- or intermolecular) was only permitted between polymers of the same component,  and all WCA, FENE and reversible bonding interactions were the same as in the other simulated systems, with the only condition that A- and B-reactive sites could not form mutual bonds and only interacted through the WCA potential. Although the breaking and formation of bonds can lead to concatenation of reversible loops in both the linear chain and ring-based systems, in the latter
intermolecular concatenation between the permanent ring backbones must be avoided. Thus, non-concatenated dilute ring-based systems were initially prepared and compressed to the target concentrations where they were further equilibrated. To prepare the linear-linear mixture with orthogonal bonding chemistry, configurations were taken from the one-component system and half of the chains were randomly assigned to each component of the mixture, which was further equilibrated with no intermolecular bonding between different components. Therefore, the final demixed state that we anticipated in the Introduction was reached spontaneously from an initially mixed state, demonstrating the robustness of this result.

The duration of the equilibration and production runs was typically $10^7$ and $8 \times 10^7$ time steps, respectively.
To improve statistics, 8 independent runs were simulated at each concentration.
In all cases the total number of polymers in the simulation box was $N_{\rm p} = 108$, with $N_{\rm m}=200$ monomers and $N_{\rm r}=20$ reactive sites per polymer, and the concentration was tuned by varying the box size. All the polymers had different random sequences of reactive sites corresponding to the case 'gap 1'.
The effective fluids were simulated by using the corresponding effective potentials obtained for the couple 1 of the case 'gap 1'. Because of the much smaller number of degrees of freedom, in the effective fluids we simulated larger boxes of $N_{\rm p}=1000$ effective particles, rescaling the box size to have the same concentrations as in the respective all-monomer systems.  Tables T1 and T2 in the SI  show the simulated box sizes for each all-monomer and effective system and the respective concentrations in absolute and reduced units.
In order to test the effect of the box size, some concentrations in the effective fluid were also simulated with $N_{\rm p} = 108$ particles. Structural properties were not changed within statistics. This is demonstrated in Figure S3 of the SI, which shows representative results of the radial distribution function $g(r)$ of the effective fluid of linear chains with reversible bonds.
Data are shown at the lowest and highest investigated concentrations, and in both cases for $N_{\rm p} = 108$ and 1000 effective particles (with the respective rescaling of the box size to produce the same concentration). No differences are found within statistics in the respective $g(r)$'s. Therefore we conclude that finite size effects are not significant (except for the phase separating system of chains wit orthogonal bonds, see below).

\subsection{Test Particle Route to Fluid Structure}\label{sec:TPR}

The test particle route (TPR) will allow us to compute the radial distribution function of the effective fluid by using the formalism of mean-field density functional theory (DFT)  for inhomogeneous fluids \cite{FoundInhomoFluids}. In our study each particle of the effective fluid represents the center-of-mass of one polymer, and interacts with the others via the effective potential $V_{\rm eff}(r)$ computed as described in section 2.2.    Within TPR, a particle is fixed at the origin of the system. As a consequence, the particle perturbs the system and the density of particles around it changes from a constant bulk value $\rho_{\rm b}$ to a spatially varying local density $\rho(\textbf{r})$. 
The external potential acting on the particle at the origin is equal to the effective potential, implying that the radial distribution function can be calculated as $g(\textbf{r}) = \rho(\textbf{r})/\rho_{\rm b}$. Following the derivation from TPR based on DFT suitable for soft potentials (see SI for details) the partial radial distribution function for the $i,j$-components of a mixture of $n$ components is given by \cite{LikosOverduin2009}:
\begin{equation}
%
g_{ij}(r) = \exp[-\beta V_{{\rm eff},ij}(r) -\beta\Sigma_{k=1}^n \rho_{{\rm b},k} (h_{ik}*V_{{\rm eff},kj})(r) ],
\label{eq:deltaRho}
\end{equation}
where $\rho_{{\rm b},i}$ is the macroscopic density of the $i$-component, $h_{ij}(r) =g_{ij}(r)-1$ is the $ij$-component of the
total correlation function,  the symbol $*$ denotes convolution, and $V_{{\rm eff},kj}(r)$ is the interaction potential
between species $k$ and $j$.

\section{Molecular conformations and computation of the effective potentials} \label{sec3}
 
\subsection{Conformations of two interpenetrated polymers}

\begin{figure}[ht!]
\centering
\includegraphics[width=0.9\textwidth]{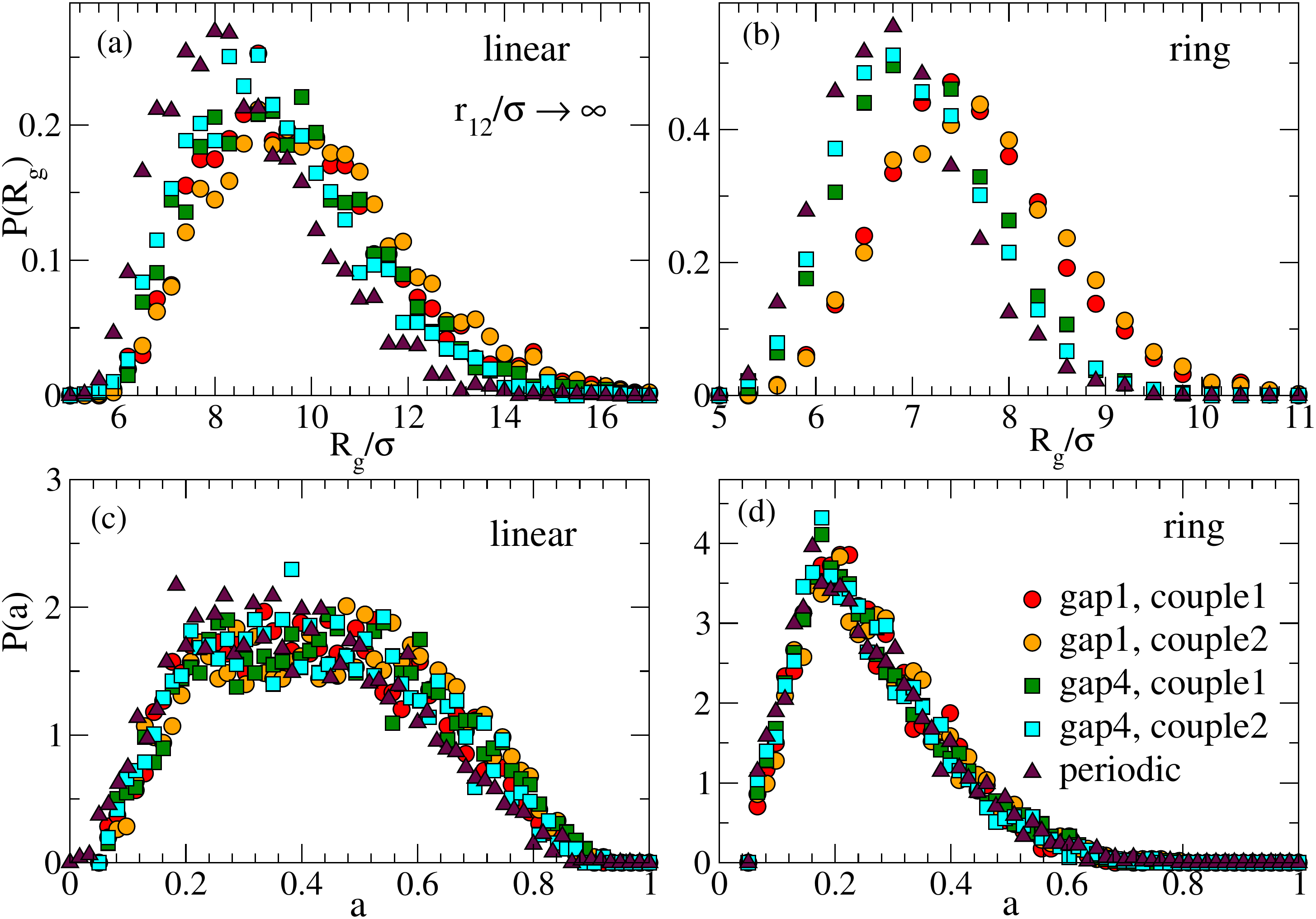}\hspace{0.001cm}
\vspace{0.001cm}
\caption{Distributions of the instantaneous values of the radius of gyration (panels (a) and (b)) 
and the asphericity (panels (c) and (d)) for isolated polymers with reversible bonds: linear chains ((a) and (c)) and rings ((b) and (d)).
Different sets correspond to different sequences of reactive sites (see panel).}
\label{fig:pRgasph}	
\end{figure}

We have investigated effective interactions between two polymers with reversible bonds, namely two linear chains (`linear-linear'), two rings (`ring-ring') and a linear chain and a ring (`linear-ring'). In all cases we have simulated two possibilities of bonding. In the first one (denoted as `all bonds') we carry out standard runs where the two polymers can form both intra- and intermolecular bonds. In the second case (denoted as `only intra') only intramolecular bonds are allowed, i.e., reactive sites belonging to different polymers only interact through the WCA potential and cannot form mutual bonds.  
Before discussing the effective interactions,  we characterize conformations of the two interacting polymers through their radius of gyration $R_{\rm g}$ and the asphericity parameter $a$. This parameter ($0 \leq a \leq 1)$ measures deviations from spherosymmetrical conformations ($a=0$).
\ref{fig:pRgasph} shows for each of the topologies (linear, ring) and sequences of reactive sites (couples 1, 2 of gap1 and gap4, and periodic) the distributions of intantaneous values of $R_{\rm g}$ and $a$ collected from the trajectories. The data are shown for isolated polymers (mimicking the case $V_{\rm eff}(r \rightarrow \infty) =0$).
Only the distributions for the first polymer of each couple of Figs.~S1 and S2 are shown. 
Figures S4 and S5 in the SI compare for each case the distributions of the two polymers of the couple.
As can be seen, the ring polymers with reversible bonds are smaller and closer to spherical than their linear counterparts.
For the same value of $n_{\rm min}$ the specific sequences (4 in total for couple 1 or couple 2) 
have at most a minor effect on the distributions $P(R_{\rm g})$ and $P(a)$. However, \ref{fig:pRgasph} shows that changing the typical distance between consecutive reactive groups does have a systematic effect on $P(R_{\rm g})$. Namely, increasing $n_{\rm min}$ leads to smaller sizes of the polymers. This is not surprising because longer distances between consecutive reactive groups promote the formation of longer loops, resulting in a stronger reduction of the molecular size with respect to the linear precursor.
No significant effect of $n_{\rm min}$ on $P(a)$ is found.

\begin{figure}[ht!]
	\centering
	\includegraphics[width=0.26\textwidth]{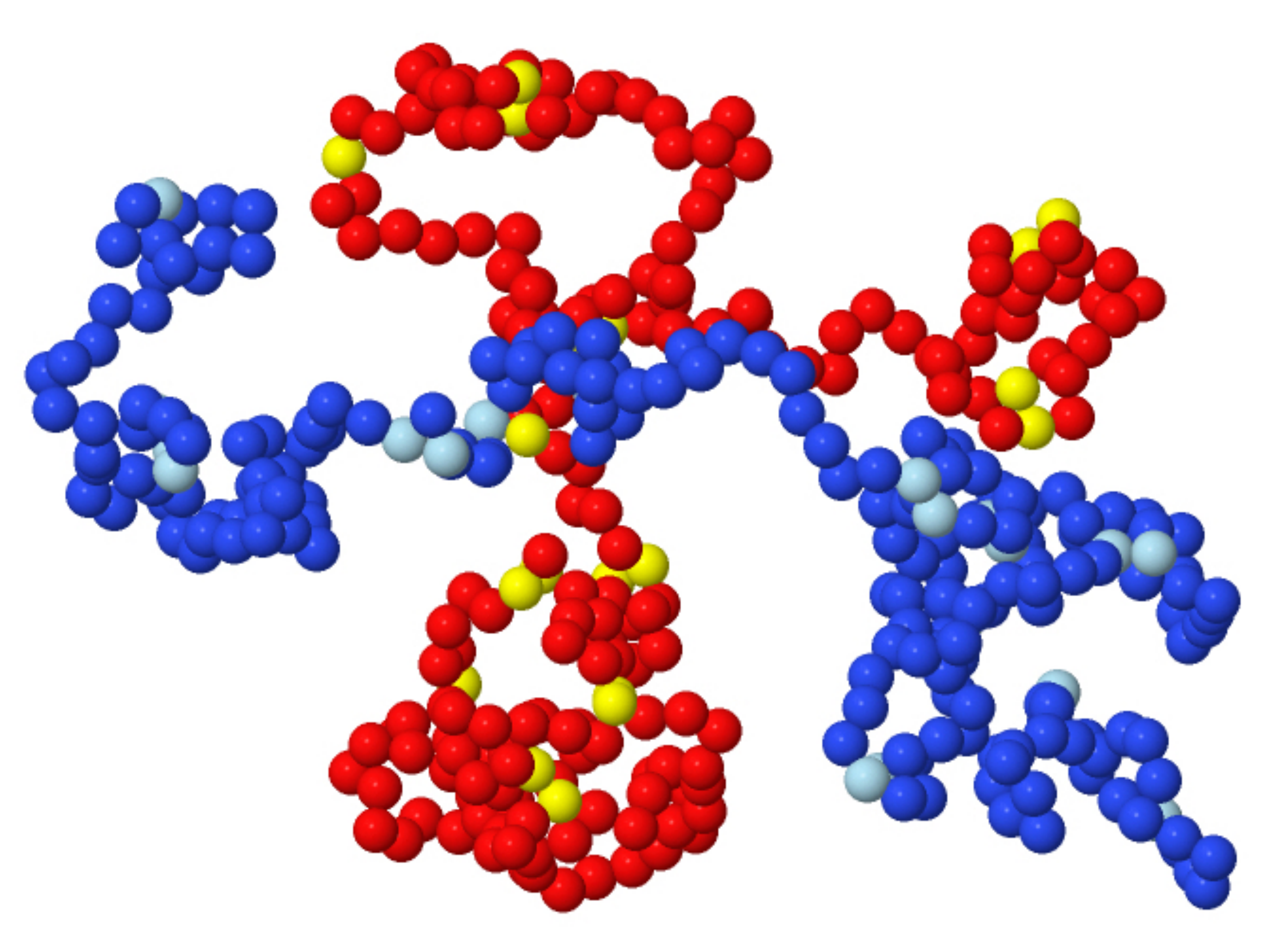}
	\qquad
	\includegraphics[width=0.26\textwidth]{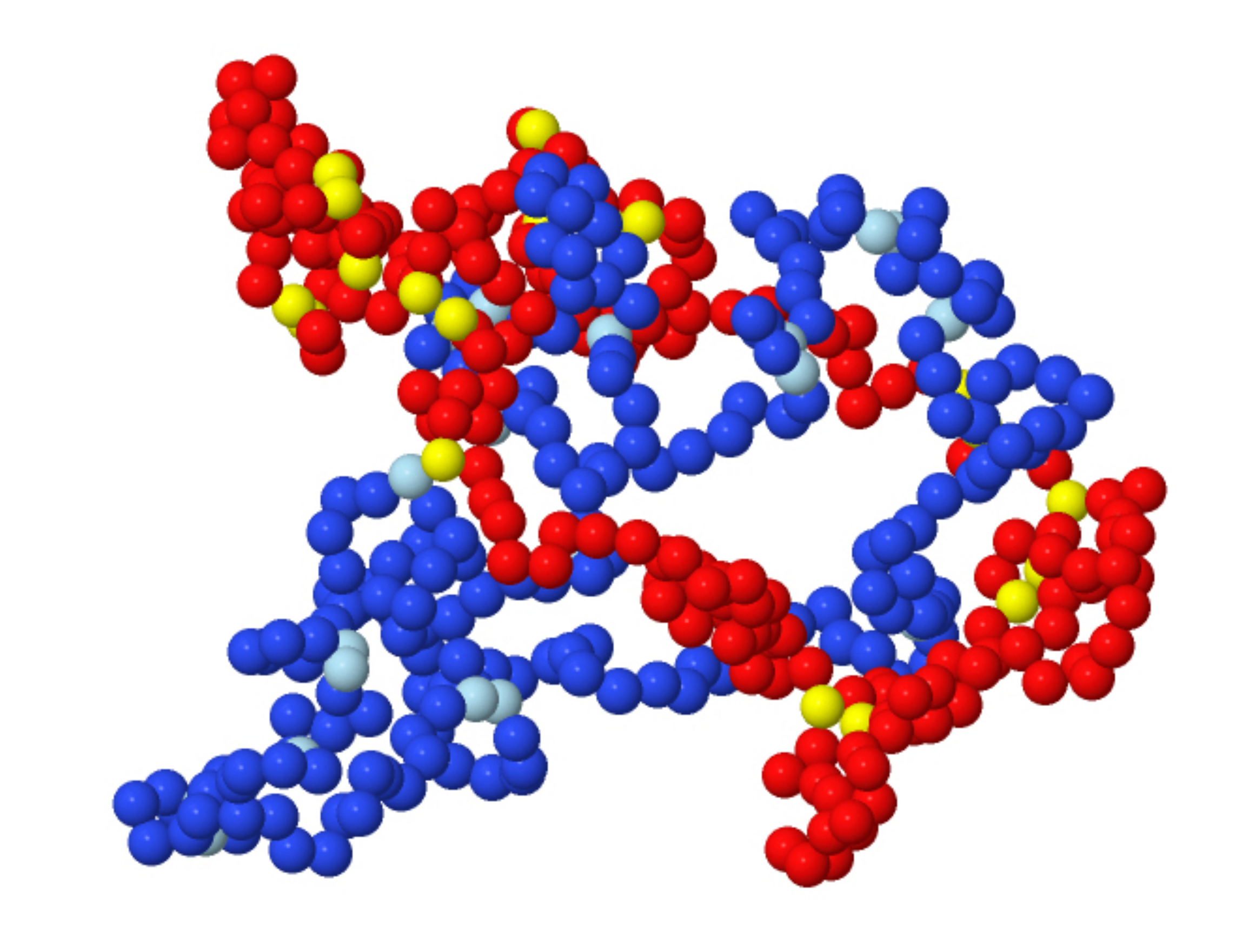}
	\qquad
	\includegraphics[width=0.26\textwidth]{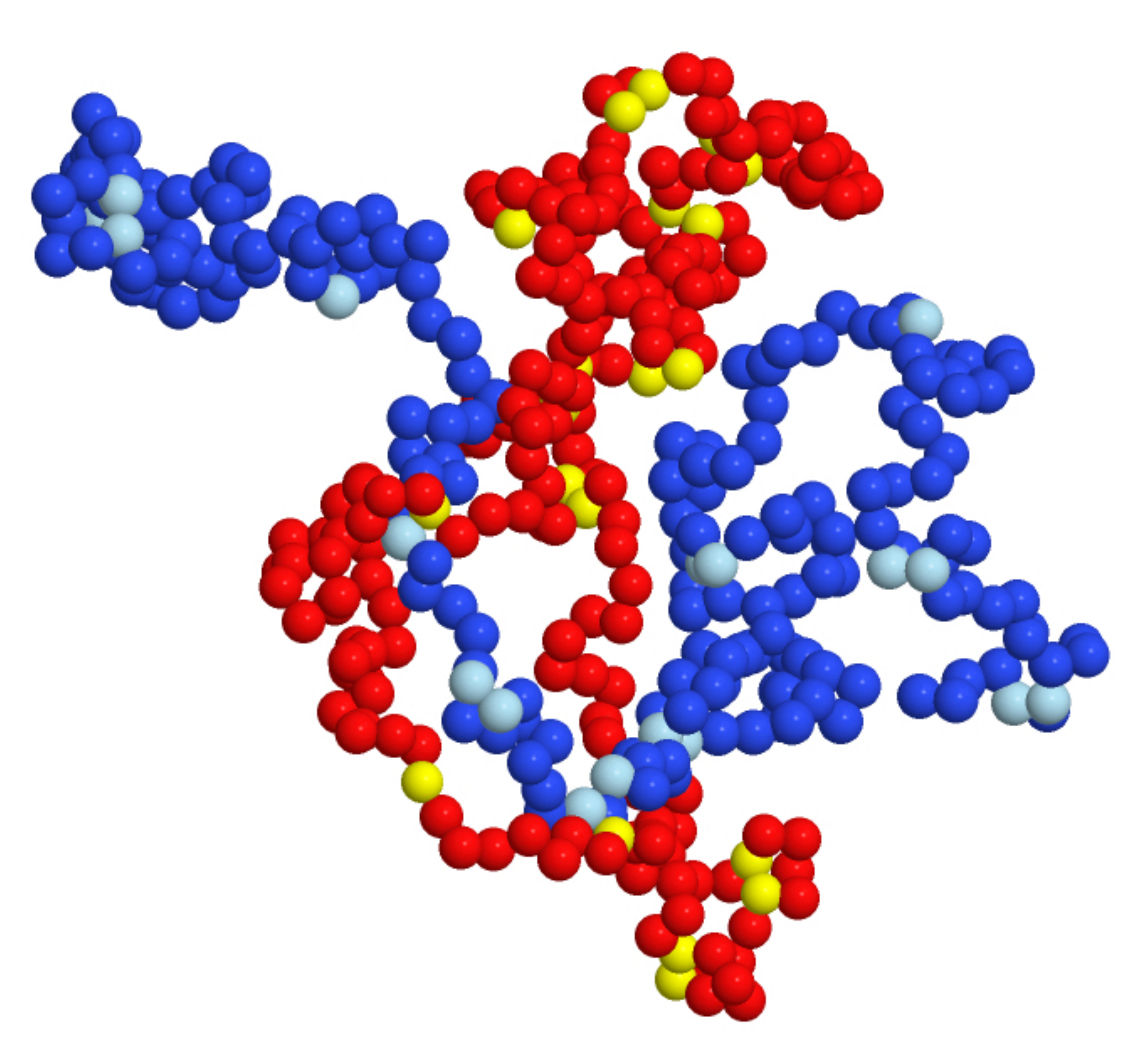}
\caption{Typical snapshots from MD runs at a fixed distance $r=0$ between centers-of-mass and with intermolecular bonding switched on. (a): linear-linear; (b): ring-ring; (c): linear (blue/cyan)-ring (red/yellow). Reactive sites are represented
by cyan and yellow beads. Threading of one ring by the other polymer is found in both (b) and (c).}
\label{fig:snap-umbrel}
\end{figure}

Figures S6 and S7 show the effect of the intermolecular interactions on the size and shape of the two polymers. The
distributions $P(R_{\rm g})$ (S6) and $P(a)$ (S7) now correspond to a distance between centers-of-mass $r= 3\sigma$ 
and allowing for intermolecular bonding. Similar results are found for other close distances. 
As can be seen in Figure S6, the mutual interaction tends to swell both polymers (the maxima of $P(R_{\rm g})$ are shifted by about 15\%) with respect to the isolated ($r \rightarrow \infty$) case. The mutual interaction also tends to increase the asphericity (Figure S7).  A remarkable effect for the ring-ring case is that the two polymers do not swell in the same way. This can be explained by the fact that at short intermolecular distances one of the rings is threaded by the other one. \ref{fig:snap-umbrel} shows typical conformations of the two polymers at mutual distance $r=0$ (from (a) to (c): linear-linear, ring-ring, and linear-ring). Panels (b) and (c) illustrate the threading of one ring by the other polymer (this also occurs in the linear-ring case). The asymmetry found in the radii of gyration of the two interpenetrated rings is also reflected in their different asphericities (Figure S7), through the effect is less pronounced than in $P(R_{\rm g})$.

\begin{figure}[ht!]
\centering
\includegraphics[width=0.496\textwidth]{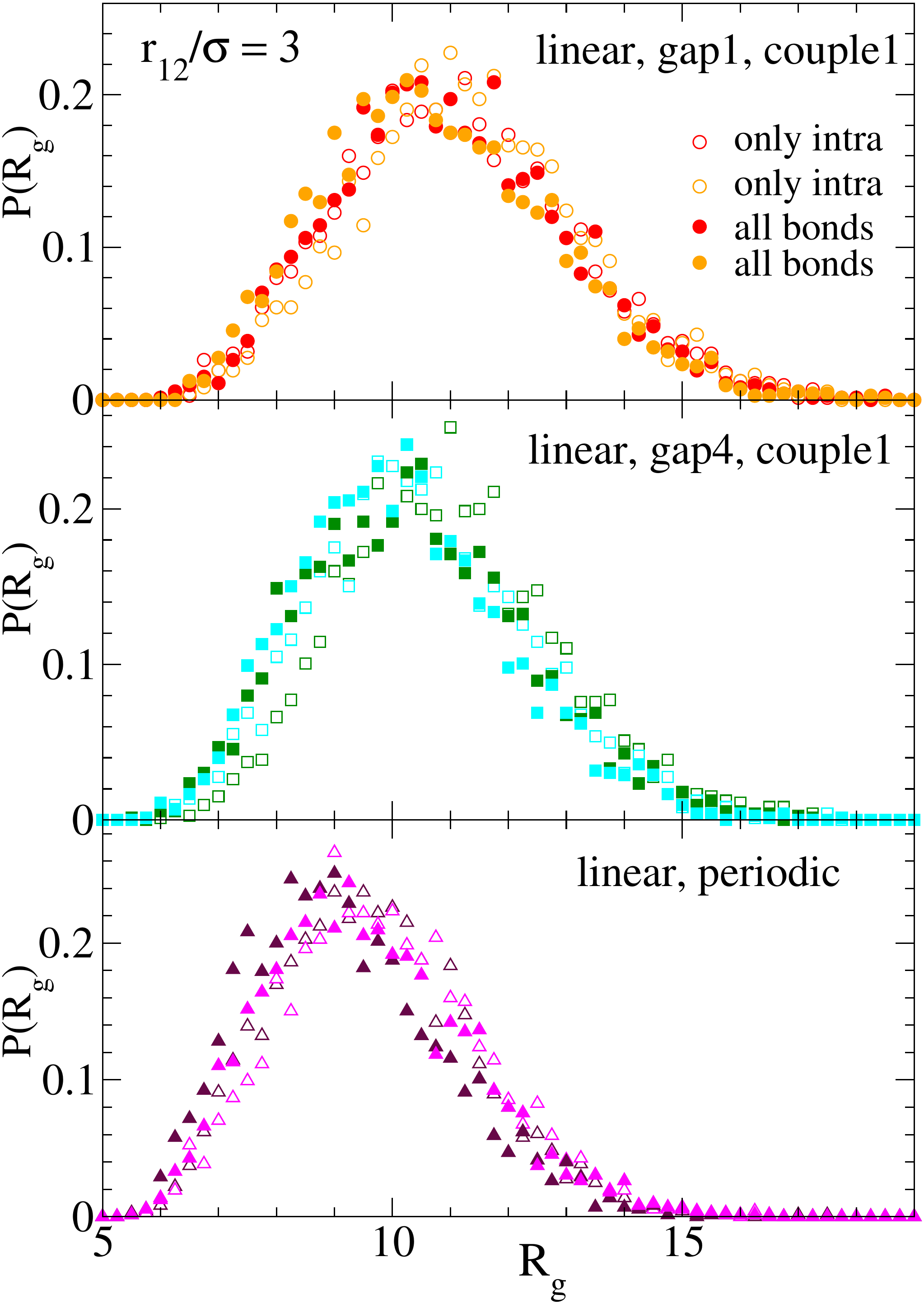} \includegraphics[width=0.496\textwidth]{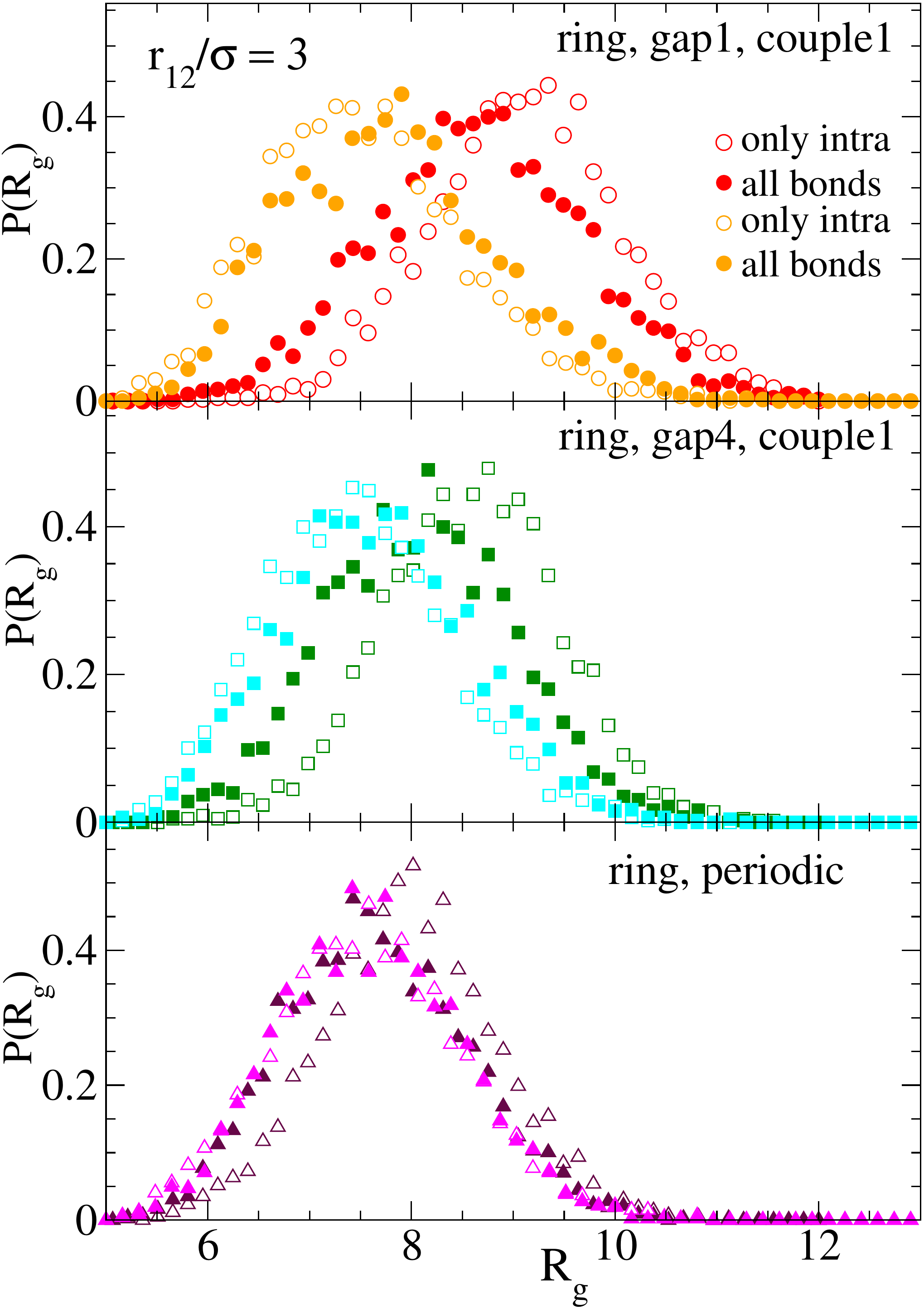}
\caption{Left column: distributions of the radius of gyration for two linear polymers with reversible bonds 
at a distance $r_{12}=3\sigma$ between centers-of-mass.
Empty symbols correspond to simulations without intermolecular bonding. Filled symbols correspond to simulations
where intermolecular bonds are allowed. Right column: as the left column, for two rings.}
\label{fig:distrg-inter-intra}	
\end{figure}

\ref{fig:distrg-inter-intra} shows the effect of switching on and off the intermolecular bonds on the conformations of the two polymers  at a close distance $r= 3\sigma$. In the case of the linear chains 
there is a tiny shrinking of the size of both polymers
when intermolecular bonding is allowed, which is presumably due to the slight reduction of the fluctuations when a few intermolecular bonds connect the two polymers. A different behavior is observed in the pair of rings, whose sizes
change in opposite directions when they form intermolecular bonds and their size 
disparity is reduced. Thus, the larger threaded ring shrinks and the smaller threading ring swells.
The combination of both effects, occurring in the asymmetric pair created by threading,
reduces distances between segments of different polymers 
and facilitates the formation of intermolecular bonds.

\subsection{Effective potentials}

\begin{figure}[ht!]
	\centering
	\includegraphics[width=0.5\textwidth]{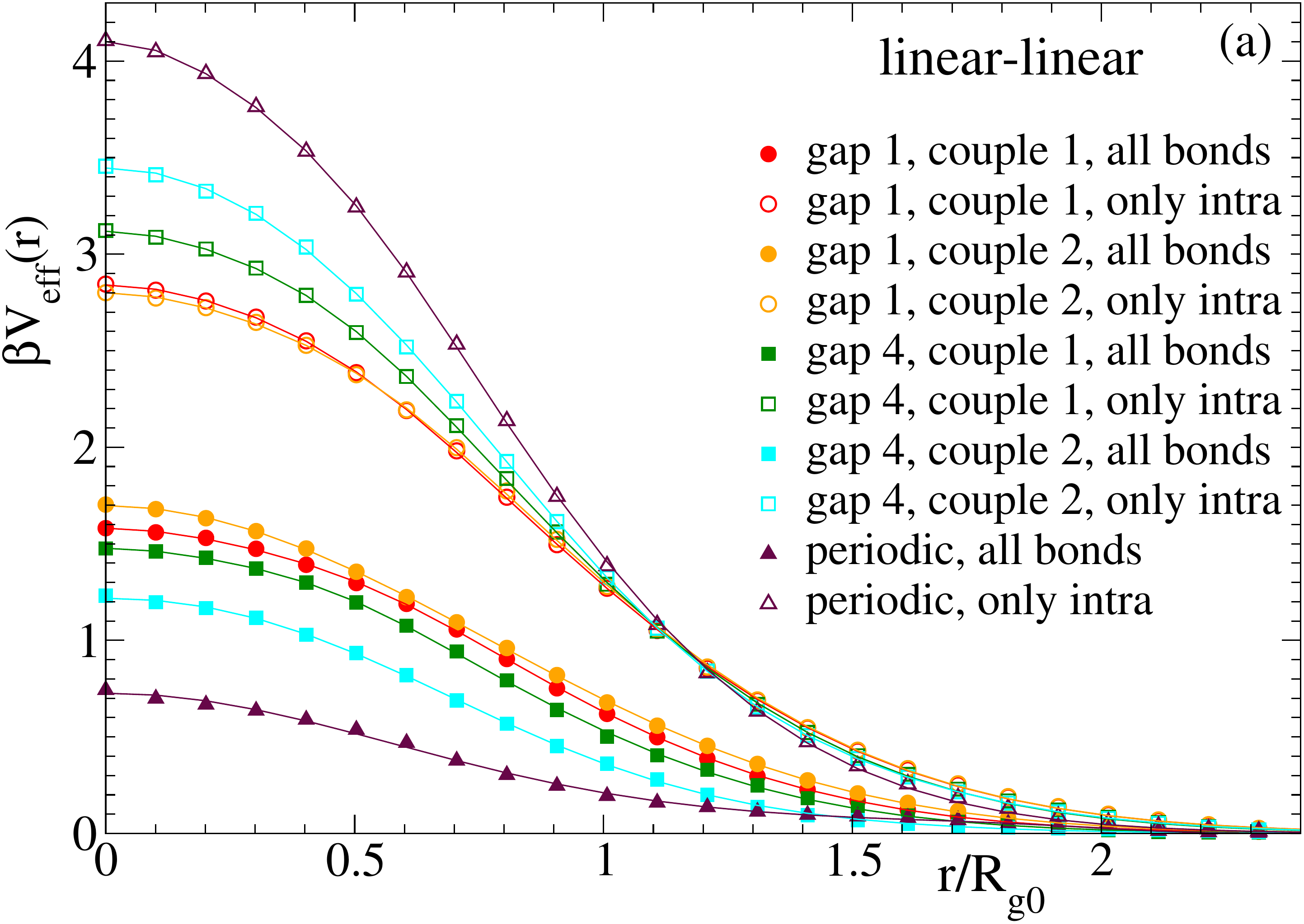}
	\qquad
	\includegraphics[width=0.5\textwidth]{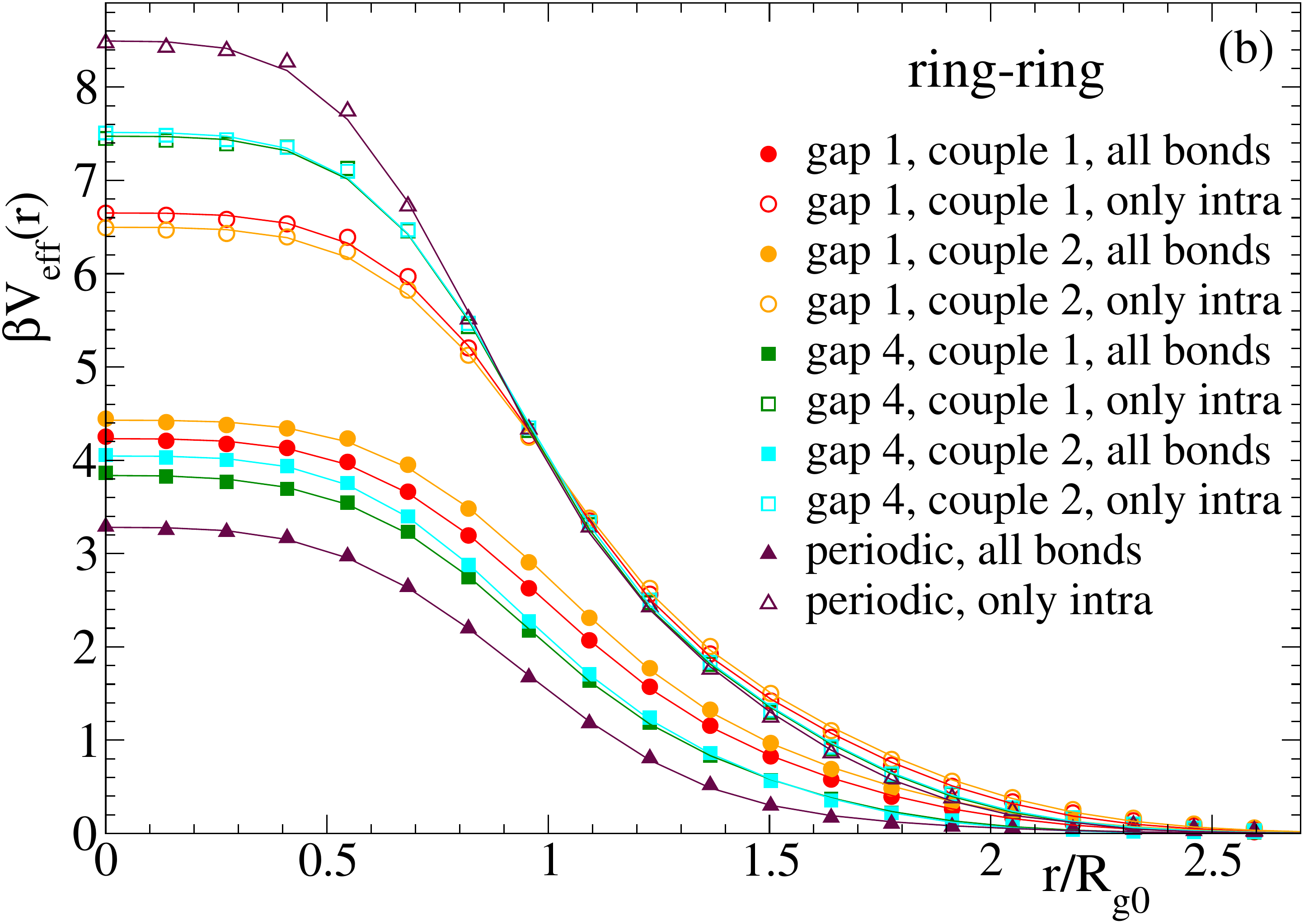}
	\qquad
	\includegraphics[width=0.5\textwidth]{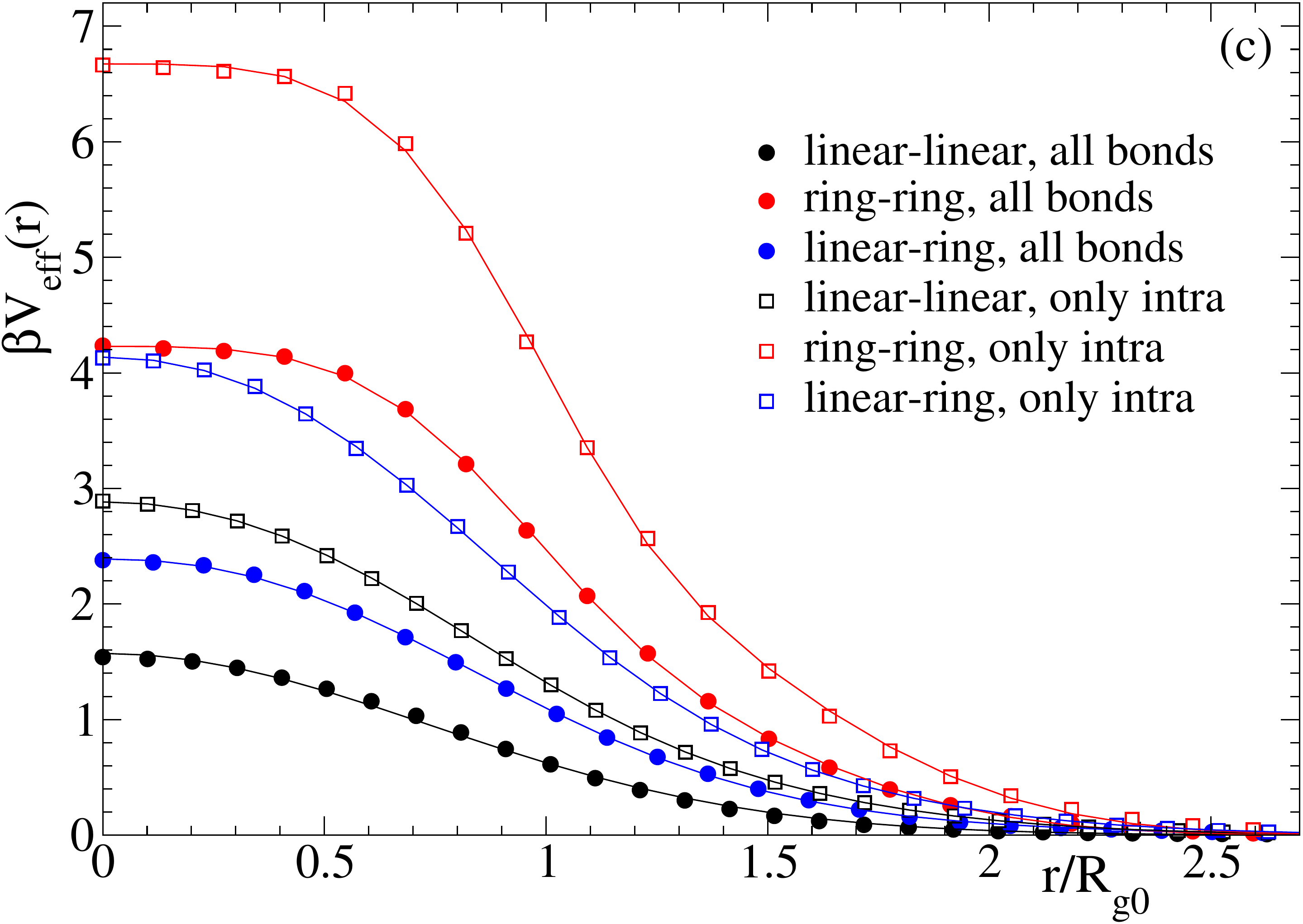}
\caption{Effective potentials (scaled by $\beta = (k_{\rm B}T)^{-1}$) for linear-linear (a) and ring-ring interactions (b). Distances are normalized by the radius of gyration $R_{\rm g0}$ of the isolated polymers. Different data sets correspond to different sequences of reactive groups (see main text). Filled and empty symbols correspond to simulations with and without intermolecular bonding. Solid lines are fits to model functions (see main text). Panel (c) compares results for the linear-ring interaction with the linear-linear and ring-ring cases. Here results for `gap1, couple1' are only included, and for the linear-ring interaction the distance is normalized by the average of the respective $R_{\rm g0}$'s of the linear chain and the ring.}
\label{fig:effpot}
\end{figure}

In panel (a) of \ref{fig:effpot}  we show the effective potentials obtained for the interaction between two linear chains with reversible bonds. The panel (b) shows the corresponding results for two rings, and the panel (c) compares results of the former cases with the effective potential between a linear chain and a ring. 
All data sets in panel (c) correspond to sequences gap1, namely the couples 1 of panels (a) and (b) for the linear-linear and ring-ring case. For the linear-ring case the simulations used the first polymer of the couples 1 of the linear-linear and ring-ring cases.
The symbols in all panels are the results obtained from the simulations. The solid lines are fits to a main function plus a tail, both given
by  generalized exponentials, $\beta V_{\rm eff}(r)= a_1\exp(-b_1r^{m_1})+a_2\exp(-b_2r^{m_2})$. The tail is added in order to obtain the best possible description
of the data sets not only for the core of the potential, but also for all distances and down to energies much lower than $k_{\rm B}T$. In general, the interactions 
between the linear chains with reversible bonds can be described by Gaussian functions (even without needing the tail), whereas exponents $m_i > 2$ 
are needed for ring-ring and ring-linear interactions. The consequences of the shape of the effective potentials on the structure of the effective fluids will be discussed in the next Section.

\ref{fig:effpot} reveals several trends. The potentials (both with intermolecular bonding switched on and off) are more repulsive for ring-ring than for linear-linear interactions, the linear-ring case being intermediate between the former two. This is consistent with the findings in the linear and ring precursors (i.e, in total the absence of both intra- and intermolecular bonding), and reveals that the topological interaction is again relevant. As can be seen in panels (a) and (b), if only intramolecular bonding is allowed (empty symbols),  the amplitudes of the potentials are systematically higher 
than in their respective precursors (about $2.5k_{\rm B}T$ and $6k_{\rm B}T$ for linear-linear and ring-ring precursors, respectively \cite{Narros2010}). This result is not surprising because the presence of intramolecular loops, even if they are transient, enhances steric hindrance and topological constraints, and creates higher effective barriers for interpenetration than in the respective precursors.

As can be seen, the effective potential becomes systematically stronger, with variations of about $30-50\%$ in its amplitude, by moving from 
the `gap 1' to the periodic sequence of the reactive groups. As mentioned before, increasing the distance between consecutive reactive groups promotes the formation of longer intramolecular loops and reduces the molecular size. 
This hinders interpenetration and leads to stronger effective repulsions. 
For a fixed value of $n_{\rm min}$ the specific sequence of reactive groups has some small, but visible effect on the effective potential (see. e.g., data for the two couples `gap 4' in \ref{fig:effpot}(a)).  \ref{fig:effpot} shows that when intermolecular bonding is switched on (filled symbols) the effective potentials experience a marked reduction with respect to the case of pure intramolecular bonding. Interestingly, the effect of the sequence of reactive sites when intermolecular bonds are allowed is the  opposite to that found when they are not: increasing the distance between consecutive reactive sites decreases the effective interaction. As a consequence, the periodic sequences of reactive groups lead to the strongest reductions of the effective potential when intermolecular bonding is switched on (with differences of 
$\Delta \beta V_{\rm eff}(r=0) \sim -3$ and $-5$ for linear-linear and ring-ring interactions. respectively).

\begin{figure}[ht!]
	\centering
	\includegraphics[width=0.5\textwidth]{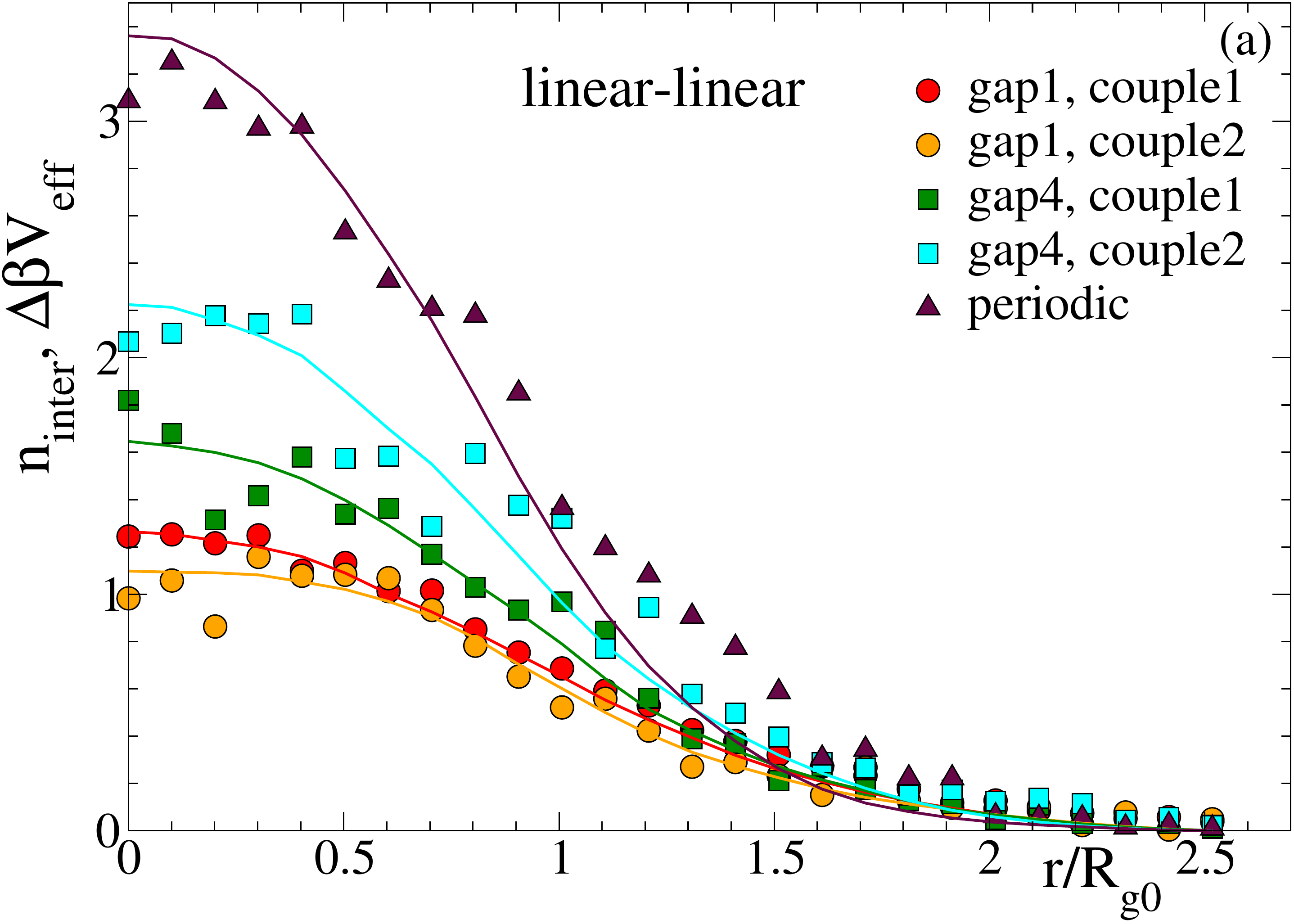}
	\qquad
	\includegraphics[width=0.5\textwidth]{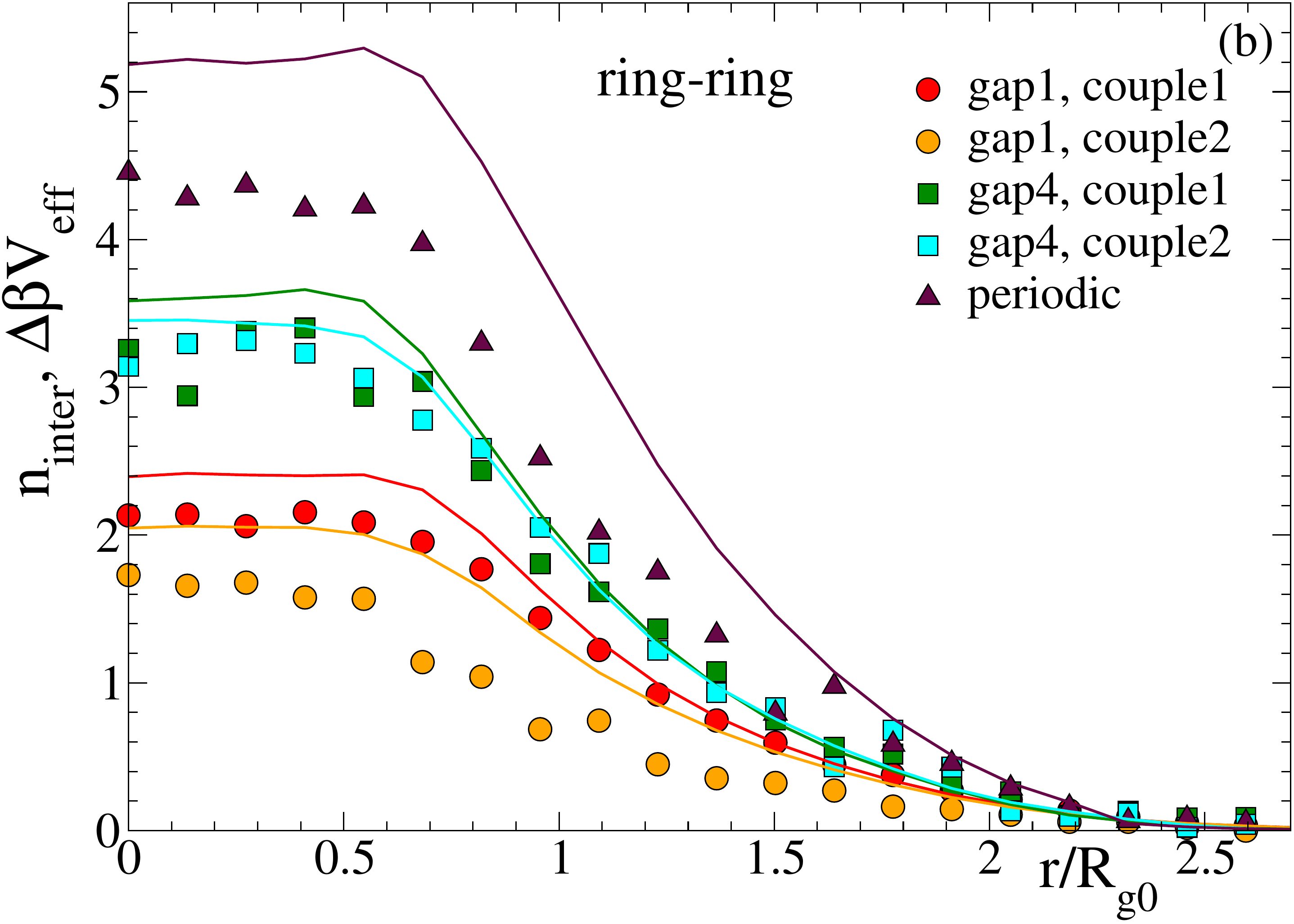}
	\qquad
	\includegraphics[width=0.5\textwidth]{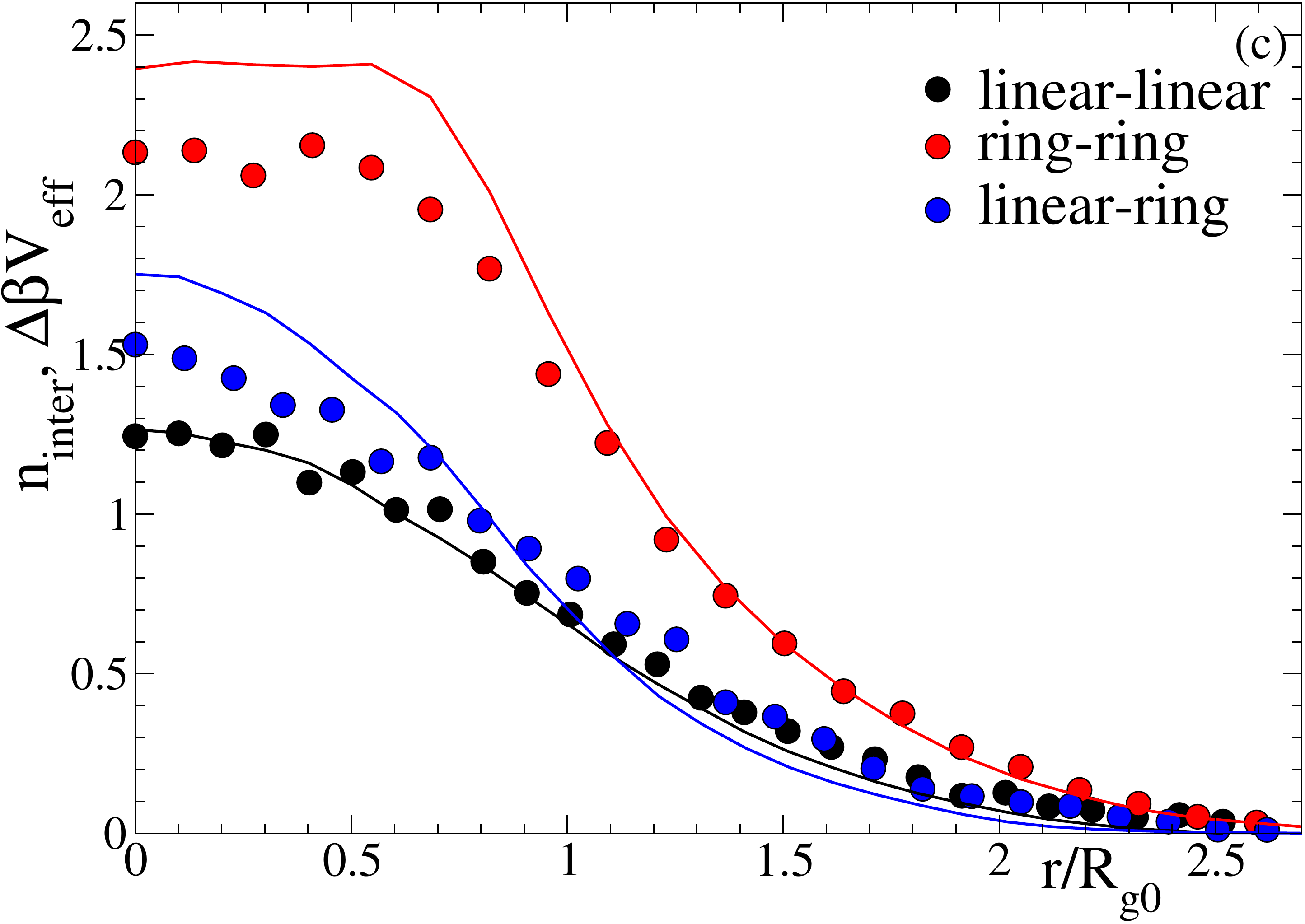}
	\caption{Symbols: as \ref{fig:effpot} for the number of intermolecular bonds vs. the distance between centers-of-mass.
	Lines (same color codes as symbols): difference between the effective potentials without and with intermolecular bonding.}
\label{fig:nbond}
\end{figure}

The analysis of the number of bonds can shed some light on the origin of
the former trends for the effective potentials. The average total number of bonds (intra- and intermolecular) 
is $n_{\rm tot} \approx 17$ in all systems, i.e., 85\% of the maximum $n_{\rm tot}=20$ that would correspond to the
fully bonded state. No differences are found within statistics and this observation 
is independent on the topology of the precursor, the sequence of reactive sites, the distance between the centers-of-mass and on intermolecular bonding being switched on or off. Although the total number of bonds is unaffected, varying the former parameters leads to a different balance between intra- and intermolecular bonds. 
\ref{fig:nbond} shows, for the cases of \ref{fig:effpot} (same symbol codes) the variation of the number of intermolecular bonds, $n_{\rm inter}$, with the distance between the centers-of-mass of the two polymers.
The number of intermolecular bonds increases by moving from gap1 to periodic sequences, i.e., by increasing the distance between consecutive reactive sites. Since increasing such a distance eliminates the shortest intramolecular loops, 
the observed conservation of the average total number of bonds is achieved by exchanging the shortest loops 
by longer ones or by forming more intermolecular bonds.
The second option is preferred, as shown by \ref{fig:nbond}. Figure S8 in the SI shows the distribution of instantaenous values
of $n_{\rm inter}$ at distance $r_{1,2}= 3\sigma$. As can be seen $n_{\rm inter}$ can fluctuate in a broad range from zero to 8-12 bonds,
and the distribution becomes more symmetric with decreasing randomness of the sequence of reactive sites.

Since the effective potential $V_{\rm eff}(r)$ is the free energy cost of changing the mutual distance from infinity to $r$, the difference
between the effective potentials without and with intermolecular bonding is 
$\Delta \beta V_{\rm eff}(r) = \beta V_{\rm eff, only\ intra}(r)- \beta V_{\rm eff, all\ bonds}(r) = \beta\Delta U(r)- k_{\rm B}^{-1}\Delta S(r)$,
with $\Delta U(r)$ and $\Delta S(r)$ the corresponding energetic and entropic changes.
For the same pair of polymers, switching intermolecular bonding on or off should not change excluded volume interactions significantly, 
and as mentioned before, it does not affect the total number of bonds. Therefore, $\Delta U(r) \approx 0$ and the difference between 
the effective potentials without and with intermolecular bonding is essentially of entropic origin, i.e.,
$\Delta \beta V_{\rm eff}(r) \approx  - k_{\rm B}^{-1}\Delta S(r)$. 
\ref{fig:nbond} shows (lines) the corresponding data for $\Delta \beta V_{\rm eff}(r)$.  Since this quantity is positive for all distances,
it is clear that forming intermolecular bonds involves an entropic gain with respect to the only intramolecularly bonded system. 
In principle intermolecular bonds limit conformational and translational fluctuations, leading to an entropic loss. 
Therefore there should be a source of entropic gain that exceeds the former loss, resulting in a net entropic gain when intermolecular bonds are formed.
As can be seen in \ref{fig:nbond} the net entropic gain is qualitatively given by $k_{\rm B}$ 
times the number of intermolecular bonds, i.e., the number of additional states of the pair of polymers
that are introduced by intermolecular bonding is essentially the exponential of the number of intermolecular bonds.

The mechanism leading to the observed entropic gain is not clear. The concept of combinatorial entropy \cite{SciortinoReview2019}, 
accounting for the different connectivities of the bonding network,  has been invoked to accurately describe a similar
effect in the case of hard nanoparticles grafted by chains with sticky ends. An expression has been proposed for the number of bonding 
patterns that can be produced by the sticky ends that can, at each distance, potentially bind to the other nanoparticle \cite{Sciortino2020ACSNano}. 
Though it is plausible that the combinatorial entropy is  a major
contribution to the $\Delta \beta V_{\rm eff}$ shown in \ref{fig:nbond}, obtaining an analytical 
accurate expression for our system is highly non-trivial \cite{footnotecomb} and is beyond the scope of this work.

\section{Crowded solutions and phase behavior}\label{sec4}

The main motivation behind the coarse-graining approach is to reduce as much as possible the degrees of freedom that define the system. 
Deriving the expression of an effective potential $V_{\rm eff}$ able to mimic the interactions between macromolecules enables
 the description of them only in terms of a few coordinates
(usually the centers-of-mass). Thus, in a dense system as a crowded solution the degrees of freedom associated to the individual monomers are wiped out 
and the whole solution is effectively described
as a fluid of particles interacting through the obtained $V_{\rm eff}$. This strategy largely reduces the computational 
cost of the all-monomer simulations, allows to investigate longer time and length scales and facilitates the applications of methods from, e.g., liquid state theory.
However, it involves a strong assumption, namely, since the effective potential has been derived for two polymers in the absence of others,
its use implicitly neglects the effective many-body interactions in the crowded solution. 
In general, this approximation is justified and works well for densities below the overlap concentration,
but it fails, even severely, as one goes deep in the semidilute and concentrated regimes \cite{Likos2001}. 
A well-known effect of the many-body interactions in dense solutions is the shrinkage found in simple linear chains, 
leading to the change from self-avoiding to Gaussian chain statistics \cite{RubinsteinColby}.

\begin{figure}[htbp!]
	\begin{minipage}[c]{.54\textwidth}
		\setlength{\captionmargin}{0pt}%
		\includegraphics[width=0.9\textwidth]{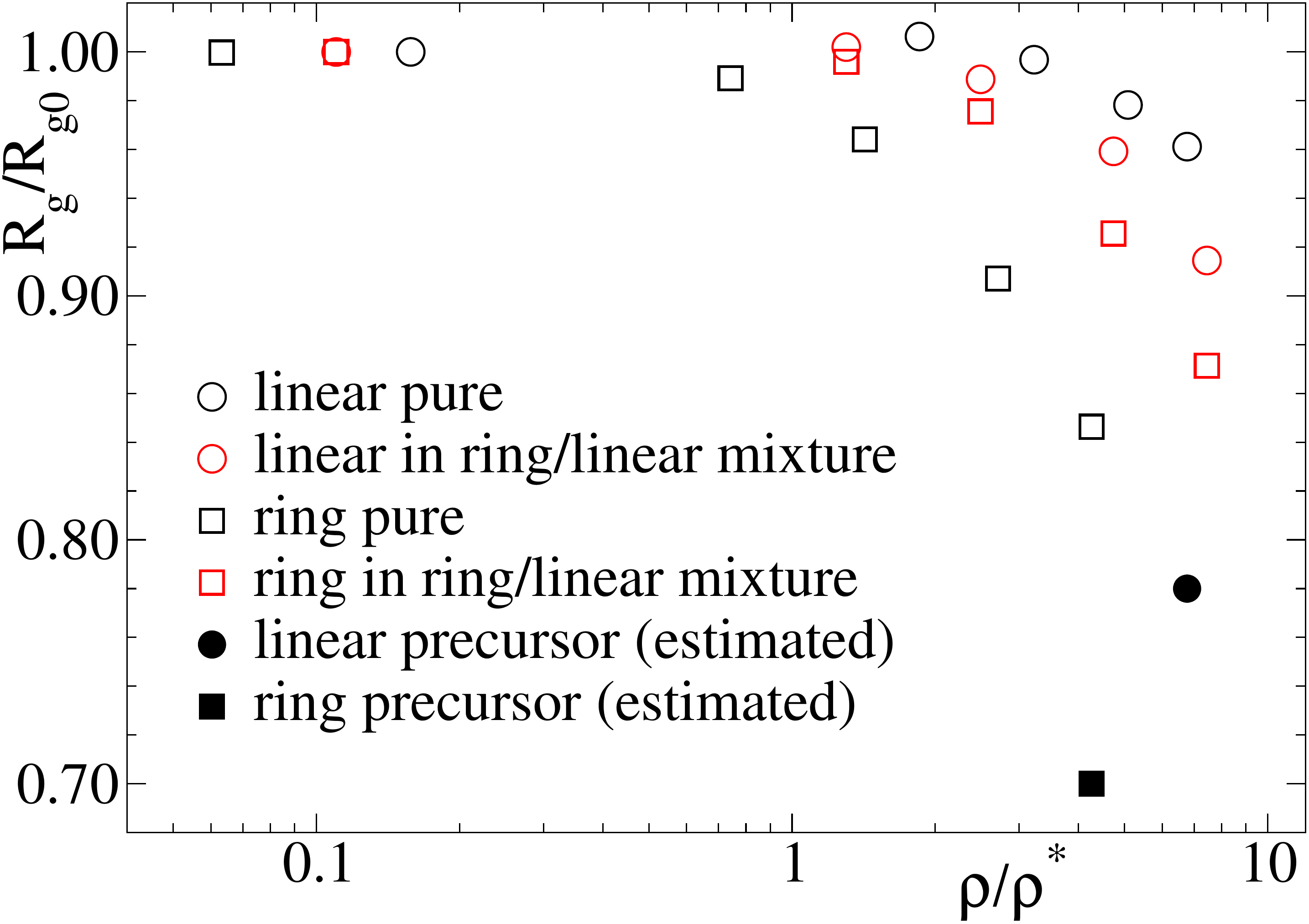}
	\end{minipage}%
\caption{Radius of gyration $R_{\rm g}$ normalized by its value at high dilution $R_{\rm g0}$ as a function of the effective density, for the pure solutions  of linear chains  and rings  with reversible bonds and for the 50/50 mixture of both polymers. For comparison we add
the values for the linear and ring precursors (no bonding) estimated at the highest simulated concentrations 
of their bonded counterparts (see text for explanation).}
\label{fig:rg}
\end{figure}

Recent simulations of solutions of reversibly cross-linking
linear chains similar to those investigated here have shown, interestingly, that the polymer size 
and shape are weakly affected by the concentration, essentially retaining
the conformational properties of high dilution \cite{Formanek2021}. 
Instead of shrinking, the chains keep such mean conformations
through forming a few intermolecular bonds with their neighbors.
This weak effect of the concentration on the molecular conformations suggests that the many-body interactions
experienced by a tagged couple
of chains are in a first approximation given by a flat energy landscape.
In such conditions the effective potential derived at high dilution may provide a good description
of the structural properties of the solution even far above the overlap concentration.

\ref{fig:rg} shows the radii of gyration, normalized by their values at $\rho=0$, as a function of the normalized 
concentration $\rho/\rho^\star$ for the linear chains and rings with reversible bonds, both in the pure systems and in the linear/ring mixture. The results for the pure linear case confirm those of the model of Ref.~\citen{Formanek2021}, 
with a shrinkage of just 4\% at  about 7 times the overlap concentration. A much steeper dependence on the concentration is found for the case of rings, with a shrinkage of 15\% at the highest simulated concentration of about 4 times the overlap concentration (for comparison at the same effective density the shrinkage of the linear chains is less than 2\%). 
This very different response of the molecular size of linear chains and rings to crowding is also found in the mixture of both molecules, though differences are less pronounced than in the pure systems. In the mixture the size of the linear chains shows a steeper dependence on the concentration than in the pure system, whereas the rings show the opposite effect.  
Having said this, in all cases the shrinkage is much weaker that in the absence of bonding. 
Solid symbols in \ref{fig:rg} are the values for the unbonded precursors at the highest effective densities of the bonded counterparts. 
Such values have been estimated through the power-laws $R_{\rm g}/R_{\rm g0} \sim (\rho/\rho^\star)^{-1/8}$ (linear chains \cite{RubinsteinColby}) and $R_{\rm g}/R_{\rm g0} \sim (\rho/\rho^\star)^{-1/4}$ (unentangled rings \cite{Narros2014}). 
Shrinkage factors of 22\% (linear) and 30\% (ring) vs. the respective aforementioned values of 4\% and 15\% are obtained, demonstrating the dramatic effect of intermolecular bonding on reducing the impact of crowding on the molecular conformations.

\begin{figure}[htbp!]
\setlength{\captionmargin}{0pt}	\includegraphics[width=0.49\textwidth]{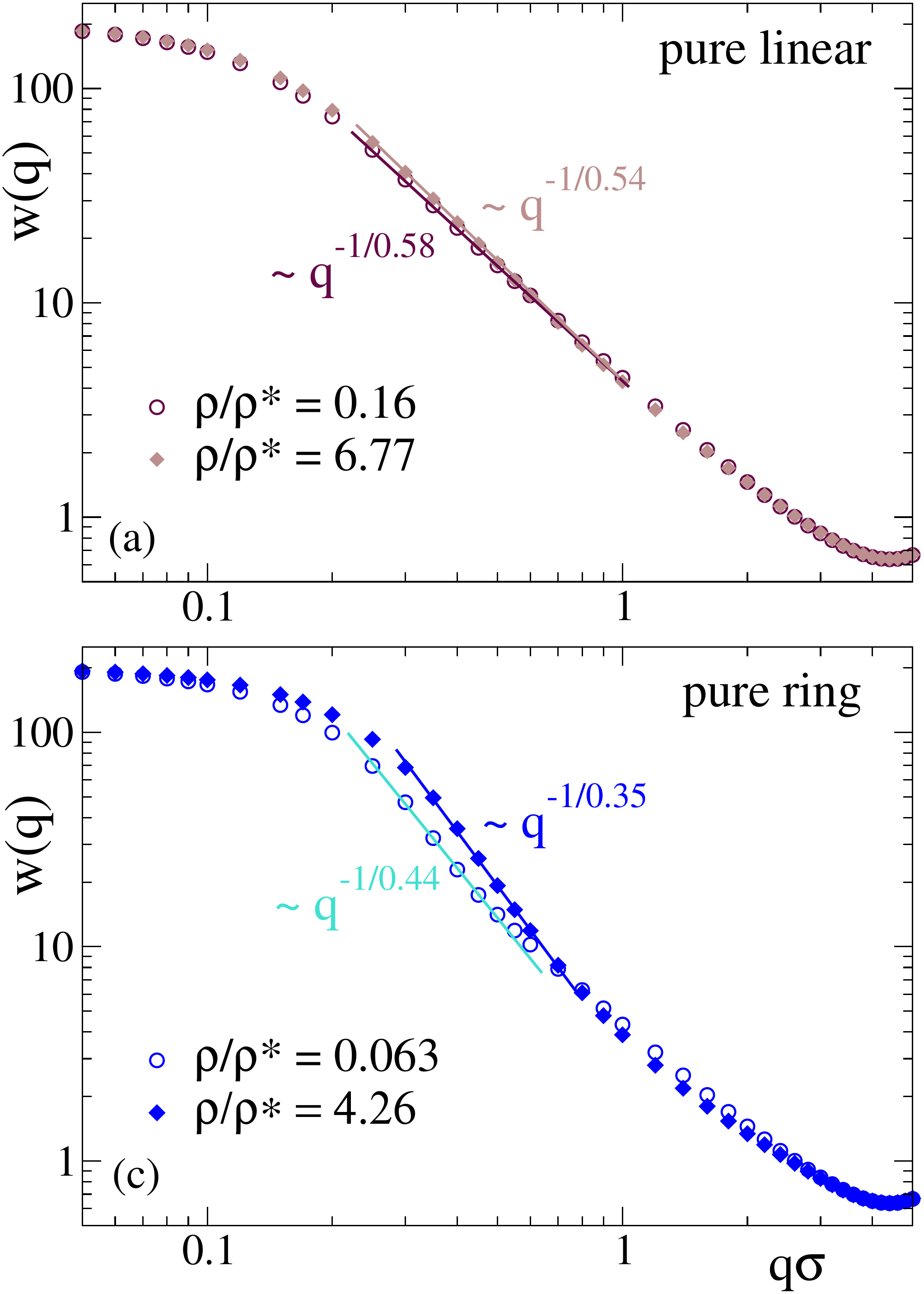}\hspace{0.1cm}\includegraphics[width=0.49\textwidth]{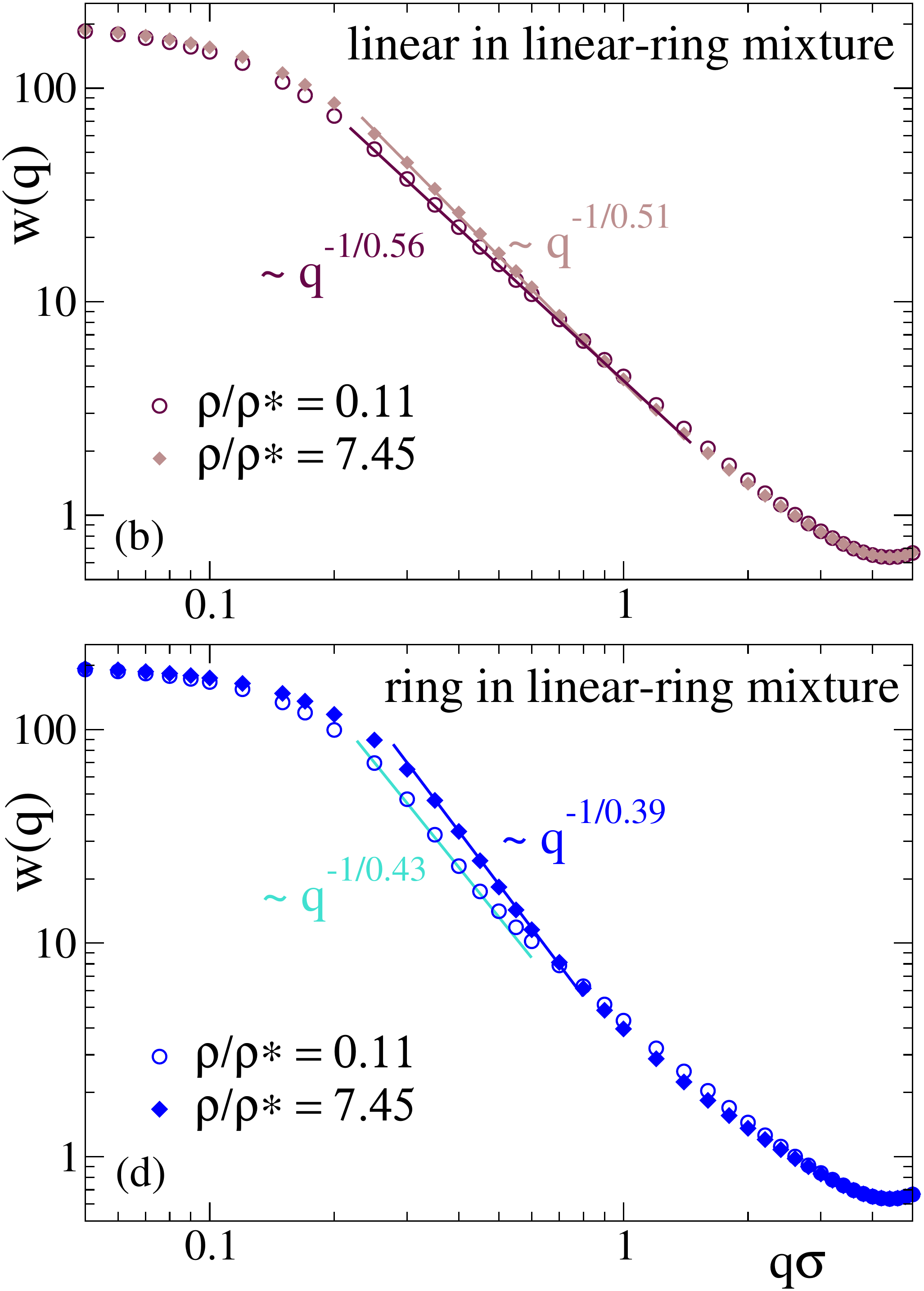}
\caption{Form factors for linear chains (a,b) and rings (c,d) with reversible bonds at two densities far below and far above the
overlap concentration. Panels (a,c) and (b,d) correspond to the pure systems and to the mixture, respectively. 
Lines are fits in the fractal regime to power laws of the form $w(q) \sim q^{-1/\nu}$.}
\label{fig:wq}
\end{figure}

Beyond the effect of the concentration on the molecular size, the scattering form factor provides more detailed information on the molecular conformations. 
The form factor is calculated as
\begin{equation}
w(q) =  \left \langle \dfrac{1}{N_{\rm m}} \sum_{j,k} \dfrac{\sin(qr_{jk})}{qr_{jk}}\right \rangle ,
\label{eq:wq}
\end{equation}
where $r_{jk}= |{\bf r}_i -{\bf r}_j|$, the sum is performed over all pairs of monomers $j,k$ belonging to the same polymer 
and is averaged over all the polymers in the solution and different configurations. 
\ref{fig:wq} shows for both linear and ring architectures the form factor  at high dilution and at the highest investigated concentration.
As $q$ grows, the form factor shows the crossover from the limit $w(q=0)= N_{\rm m}$
to the fractal regime \cite{RubinsteinColby}, $w(q) \sim q^{-1/\nu}$, which originates from the scaling of the intramolecular distances with the contour length. 
Similar to the overall molecular size, we find a tiny effect of crowding on the effective exponent of the linear chains, which changes from $\nu = 0.58$ to 0.54 from high dilution to $\rho/\rho^{\star} \sim 7$, i.e., a narrow range between the  the Flory value ($\nu =0.59$) for self-avoiding chains and $\nu= 1/2$ for Gaussian chains.
The more pronounced effect of crowding on the molecular size of rings is also reflected in the scaling behavior, with a change from $\nu =0.44$ at high dilution to $\nu =0.35$ at $\rho/\rho^{\star} \sim 4$, resembling crumpled globule behavior \cite{Halverson2011,Halverson2014} ($\nu = 1/3$).
Similar trends are found in the 50/50 mixture of linear chains and rings. Consistently with the observations for the molecular size, the conformations of the linear chains in the mixture are slightly more affected by crowding than in the pure system, and the opposite effect is found for the rings. 

\begin{table}[h!]
	\centering
	\begin{tabular}{|c | c|} 
	       \hline 
		& \\
		linear-linear, all bonds  &  $\beta V_{\rm eff}(r) = 1.57e^{-(\frac{r}{1.04R_{\rm g0 }})^2} $ \\[0.5ex]		
		\hline
		 & \\ 
		ring - ring, all bonds  &   $\beta V_{\rm eff}(r) = 1.87e^{-(\frac{r}{1.009R_{\rm g0}})^{3.997}} + 2.35 e^{-(\frac{r}{1.48R_{\rm g0}})^{3.105}} $ \\[0.5ex]   
		\hline
		& \\
				linear - ring, all bonds & $\beta V_{\rm eff}(r) = 1.66 e^{-(\frac{r}{1.005R_{\rm g0}})^{2.35}} + 0.72 e^{-(\frac{r}{1.46R_{\rm g0}})^{2.31}} $ \\[0.5ex]  
	    \hline
	    & \\
	    		linear - linear, only intra & $\beta V_{\rm eff}(r) = 2.32 e^{-(\frac{r}{1.18R_{\rm g0}})^{2}} + 0.56 e^{-(\frac{r}{0.97R_{\rm g0}})^{3.012}} $ \\ [0.5ex] 
	    		\hline
	\end{tabular}
	\caption{Effective potentials used in the effective fluids (see main text for explanation). As mentioned before, $R_{\rm g0}$
	is the radius of gyration of the isolated polymer, and in the case of the linear-ring interaction we use the average of the
	respective $R_{\rm g0}$'s of the isolated linear and ring polymers.}
	\label{table:1}
\end{table}

In summary, \ref{fig:rg} and \ref{fig:wq} show that the typical conformations of the linear chains are weakly  
distorted by crowding and hence the two-body approximation under which the effective potential is derived might work 
reasonably even at unusually high densities, far above the overlap concentration. Comparatively, crowding has a stronger 
effect on  the conformations of the rings, and their effective potential 
is expected to work in a narrower range of concentrations than in their linear counterparts.
In what follows we test these expectations by comparing the results for the all-monomer solutions with those for the corresponding effective fluids. Moreover, we test the validity of mean-field DFT in our systems through calculations from test particle route (TPR). As mentioned in \ref{sec2}, all the simulated solutions corresponds to sequences of type `gap1'. The interactions in the all-monomer simulations are given by \ref{eq:WCA},\ref{eq:fene},\ref{eq:SS},\ref{eq:3BODY}. The data for the corresponding effective potentials of \ref{fig:effpot} were fitted by the functions of \ref{table:1}, and  these functions were used in the simulations and TPR calculations of the effective fluids. Namely, the `linear-linear, all bonds' and the `ring-ring, all bonds' potentials were used in the effective fluids of the pure (one-component) systems of linear and ring polymers with reversible bonds. They were also used for the linear-linear and ring-ring interactions in the effective linear-ring mixture, while the `linear-ring, all bonds'
potential was used for the linear-ring interactions. In the mixture (A/B) of linear chains with orthogonal chemistry, the `linear-linear, all bonds' potential was used for the A-A and B-B interactions. Since by construction there were no intermolecular A-B bonds in the all-monomer simulations, the `linear-linear, only intra' potential was used for the A-B interactions in the effective fluid.




\begin{figure}[htbp!]
	\begin{minipage}[c]{.46\textwidth}
		\setlength{\captionmargin}{0pt}%
		\includegraphics[width=1.05\textwidth]{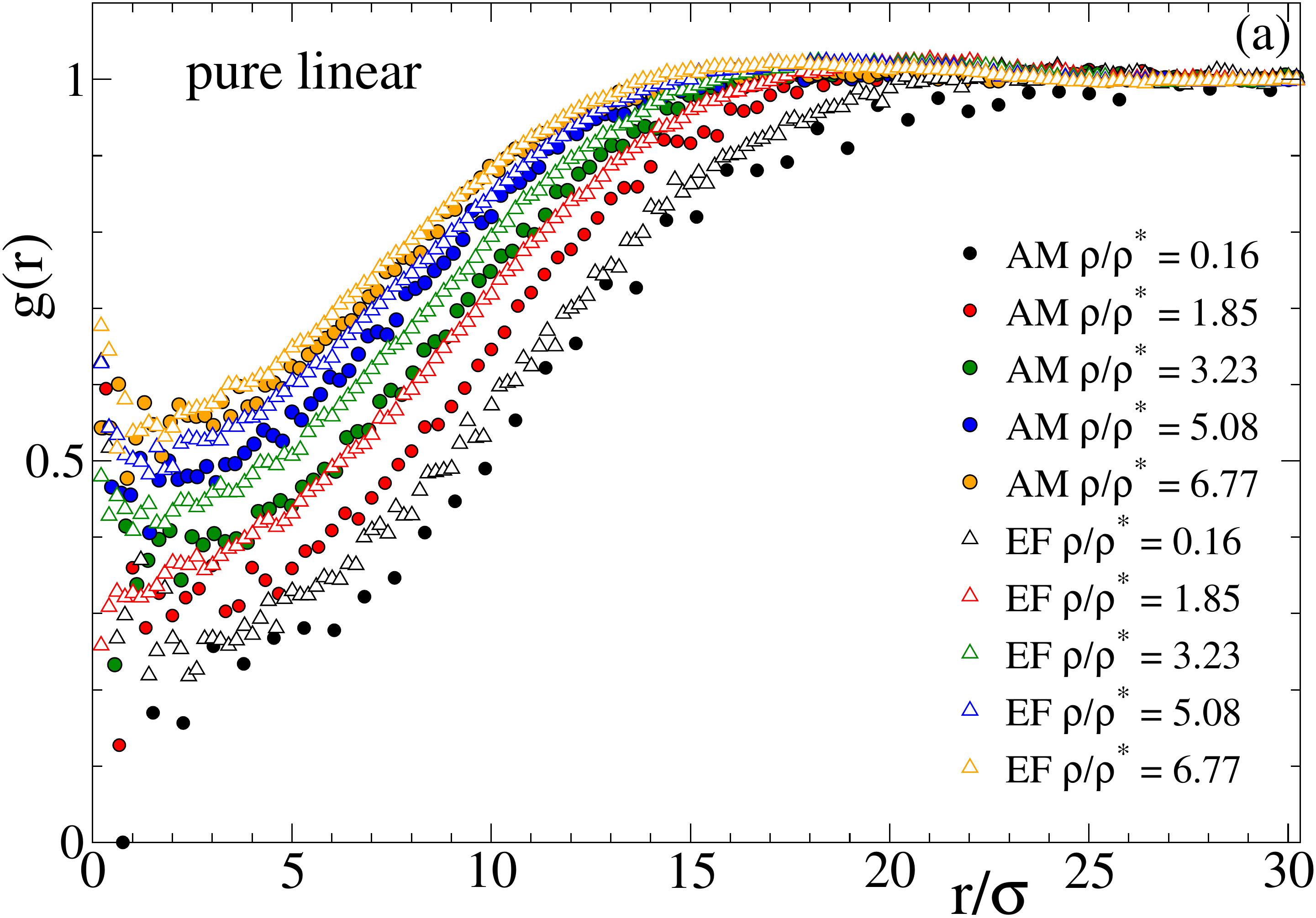}
	\end{minipage}%
	\hspace{5mm}%
	\begin{minipage}[c]{.46\textwidth}
		\setlength{\captionmargin}{0pt}%
		\includegraphics[width=1.05\textwidth]{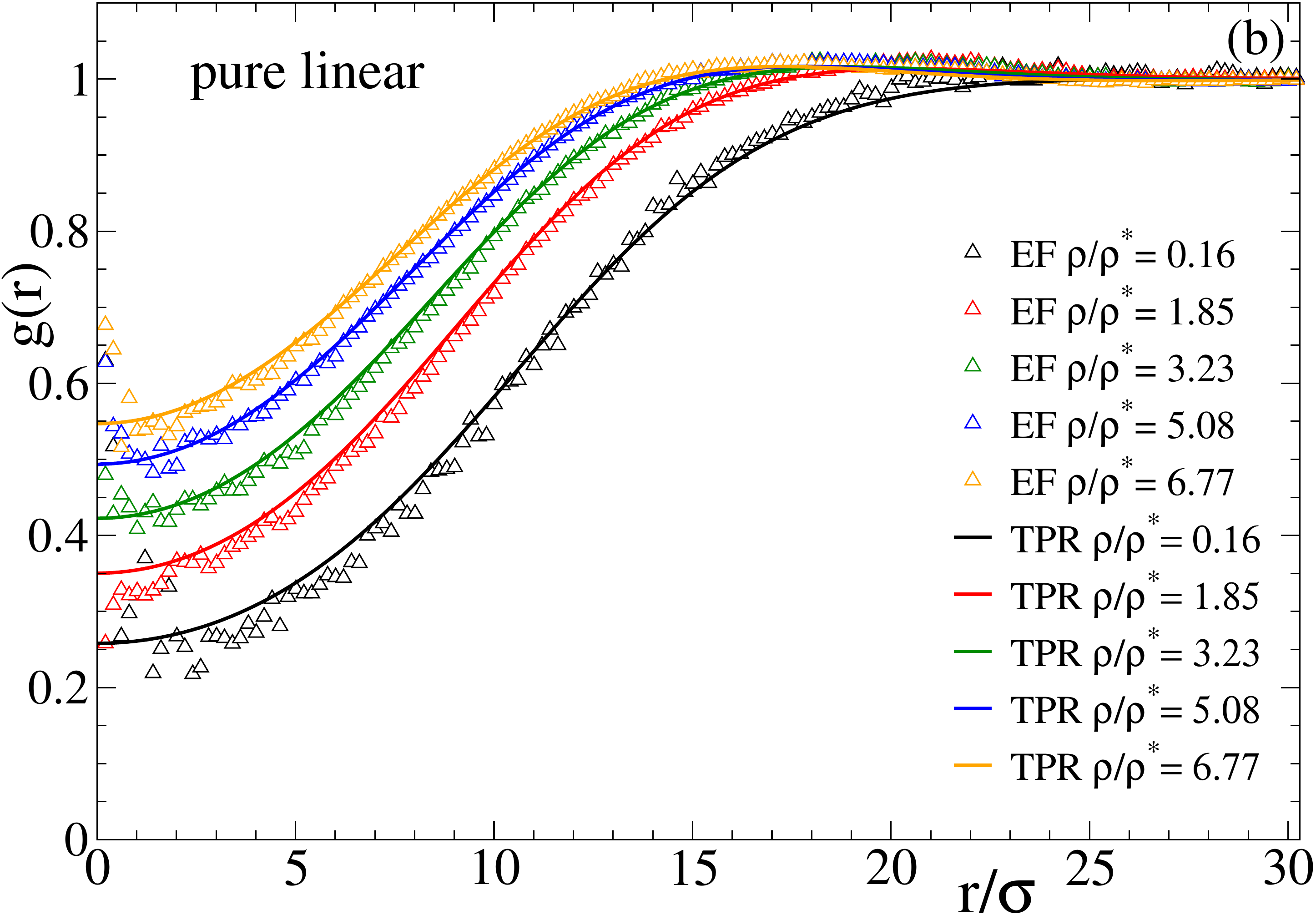}
	\end{minipage}		
	\caption{Radial distribution function of solutions of linear chains with reversible bonds, in a broad range of densities from high dilution
	to far above the overlap concentration. Panel (a) compares the results for the molecular centers-of-mass in the all-monomer simulations (AM, full symbols) with
	the results for the particles of the effective fluid simulations (EF, empty symbols). Panel (b) compares the EF simulations with the 
	theoretical predictions of the test particle route (TPR, lines).}
	\label{fig:grlin}
\end{figure}

\ref{fig:grlin} shows the radial distribution function $g(r)$ of the centers-of-mass in the pure solutions of linear chains with reversible bonds. 
Panel (a) compares the correlations for the centers-of-mass of the real all-monomer (AM) system with those for the particles of the effective fluid (EF).
Panel (b) compares the results for the effective fluid with the calculations from TPR. 
An excellent agreement between effective fluid and TPR is obtained,
demonstrating the validity of the mean-field approximation for the effective fluid even at low densities. 
The comparison between the all-monomer and effective fluid reveals some interesting trends.
Contrary to the usual observations in macromolecular systems, the effective potential provides a very good description of the real system at $\rho/\rho^\star >5$, i.e, far above the overlap concentration,  where many-body effects are usually expected. This finding confirms
that the many-body effects are basically averaged out and lead to a flat energy landscape. Again contrary to the usual observations,
there are systematic differences between the all-monomer and effective fluid at densities below the overlap concentration,
even at values as low as $\rho/\rho^\star \sim 0.1$, for which one might expect an excellent accuracy of the two-body approximation. As can be seen in panel (a), the $g(r)$ for the all-monomer system is shifted to longer distances, indicating less interpenetration than prediced by the effective fluid. The reason for this small but significant disagreement is likely the significant number of clusters of three polymers found at low concentrations in the real system. Figure S9 in the SI shows the cluster size distribution $P(n)$ at the lowest investigated concentration, where $n$ is the number of polymers
in a cluster and two polymers belong to a same cluster if they are mutually linked by at least one intermolecular bond. As can be seen the ratio of clusters of $n=3$ vs. those of $n=2$ is non-neglible (about 0.1).
In these clusters (which do not exist in simple systems with no bonds) the 3-body interaction cannot be oversimplified by a flat landscape, and the two-body approximation just gives a semiquantitative description of the static correlations. 


\begin{figure}[htbp!]
	\begin{minipage}[c]{.46\textwidth}
		\setlength{\captionmargin}{0pt}%
		\includegraphics[width=1.05\textwidth]{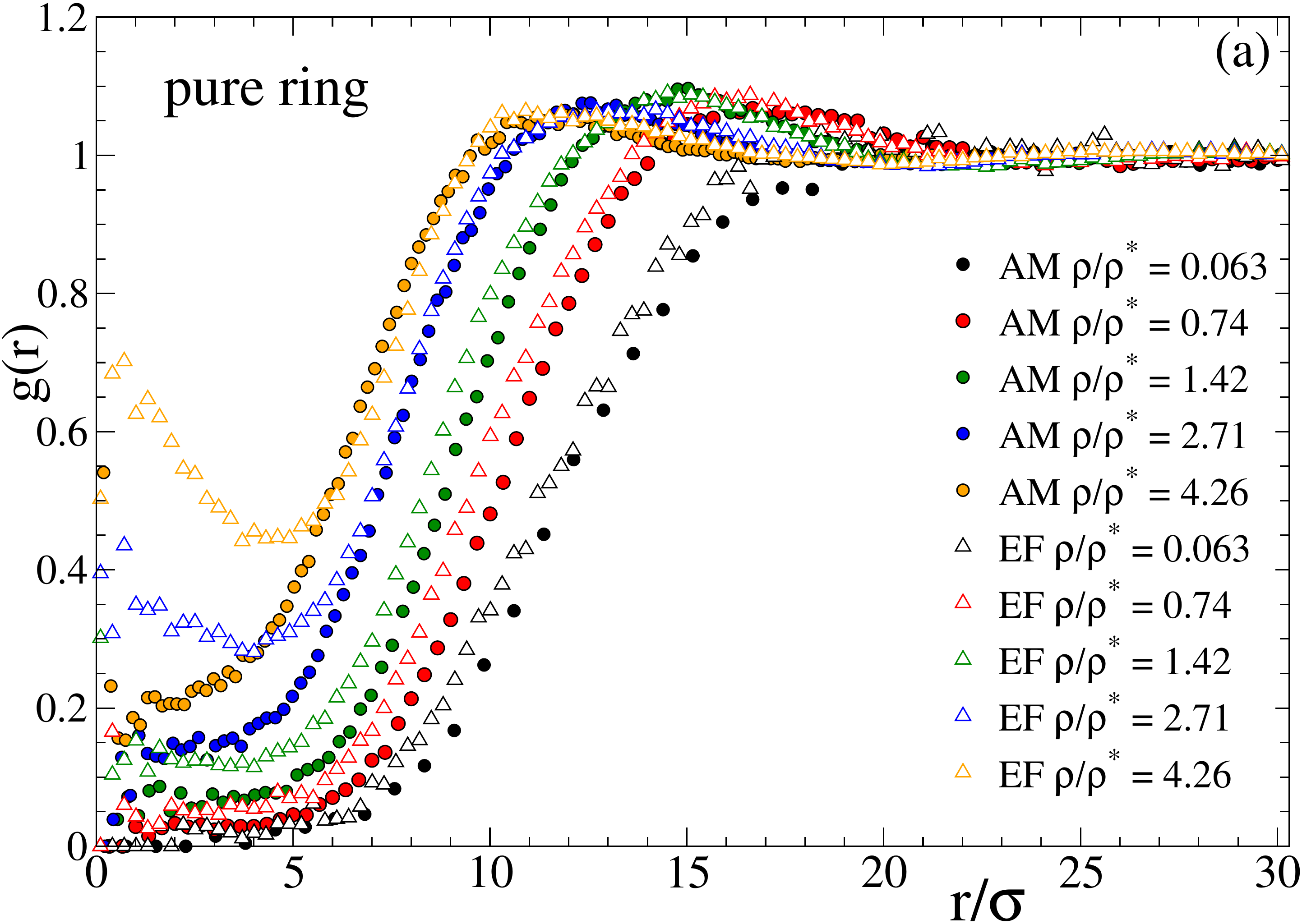}
	\end{minipage}%
	\hspace{5mm}%
	\begin{minipage}[c]{.46\textwidth}
		\setlength{\captionmargin}{0pt}%
		\includegraphics[width=1.05\textwidth]{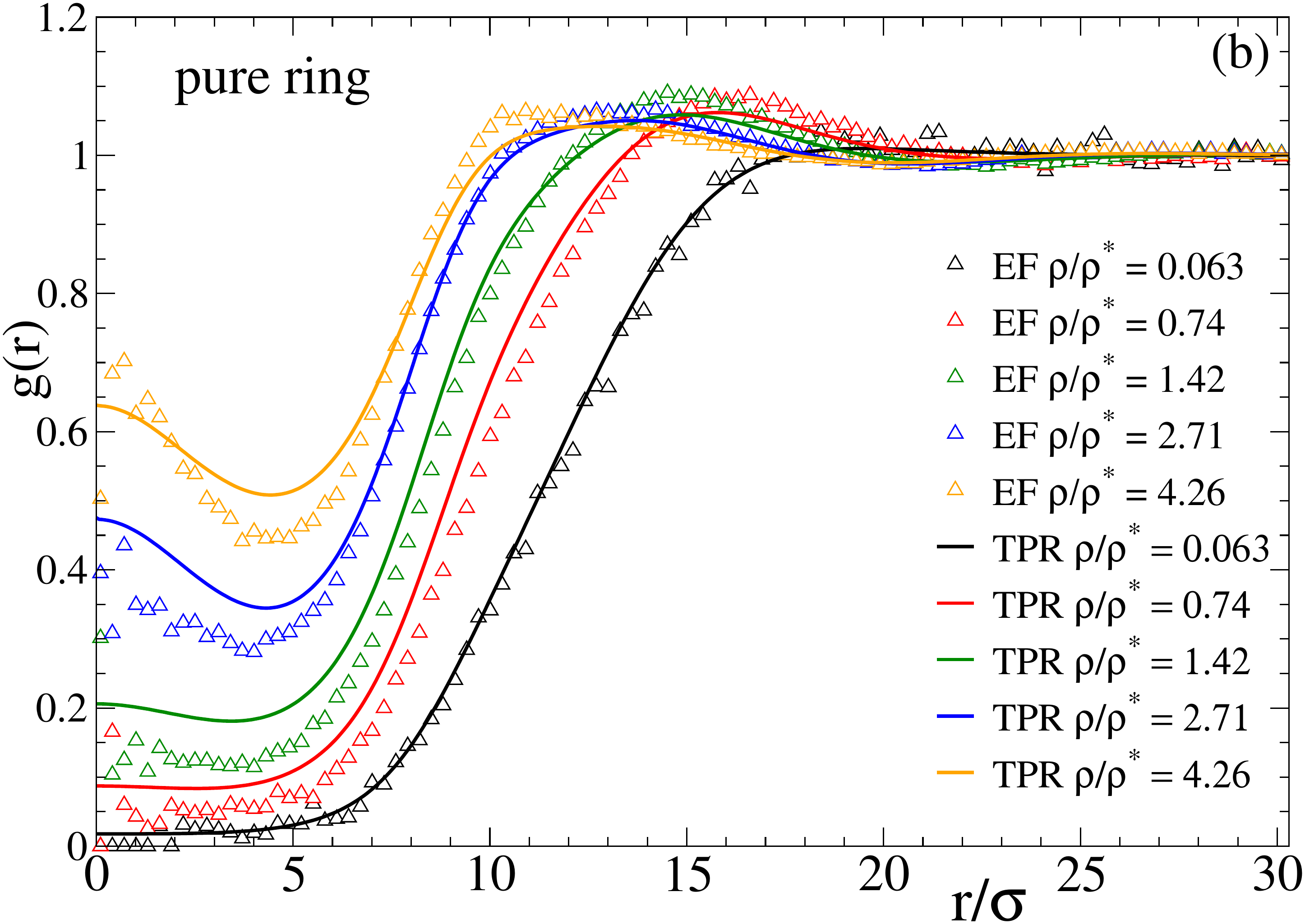}
	\end{minipage}		
	\caption{As \ref{fig:grlin} for the solutions of rings with reversible bonds.}
	\label{fig:grring}
\end{figure}

Results for the solutions of rings with reversible bonds are shown in \ref{fig:grring}. In comparison with the linear case, the all-monomer rings show a larger correlation hole and therefore a weaker interpenetration. This is consistent with the observed
stronger response of their  conformations to crowding (\ref{fig:rg} and \ref{fig:wq}), which leads to objects similar to crumpled globules ($\nu \sim 1/3$) and therefore less penetrable than their linear counterparts ($\nu \sim 0.5$). Although at low concentrations there is still a systematic small disagreement
between the $g(r)$ of the effective fluid and the all-monomer system, this effect is weaker than for the linear counterparts. This is consistent with the smaller number of 3-body clusters found for the rings (Figure~S9 in the SI). In this case the ratio of $n=3$ vs. $n=2$-clusters is about 0.05. For concentrations higher than $\rho/\rho^{\star}$ the effective fluid provides a much worse description than in the 
linear system, and indeed the all-monomer solution of rings does not show the peak at $r=0$ found in the effective fluid. 
In a similar fashion to the simple case of rings without bonds, the peak formed at $r=0$ and growing with the concentration 
is the signature of a fluid of clusters formed by strongly interpenetrated particles. The effective fluid will ultimately show a transition to a cluster crystal phase, where the clusters are arranged in the nodes of a regular lattice 
that is sustained through incesant hopping of the particles between the clusters.
The existence of cluster crystal phases is predicted within mean field DFT for potentials that are bounded and show  
negative values in their Fourier transform \cite{Likos2001PRE}.
Both conditions are fulfilled by the effective potentials of the rings with reversible bonds. Indeed they can be described by generalized exponential functions (\ref{table:1}),
which for exponents higher than 2 have negative Fourier components \cite{LIKOS2007}. Moreover the mean field approximation is justified, 
as can be seen in \ref{fig:grring}b by the good agreement between the TPR and
the simulations of the effective fluid. However, the cluster fluid is not found in the all-monomer system. 
As found for simple rings without bonding interactions \cite{Narros2010}, the preferred crumpled globular conformations prevent 
the degree of nesting and threading needed to form the characteristic peak at $r=0$.

\begin{figure}[htbp]
	\begin{minipage}[c]{.46\textwidth}
		\setlength{\captionmargin}{0pt}%
		\includegraphics[width=1.06\textwidth]{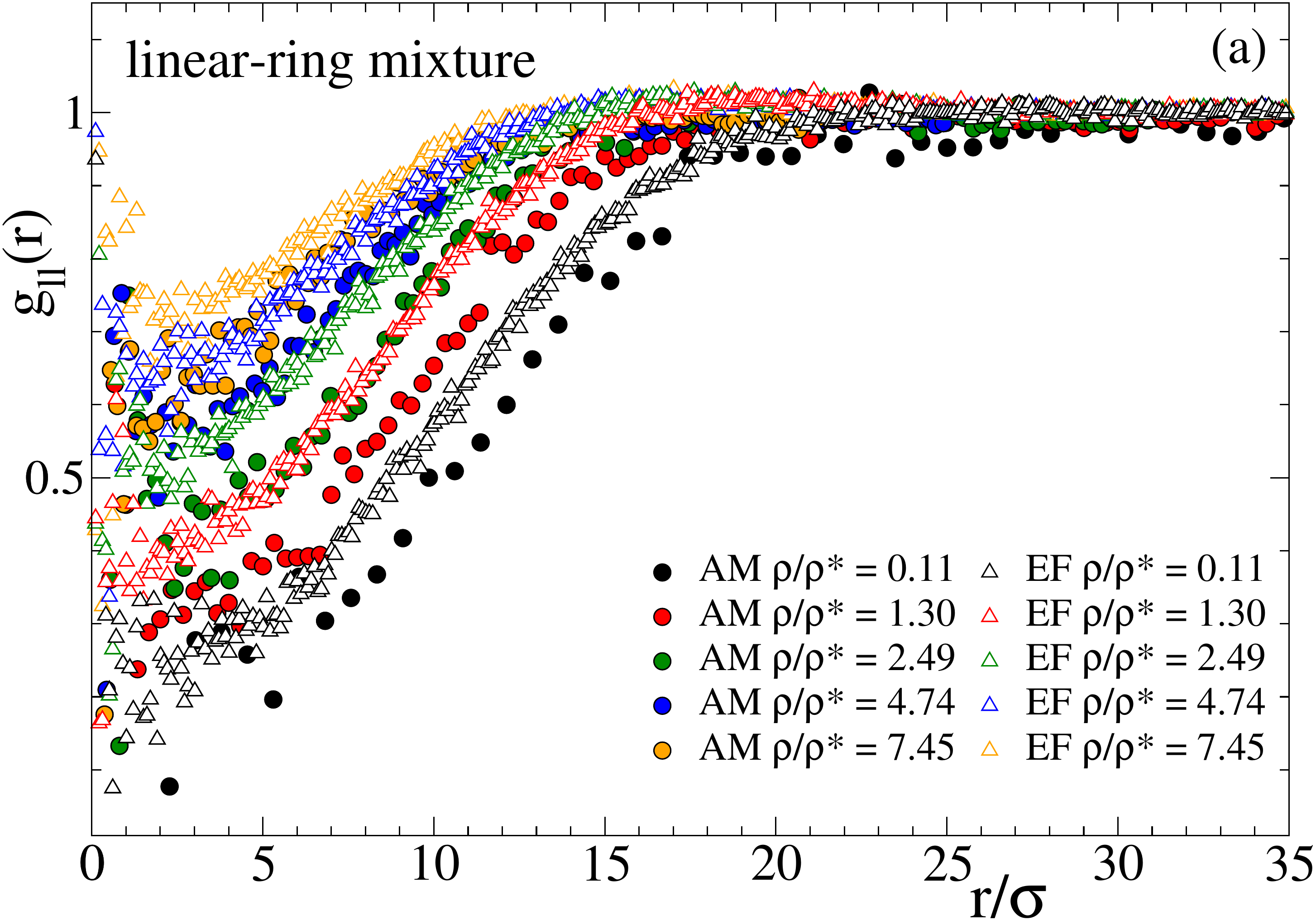}
		\end{minipage}%
		\hspace{5mm}%
		\begin{minipage}[c]{.46\textwidth}
			\setlength{\captionmargin}{0pt}%
			\includegraphics[width=1.06\textwidth]{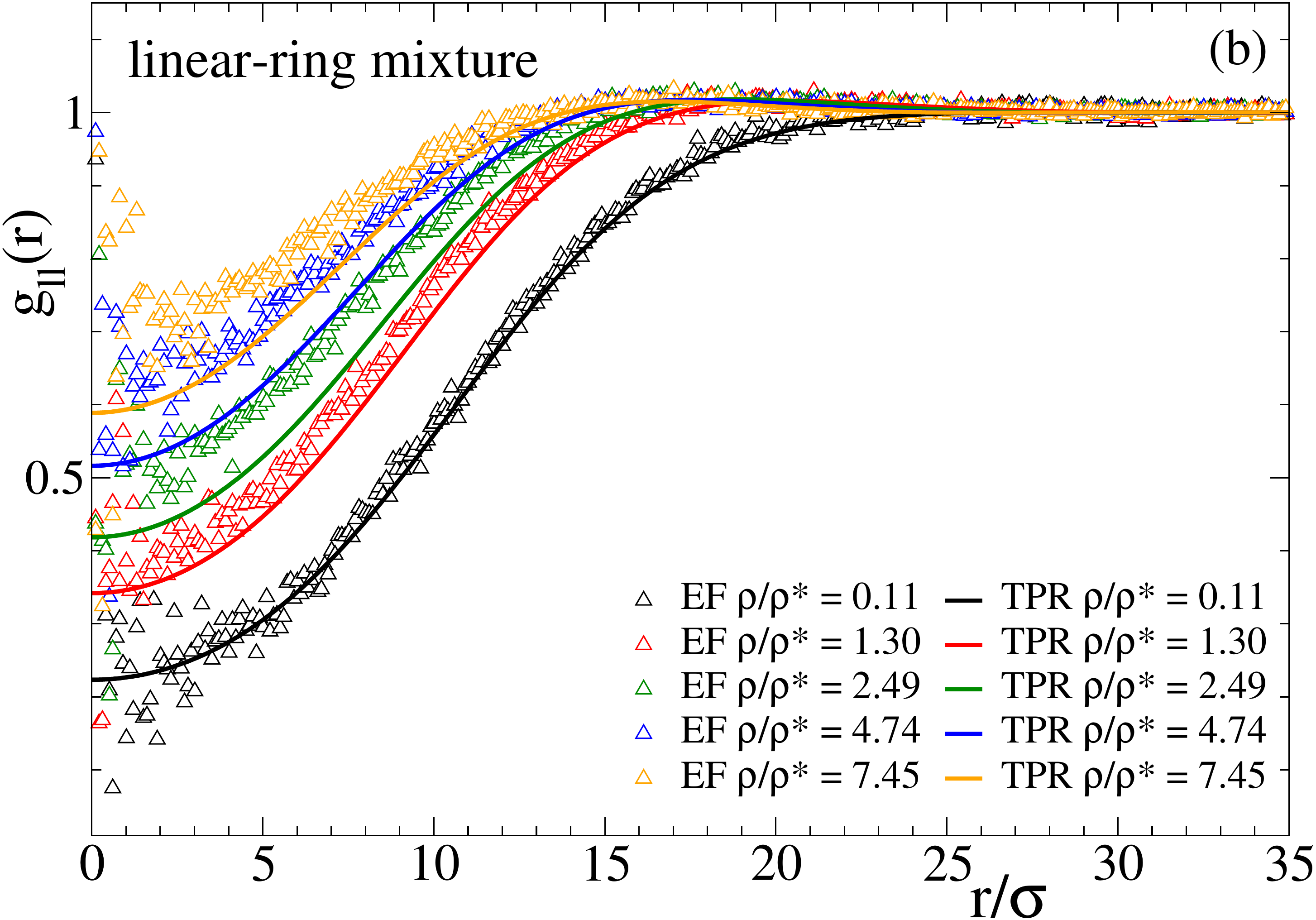}
			\end{minipage}			
				\begin{minipage}[c]{.46\textwidth}
					\setlength{\captionmargin}{0pt}%
					\includegraphics[width=1.06\textwidth]{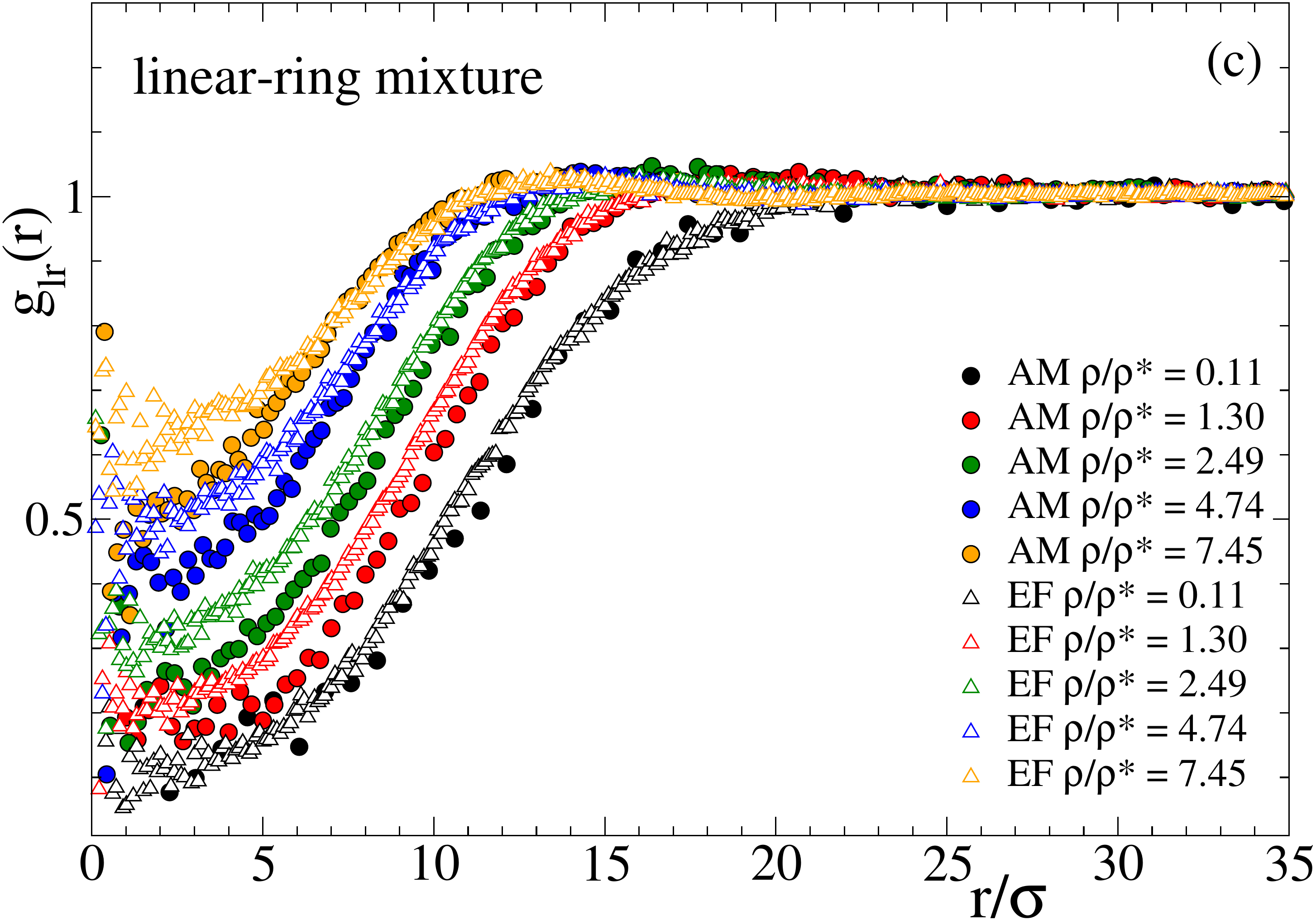}
				\end{minipage}%
				\hspace{5mm}%
				\begin{minipage}[c]{.46\textwidth}
					\setlength{\captionmargin}{0pt}%
					\includegraphics[width=1.06\textwidth]{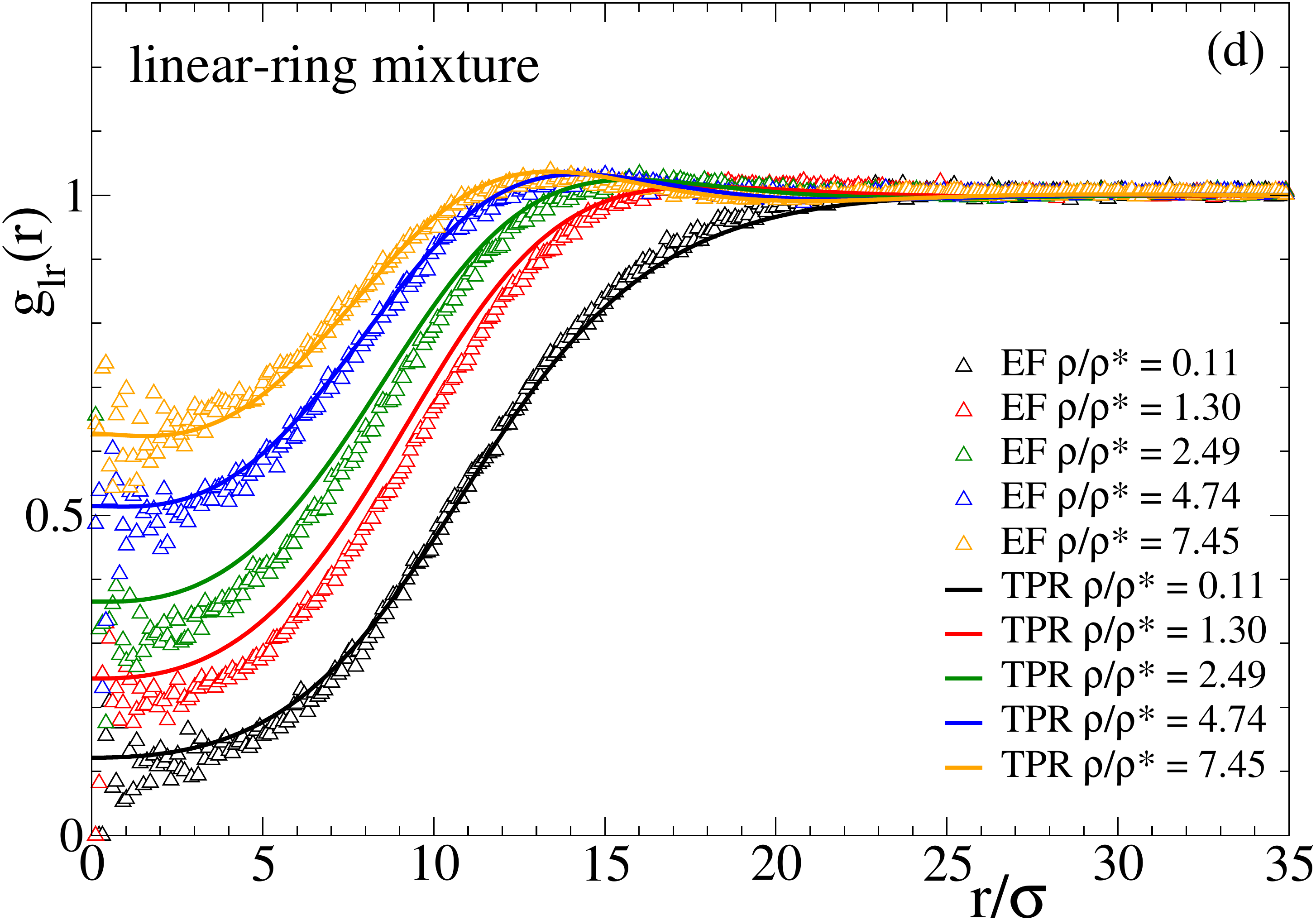}
				\end{minipage}
					\begin{minipage}[c]{.46\textwidth}
						\setlength{\captionmargin}{0pt}%
						\includegraphics[width=1.06\textwidth]{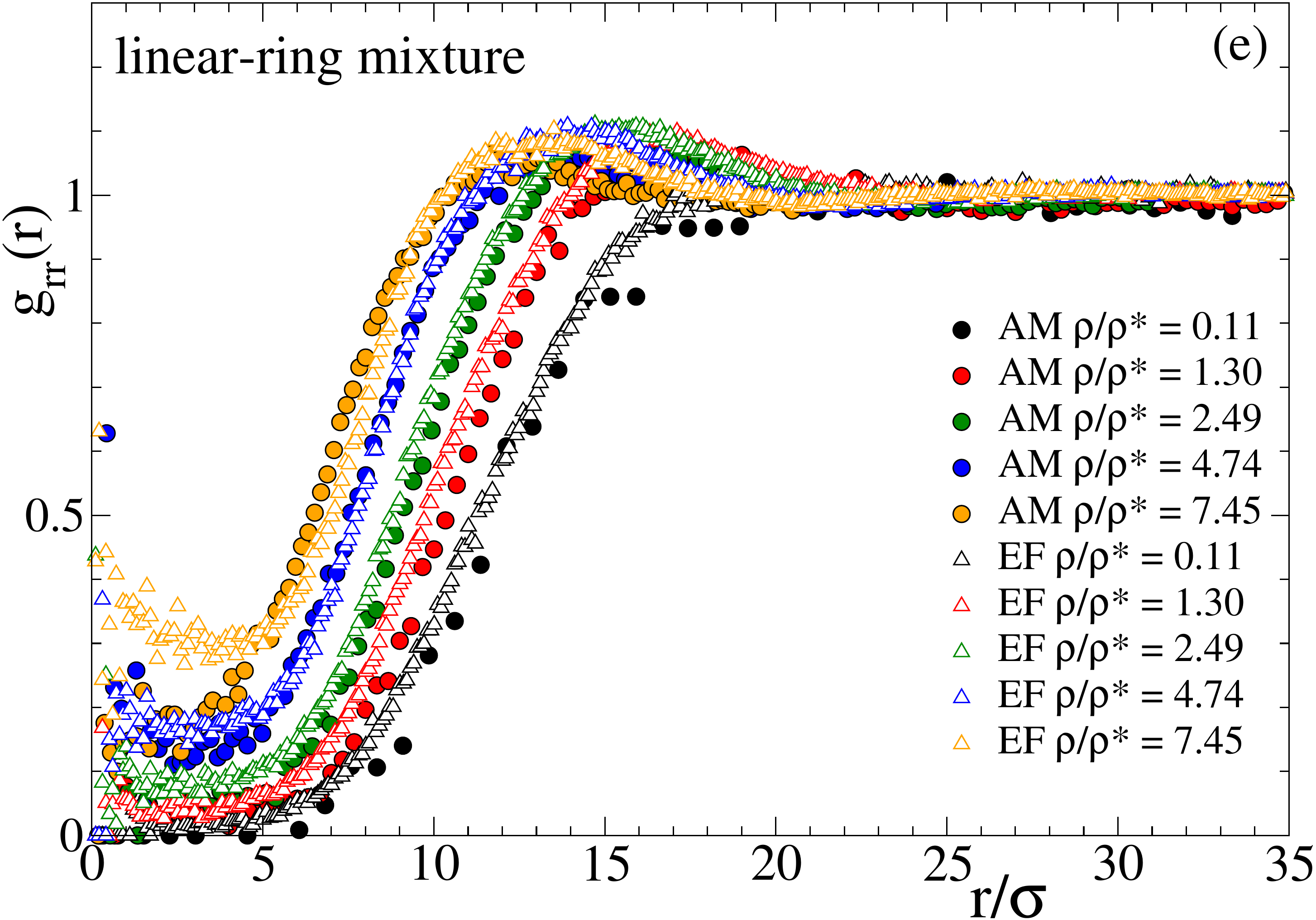}
					\end{minipage}%
					\hspace{5mm}%
					\begin{minipage}[c]{.46\textwidth}
						\setlength{\captionmargin}{0pt}%
						\includegraphics[width=1.06\textwidth]{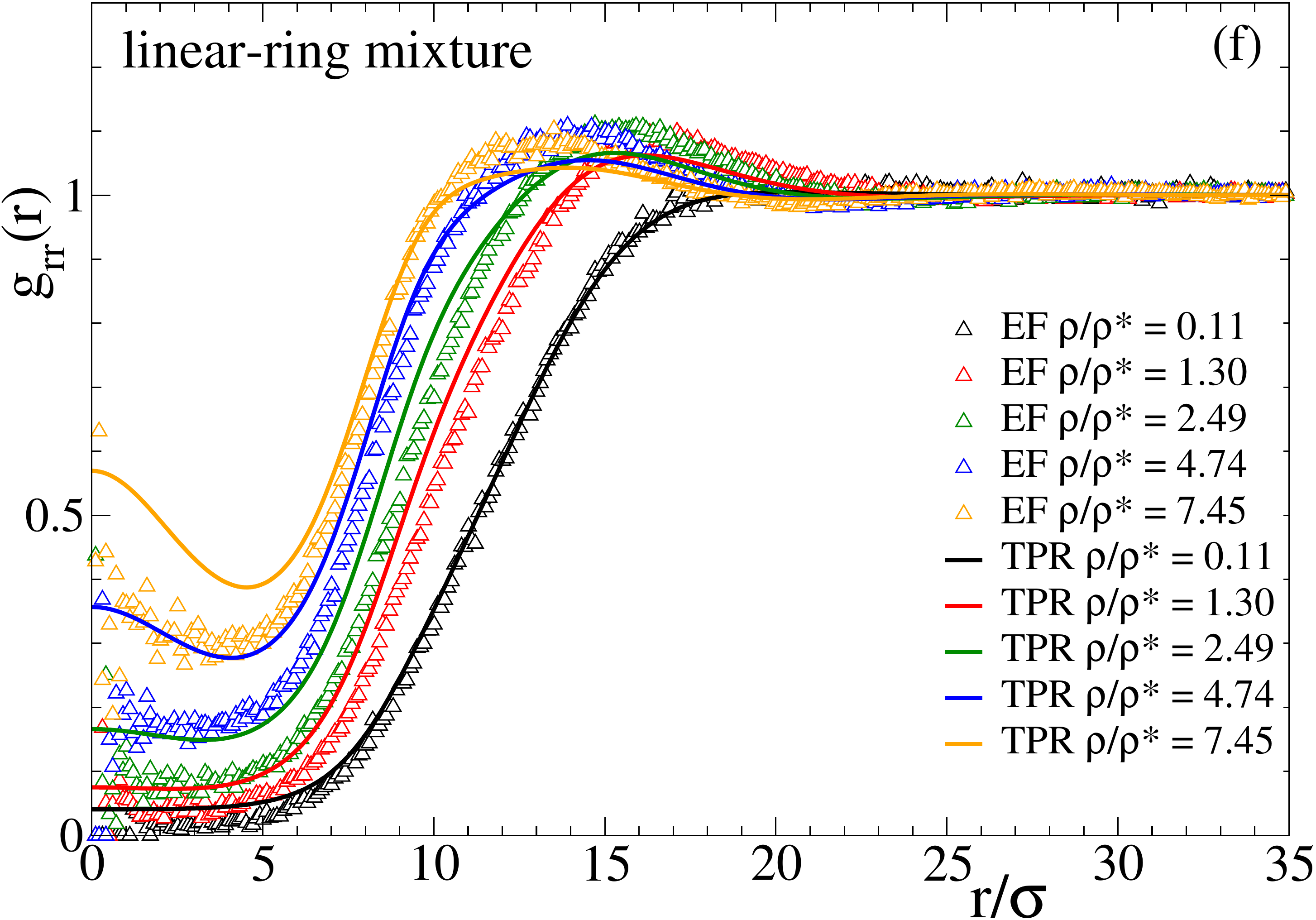}
					\end{minipage}
\caption{Radial distribution function of the 50/50 mixture of linear chains and rings with reversible bonds. 
Panels (a,c,e) compare results for the AM and EF simulations. Panels (b,d,f) compare the EF simulations with the predictions of TPR. 
Panels (a,b), (c,d) and (e,f) show such comparisons for the partial linear-linear, linear-ring and ring-ring correlations, respectively}
\label{fig:grlinring}
\end{figure}

\ref{fig:grlinring} shows the partial correlations of the radial distribution function for the 50/50 ring-linear mixture. 
As in \ref{fig:grlin} and \ref{fig:grring} the left column compares AM and EF simulations, 
whereas the right column compares the EF simulations with the predictions of TPR.  
The top panels (a,b) display the partial correlations between the linear chains, $g_{\rm ll}(r)$. Middle panels (c,d) show the cross-correlations (linear-ring, $g_{\rm lr}(r)$), and the correlations between the rings ($g_{\rm rr}(r)$) are displayed in the bottom panels (e,f).
At low and moderate concentrations, the small but systematic deviations between the AM and EF for the linear-linear correlations in the mixture are similar to those found in the pure system. However, whereas at large concentrations ($\rho/\rho^\star > 4$) there is a very  good agreement  between the 
AM simulations and EF in the pure linear system, significant differences are observed in the mixture. This suggests that the picture of an effective flat energy landscape describing the many-body interactions in the pure linear system is an oversimplification when the linear neighbours are partially substituted by
rings adopting less penetrable crumpled globular conformations and hence leading to an heterogenous landscape. This is consistent with the found deviations between the EF and TPR (see panel (b)), in contrast to the excellent agreemnt observed in the pure system.
The description of the ring-ring AM correlations by the effective potentials is improved in the mixture with respect to the pure solutions. 
Indeed the presence of a 50\% of particles (representing the linear chains) in the EF interacting through Gaussian potentials (which do not lead to cluster phases) reduces the tendency to the cluster phase of the particles representing the rings, and the EF becomes closer to the real system where no peak at $r=0$ is found.
On the other hand the TPR provides a worse description of the ring-ring correlations in the EF of the mixture than in the EF of the pure system of rings. Again this might be related to the structural heterogeneity of the EF of the mixture that worsens the mean field approach of TPR, though surprisingly, TPR does provide a very good description of the cross correlations (linear-ring) in the EF. A reasonably good agreement is also found between the cross-correlations in the AM and EF systems.
The results for the cross-correlations in panel (c) show a good mixing of the linear chains and rings with reversible bonds, with no signatures of segregation or incoming phase
separation. Indeed the correlation holes are just intermediate between those for the self-correlations.

\begin{figure}[htbp!]
	\centering
	\includegraphics[width=0.5\textwidth]{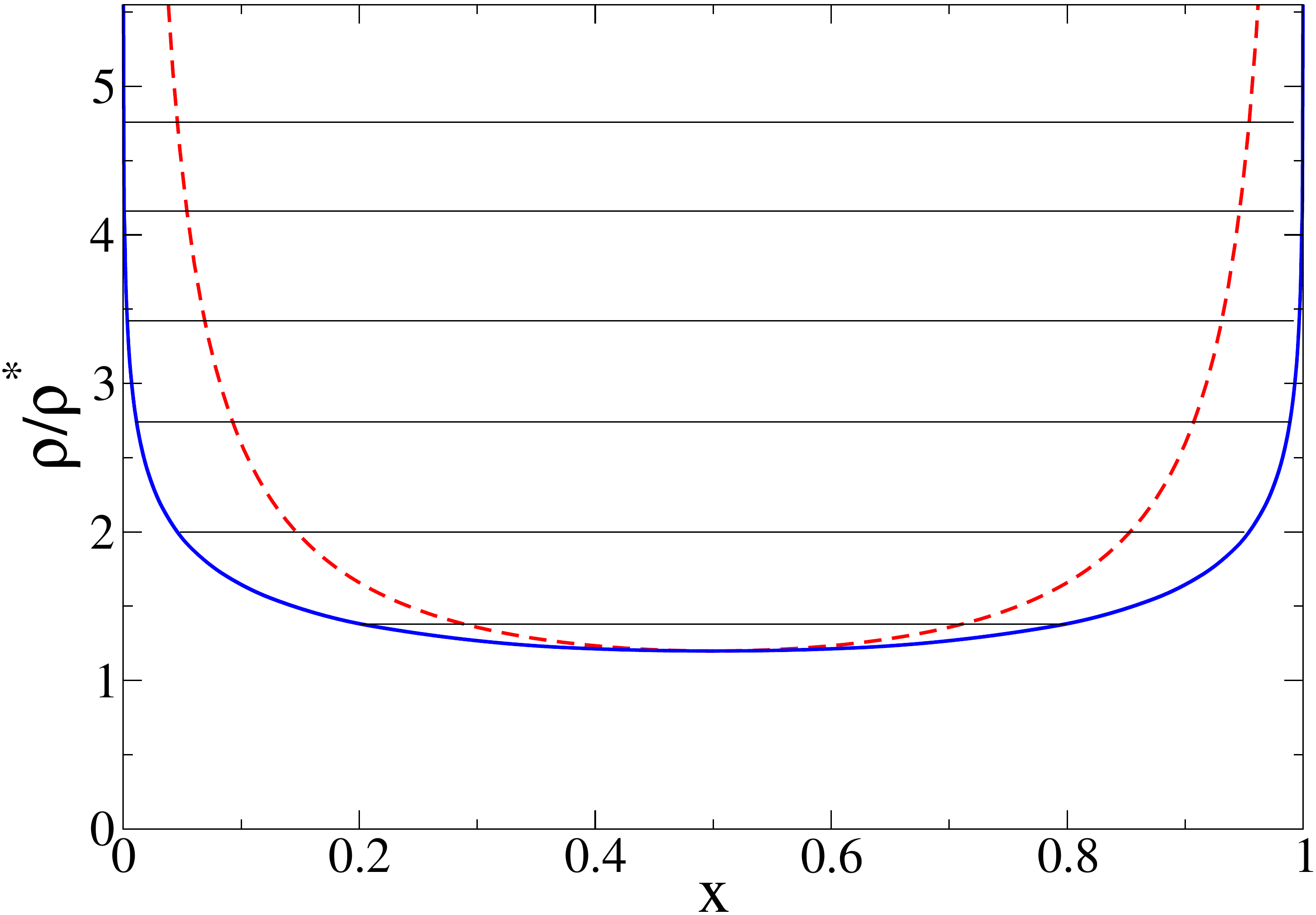}
	\caption{Theoretical phase diagram (reduced concentration $\rho/\rho^{\star}$ vs. composition)  for the binary mixture of linear chains with orthogonal reversible bonds  The dashed and thick lines are the spinodal and binodal lines, respectively. The thin straight lines join the coexistence points.}
\label{fig:phasediag}
\end{figure}

Finally we push further our investigation on the validity of effective potentials to describe correlations in crowded solutions of polymers with reversible bonds. Our last question is whether it is possible in our model to form interpenetrated networks (IPNs) from two polymers with reversible bonds and orthogonal chemistry. For this purpose, we consider a linear binary mixture with the same fraction $x = 50 \%$  for both components. As mentioned before, in the all-monomer simulations the WCA, FENE and reversible bonding interactions are identical for both components, with the only difference that intermolecular bonding is switched off between
chains of different components. In the effective fluid, the interactions between particles of the same component are the same as in the EF of the pure linear case (`linear-linear, all bonds' in \ref{table:1}), whereas for the cross-interactions we use the effective potentials 
derived in the absence of intermolecular bonding (`linear-linear, ony intra').
To have a first idea of the emerging scenario for this system we obtain the theoretical phase diagram for the effective fluid 
in the plane of reduced concentration ($\rho/\rho^{\star}$) vs. composition ($0 \leq x \leq 1$) of the mixture,
by using the the random phase approximation for the partial correlations as a closure to the Ornstein-Zernike relation
\cite{Gotze2006, Gotze2004, LikosOverduin2009, LouisHansen2000}. 
This should be a reasonable approximation on the basis of the observed quality of the mean field-TPR. 
We find a spinodal line (dashed line in \ref{fig:phasediag}) attesting  the existence of a region with macrophase separation (demixing) which, because self-interactions for both components are identical, becomes symmetric with respect to the composition. 
Forming a pair of interpenetrated networks (IPN) first requires percolation of both components of the mixture, which does not occur if the composition
is very asymmetric. On the other hand  we find that,  except for very asymmetric compositions, the system demixes when the density is increased slightly above the overlap concentration. Since the onset of network percolation occurs at such concentrations or above them \cite{Formanek2021}, the theoretical phase diagram of  \ref{fig:phasediag}
suggests that it is not possible to form an IPN in the mixture of chains with reversible bonds, this being frustrated by the demixing of both components.


\begin{figure}[htbp!]
\centering
\includegraphics[width=1.0\textwidth]{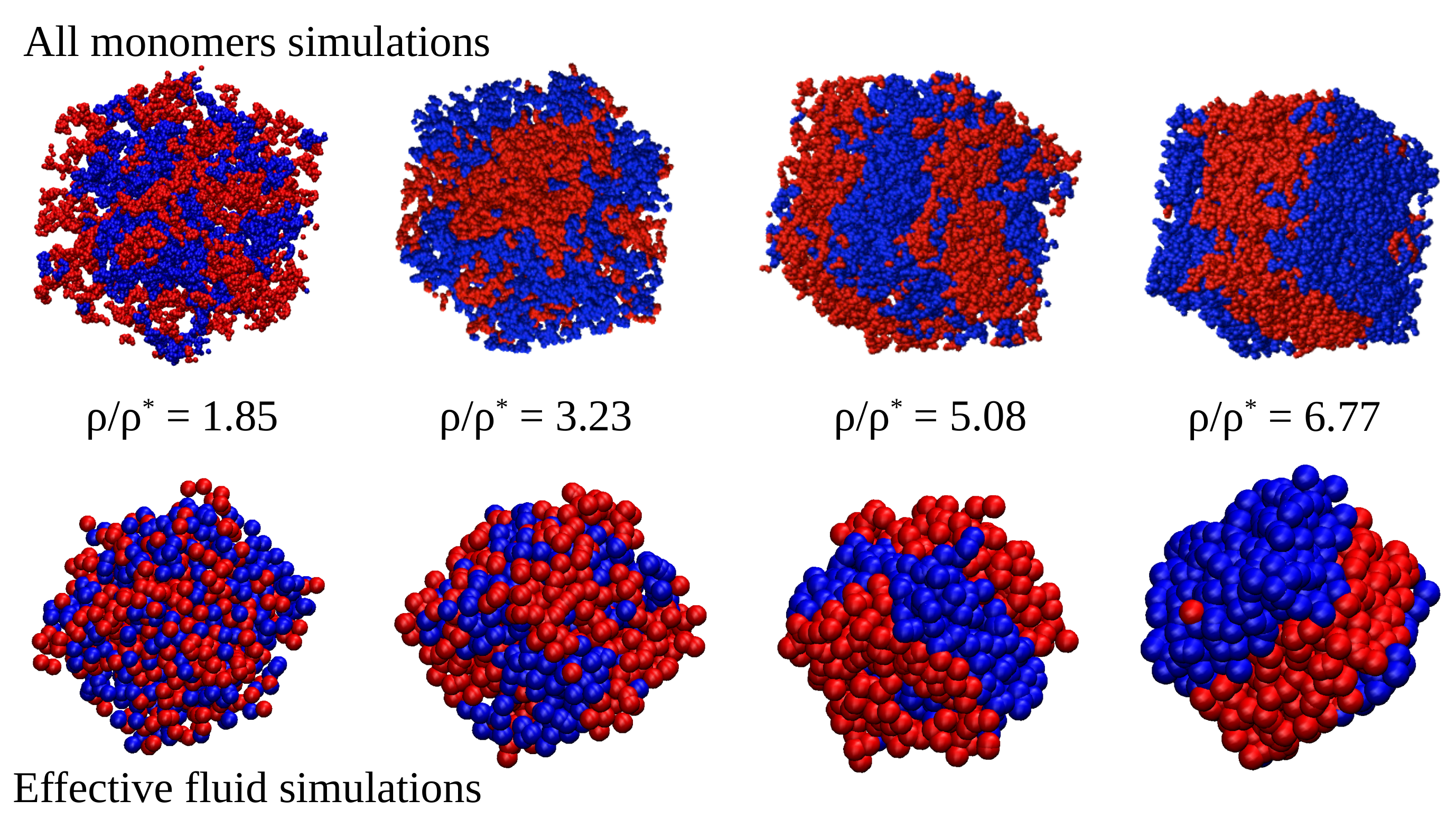}
\caption{Snapshots from the all monomers simulations (upper row) and the effective fluid simulations (bottom row) at different concentrations
of the binary mixture of linear chains with orthogonal chemistry of bonding. The beads represent the actual monomers (21600 in total) in the AM case 
and the effective ultrasoft particles (1000) in the EF. Molecules belonging to different components of the mixture are represented by different colors.
Demixing is evident in both the AM and EF simulations.}
\label{fig:snapmix}
\end{figure}

\begin{figure}[htbp!]
\includegraphics[width=0.8\textwidth]{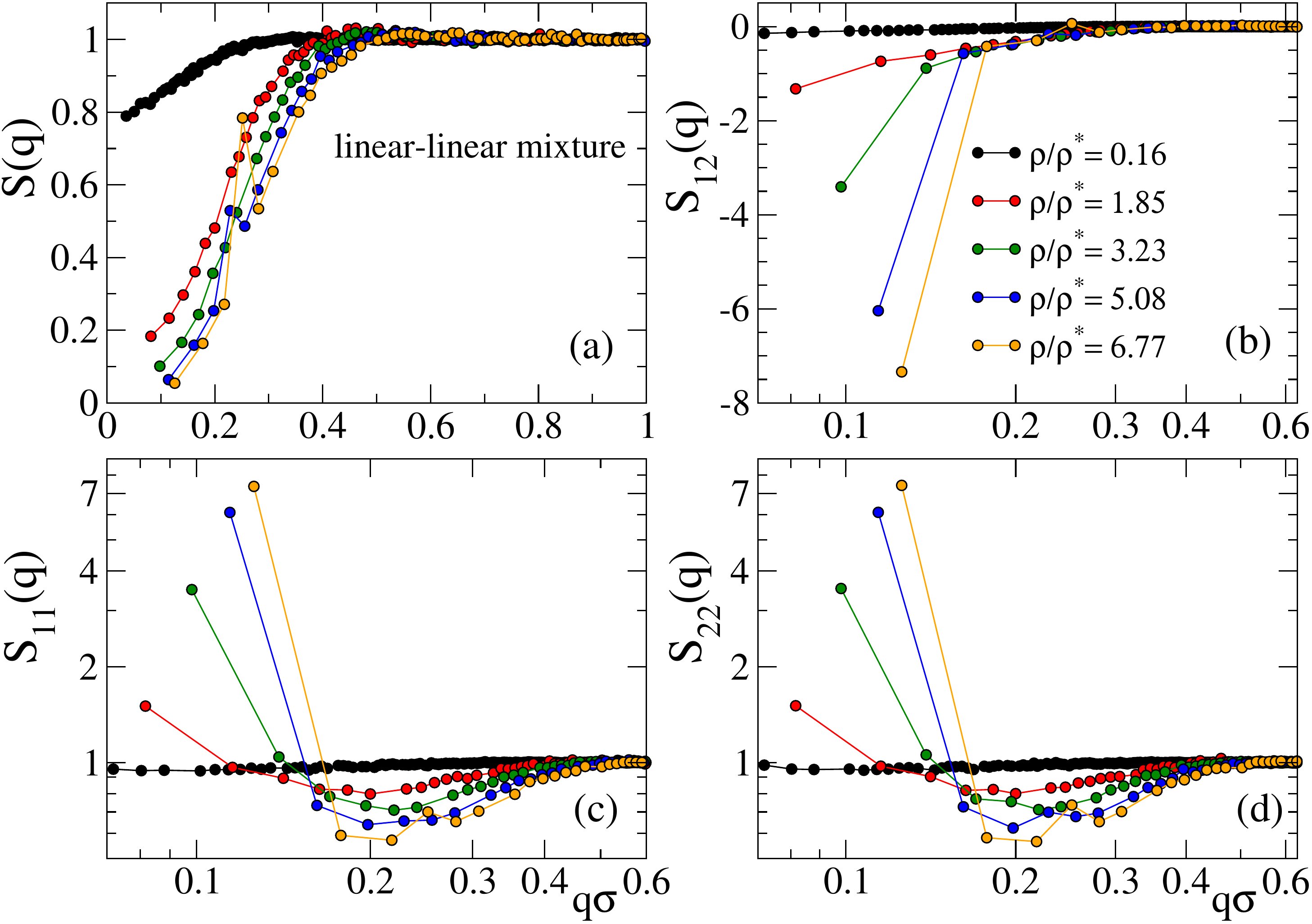}
\caption{Total (a) and partial static structure factors (b,c,d) of the centers-of-mass in the AM binary mixture of linear chains with reversible bonds and orthogonal chemistry. Because the fraction and the self-interactions 
of each component are identical, $S_{11}(q)=S_{22}(q)$.}
\label{fig:sq}
\end{figure}

\ref{fig:snapmix} shows AM and EF simulation snapshots at different concentrations for the 50/50 mixture of linear chains with reversible bonds and orthogonal chemistry. The beads represent the monomers and the effective particles in the AM and EF systems respectively. By depicting the two components with different colors, demixing (which as anticipated in \ref{sec2} occurs spontaneously by evolution from an initially mixed state) is evident and confirms the
expectation from the theoretical phase diagram. This is quantitatively reflected in the partial components of the total static structure factor of the
molecular centers-of-mass, $S_{\alpha\beta}(q)$, 
where $\alpha, \beta$ refer to the components (1,2) of the mixture, so that $S_{11}(q)$ and $S_{22}(q)$ represent correlations within a same component and $S_{12}(q)$
represents cross-correlations between chains of different components. These quantities are calculated as:
\begin{equation}
S_{\alpha\beta}(q) = (N_\alpha N_\beta)^{-1/2}\langle  \sum_{j,k} \exp[i{\bf q}\cdot({\bf r}^{\alpha}_j-{\bf r}^{\beta}_k)]\rangle .
\label{eq:sq}
\end{equation}
In this equation $N_\alpha$ is the number of relevant coordinates of the $\alpha$-component in the simulation box (the molecular centers-of-mass in the AM and all the effective 
particles in the EF), and $r^{\alpha}_j$ denotes the coordinate of the $j$th molecule of the $\alpha$-component. 
The average is performed over several realizations of the box and different runs at the same
concentration.  The total structure factor, $S(q)$, accounting for all the correlations without distinguishing components of the mixture, is just obtained by running the sum over 
all pairs of coordinates in the box  (irrespective of their respective components) and normalizing the sum by the inverse of the total of number molecules $N_{\rm A} +N_{\rm B}$. \ref{fig:sq} shows the total $S(q)$ (panel (a)) and the partial components
$S_{\alpha\beta}(q)$ (panels (b,c,d)) of the molecular centers-of-mass in the AM system. It should be noted that, because the composition is equimolar and the self-interactions of the two components are identical, $S_{11}(q)=S_{22}(q)$. 
No signatures of growing length scales are observed in the total $S(q)$, which shows the typical behavior of a homogeneous fluid with increasing the concentration. The growing length scales of the two separating phases are evidenced by the growing peaks of the partial $S_{11}(q) = S_{22}(q)$ at $q \rightarrow 0$, with the corresponding anticorrelation for $S_{12}(q)$.

Phase separation has been found in a simplified mixture where all non-bonded and bonded interactions are identical, with the only constraint that intermolecular bonds between different components of the mixture are not allowed. A more realistic model should at least introduce different activation energies for the two kinds of orthogonal reactive sites. This would likely break the symmetry of the phase diagram with respect to the mixture composition, but the qualitative emerging scenario (demixing and impossibility of forming the IPN in equilibrium) is robust. Indeed, demixing is inherently connected to the more repulsive character of the cross-interactions than of the self-interactions. Introducing different bond activation energies will lead to different bonding probabilities and hence different self-interactions of the two components, but the cross-interaction should still be much more repulsive than the self-interactions because, as discussed in \ref{sec3}, the latter contain the combinatorial entropic gain associated to intermolecular bonding, this being absent between polymers of different components of the mixture. Having said this, there is plenty of evidence in the literature on formation of IPNs with purely reversible cross-links 
\cite{Chen2015,Wang2017,Foster2017,Yang2020,Hammer2021}. Our results suggests that such IPNs are kinetically trapped states. The typical activation energies of the dynamic bonds in these IPNs are of several hundreds of kJ/mol \cite{Hammer2021}, i.e., of the order of $100 k_{\rm B}T$ whereas in our simulations they are about $10k_{\rm B}T$. Moreover the experimental chains are much longer than the unentangled chains used here (the entanglement monomer density for our linear precursors is \cite{Formanek2021} $\rho_{\rm e} \gtrsim 0.42 \sigma^{-3}$). Thus, our results suggest that, in real systems, the combination of both high molecular weights and long lifetimes of the bonds creates large barriers impeding relaxation to the equilibrium demixed state, and the IPNs (created in out-of-equilibrium conditions) remain stable.

\section{Conclusions}\label{sec5}

We have systematically investigated effective potentials between polymeric molecules functionalized with groups
that can form intra- and intermolecular reversible bonds.
A rich scenario emerges for the dependence of the effective potential on the relevant control parameters.
In spite of the additional complexity introduced by the high number of instantaneous intramolecular loops originated by the reversible cross-links, 
the topological interaction of the unbonded precursor (linear or ring) still has a dominant contribution in the bonded state, leading to very different strengths of the effective interaction (being more repulsive for the ring-based system). Even if the molecular weight and the fraction of reactive sites are fixed, the effective potentials exhibit a significant dependence on the degree of randomness of the sequence of reactive sites (from fully random to periodic). 
If the reactive sites of the two polymers are orthogonal, so that only intramolecular bonds are formed, decreasing randomness leads to longer intramolecular loops, which hinders interpenetrability and leads to a stronger effective intermolecular repulsion. The opposite effect is found if reactive sites of both polymers are identical and both intra- and intermolecular bonding occurs. We suggest that the free energy loss caused by the intermolecular bonds is mainly given by combinatorial entropy arising from the exponential number of bonding patters that the two intermolecularly bonded polymers can adopt.

We have explored the accuracy of the effective potentials to describe the equilibrium correlations 
betwen centers-of-mass in the crowded solutions. In the case of the linear chains a very good agreement between the effective fluid and the all-monomer simulations is found ever far above the overlap concentration. This is consistent with the fact that shrinking is highly prevented by forming intermolecular bonds with neighboring chains, which makes the conformations at high dilution weakly sensitive to crowding, and many-body effects basically contribute as a flat energy landscape. In a similar fashion to the case of rings with no bonds, the comparison with the effective fluid is less satisfactory in the system of rings with reactive sites, which does not show the cluster phase predicted by the effective fluid. This is consistent with the crowding-driven collapse to crumpled globule-like conformations, reflectig the relevance of the many-body interactions.
We have further extended our investigation to a 50/50 mixture of the former types of polymers. The results for the partial correlations  
are qualitatively similar to those of the pure polymers and the system is fully miscible.

Finally, we have explored the possibility of forming two interpenetrated networks in a linear-linear mixture where the reactive sites of the two components are orthogonal, i.e., intermolecular bonds only occur between chains of the same component. In agreement with the energetic penalty found for the effective cross-interaction potential, the simulations of the effective fluid, and the phase diagram obtained by the test particle route, no interpenetrated networks are found and the two components demix. This result suggests that real interpenetrated networks, where the lifetimes of the reversible bonds are much longer than in our simulations, are kinetically trapped states with large entropic barriers impeding the relaxation to the equilibrium demixed state (arrested demixing). Our results may motivate future experimental tests in mixtures of oligomers with low bond energies. 
On the other hand, an interesting problem  to address in the future is  the accuracy of the effective fluid approach in dual networks, where both types of orthogonal reactive sites are present in all the chains, including the determination of the phase behavior in mixtures with different fractions of both sites in each component. Work in this direction is in progress.

\begin{acknowledgement}

We acknowledge the Grant PGC2018-094548-B-I00 funded by MCIN/AEI/10.13039/501100011033 and by `ERDF A way of making Europe' and the Grant IT-1175-19 funded by Eusko Jaurlaritza (Basque Government).
We thank F. Sciortino and L. Rovigatti for valuable discussions.
Part of this work was carried out during M.P's secondment at University of Vienna. M.P. acknowledges  the travel grant
from Materials Physics Center MPC and thanks the staff of University of Vienna for their hospitality.

\end{acknowledgement}


\providecommand{\latin}[1]{#1}
\providecommand*\mcitethebibliography{\thebibliography}
\csname @ifundefined\endcsname{endmcitethebibliography}
  {\let\endmcitethebibliography\endthebibliography}{}

\newpage
\noindent\LARGE SUPPORTING INFORMATION
\normalsize

It can be shown \cite{FoundInhomoFluids-s} that the grand potential of an inhomogeneous fluid in equilibrium with a bulk reservoir that fixes the chemical potential $\mu$ is a unique functional of the equilibrium single-particle density $\rho(\textbf{r})$.
DFT enters  in the computation of $\rho(\textbf{r})$, which can be obtained by minimising the grand potential functional\cite{ChubakLocatelli2018-s}:
\begin{equation}
\Omega[\rho] = F_{\rm ideal}[\rho] + F_{\rm ex}[\rho] + \int \rho(\textbf{r}) V_{\rm eff}(\textbf{r}) d^{3}r - \mu \int \rho(\textbf{r}) d^{3}r ,
\label{eq:omega}
\end{equation}
The first sum on the right side is the Helmholtz free energy $F[\rho]$ with a well-known ideal contribution equal to:
\begin{equation}
F_{\rm ideal}[\rho] = \beta^{-1} \int \rho(\textbf{r})[\ln (\rho(\textbf{r})\Lambda^{3})-1] d^{3}r .
\label{eq:f_ideal}
\end{equation}
In this expression $\beta = 1/k_{\rm B}T$ and $\Lambda = \sqrt{2\pi\beta\hslash^{2}/ m}$ is the thermal de Broglie wavelength. In order to compute the excess contribution to the free energy functional we use the mean field approximation \cite{LikosOverduin2009-s,LouisHansen2000-s}:
\begin{equation}
F_{\rm ex}[\rho] = \dfrac{1}{2}\iint \rho(\textbf{r}) \rho(\textbf{r}') V_{\rm eff}(|\textbf{r}-\textbf{r}'|) d^{3}r\hspace{1mm}d^{3}r'.
\label{eq:f_ext}
\end{equation}
By substituting \ref{eq:f_ideal} and \ref{eq:f_ext} in \ref{eq:omega}, and by performing the functional derivative, we find: 
\begin{equation}
\dfrac{\delta(\beta\Omega[\rho])}{\delta\rho(\textbf{r})} = \int \rho(\textbf{r}') \beta V_{\rm eff}(|\textbf{r}-\textbf{r}'|) d^{3}r' + \ln (\rho(\textbf{r})\Lambda^{3}) +  \beta V_{\rm eff}(|\textbf{r}|) - \beta \mu ,
\label{eq:deltaOmega}
\end{equation}
The expression for the chemical potential is obtained by computing the Helmholtz free energy in the bulk density $\rho_{\rm b}$:
\begin{equation}
\mu = \dfrac{\partial (\beta F[\rho_{\rm b}])}{\partial N} = \ln(\rho_{\rm b}\Lambda^{3}) + \rho_{\rm b} \int \beta V_{\rm eff}(|\textbf{r}|) d^{3}r ,
\label{eq:mu}
\end{equation}
By substituting \ref{eq:mu} in \ref{eq:deltaOmega} and noting that the integral involves a convolution of $V_{\rm eff}$ and $\rho$, we obtain
\begin{equation}
\ln(\rho(|\textbf{r}|)/\rho_{\rm b}) = -\beta(V_{\rm eff}*\rho)(|\textbf{r}|) - \beta V_{\rm eff}(|\textbf{r}|)+ \rho_{\rm b} \int \beta V_{\rm eff}(|\textbf{r}|) d^{3}r ,
\label{eq:rho}
\end{equation}
where the symbol $*$ denotes the convolution.
By introducing a new variable\cite{ChubakLocatelli2018-s} $\Delta\rho(\textbf{r}) = \rho(\textbf{r}) - \rho_{\rm b}$,  which decays to 0 at long distances, 
and by substituting it in \ref{eq:rho}, the final expression for the density profile is obtained:
\begin{equation}
\Delta\rho(|\textbf{r}|) = \rho_{\rm b}(\exp[-\beta(V_{\rm eff}*\Delta\rho)(|\textbf{r}|) - \beta V_{\rm eff}(|\textbf{r}|)]-1) ,
\label{eq:deltaRho_s}
\end{equation}
and as mentioned in the article the radial distribution function is just calculated as $g(|\textbf{r}|)= \rho(|\textbf{r}|)/\rho_{\rm b}$.

In the case of a binary mixture, there are 2 species with densities $\rho_{i}(\textbf{r})$ that interact through $V_{ij}(|\textbf{r}|)$, with $i,j = 1,2$. The corresponding Helmholtz free energy for this system is given by: 
\begin{equation}
F[\rho_1, \rho_2] = \sum_{i = 1}^{2} \int \rho_i(\textbf{r}) [\ln(\Lambda_i^3\rho_i(\textbf{r}))-1]d^{3}r + \dfrac{1}{2} \sum_{i = 1}^{2} \sum_{j = 1}^{2} \int d^{3}r \int d^{3}r'\rho_i(\textbf{r})\rho_j(\textbf{r}')V_{ij}(|\textbf{r}-\textbf{r}'|) .
\label{eq:deltaRho2}
\end{equation}
Thus, if a particle of the species 1 is fixed in the origin, the grand potential functional reads as:
\begin{equation}
\Omega[\rho_{1}, \rho_{2}] = F[\rho_{1}, \rho_{2}] + \int \rho_1(\textbf{r}) V_{11}(|\textbf{r}|) d^{3}r+ \int \rho_2(\textbf{r}) V_{12}(|\textbf{r}|) d^{3}r - \mu_1\int \rho_1(\textbf{r}) d^{3}r - \mu_2\int \rho_2(\textbf{r}) d^{3}r .
\label{eq:omega2}
\end{equation}
Then by using the same procedure as before it is possible to obtain 2 equations that need to be solved iteratively for determining $g_{11}(|\textbf{r}|)$ and $g_{12}(|\textbf{r}|)$. Equivalently, by fixing in the origin a particle of species 2 one obtains the expressions for $g_{22}(|\textbf{r}|)$ and $g_{21}(|\textbf{r}|)$. 

\providecommand{\latin}[1]{#1}
\makeatletter
\providecommand{\doi}
  {\begingroup\let\do\@makeother\dospecials
  \catcode`\{=1 \catcode`\}=2 \doi@aux}
\providecommand{\doi@aux}[1]{\endgroup\texttt{#1}}
\makeatother
\providecommand*\mcitethebibliography{\thebibliography}
\csname @ifundefined\endcsname{endmcitethebibliography}
  {\let\endmcitethebibliography\endthebibliography}{}

\newpage

\begin{table}[h!]
\small
	\begin{tabular}{| c | c |}	
		\hline
	        & Linear  \\[0.5ex]
		& \\[0.5ex]
		\hline
		all monomers (AM) &  $L_{\rm box} = 175,\hspace{0.4cm} N_{\rm lin} = 108,\hspace{0.4cm}  \rho = 0.004,\hspace{0.4cm} \rho/\rho^* = 0.16$  \\[0.5ex] 
			&  $L_{\rm box} = 77,\hspace{0.4cm} N_{\rm lin} = 108,\hspace{0.4cm}   \rho = 0.047,\hspace{0.4cm} \rho/\rho^* = 1.85$  \\	[0.5ex] 
			&  $L_{\rm box} = 64,\hspace{0.4cm} N_{\rm lin} = 108,\hspace{0.4cm}   \rho = 0.082,\hspace{0.4cm} \rho/\rho^* = 3.23$  \\	[0.5ex] 
			&  $L_{\rm box} = 55,\hspace{0.4cm} N_{\rm lin} = 108,\hspace{0.4cm}   \rho = 0.129,\hspace{0.4cm} \rho/\rho^* = 5.08$   \\	[0.5ex] 
			&  $L_{\rm box} = 50,\hspace{0.4cm} N_{\rm lin} = 108,\hspace{0.4cm}   \rho = 0.173,\hspace{0.4cm} \rho/\rho^* = 6.77$  \\	[0.5ex] 
		\hline
			effective fluid (EF) &  $L_{\rm box} = 367.5,\hspace{0.4cm} N_{\rm lin} = 1000,\hspace{0.4cm} \rho/\rho^* = 0.16$  \\[0.5ex] 
			&  $L_{\rm box} = 161.7,\hspace{0.4cm} N_{\rm lin} = 1000,\hspace{0.4cm} \rho/\rho^* = 1.85$  \\	[0.5ex] 
			&  $L_{\rm box} = 134.4,\hspace{0.4cm} N_{\rm lin} = 1000,\hspace{0.4cm} \rho/\rho^* = 3.23$  \\	[0.5ex] 
			&  $L_{\rm box} = 115.5,\hspace{0.4cm} N_{\rm lin} = 1000,\hspace{0.4cm} \rho/\rho^* = 5.08$  \\	[0.5ex] 
			&  $L_{\rm box} = 105.0,\hspace{0.4cm} N_{\rm lin} = 1000,\hspace{0.4cm} \rho/\rho^* = 6.77$  \\		[0.5ex]  
		\hline
	        & Ring  \\[0.5ex]
		& \\[0.5ex]
		\hline
				all monomers (AM) &  $L_{\rm box} = 175,\hspace{0.4cm} N_{\rm lin} = 108,\hspace{0.4cm}  \rho = 0.004,\hspace{0.4cm} \rho/\rho^* = 0.063$  \\[0.5ex] 
				&  $L_{\rm box} = 77,\hspace{0.4cm} N_{\rm ring} = 108,\hspace{0.4cm}  \rho = 0.047,\hspace{0.4cm} \rho/\rho^* = 0.74$  \\	[0.5ex] 
				&  $L_{\rm box} = 62,\hspace{0.4cm} N_{\rm ring} = 108,\hspace{0.4cm}  \rho = 0.090,\hspace{0.4cm} \rho/\rho^* = 1.42$  \\	[0.5ex] 
				&  $L_{\rm box} = 50,\hspace{0.4cm} N_{\rm ring} = 108,\hspace{0.4cm}  \rho = 0.173,\hspace{0.4cm} \rho/\rho^* = 2.71$  \\	[0.5ex] 
				&  $L_{\rm box} = 43,\hspace{0.4cm} N_{\rm ring} = 108,\hspace{0.4cm}  \rho = 0.272,\hspace{0.4cm} \rho/\rho^* = 4.26$  \\		[0.5ex] 
				\hline
				effective fluid (EF) &  $L_{\rm box} = 367.5,\hspace{0.4cm} N_{\rm ring} = 1000,\hspace{0.4cm} \rho/\rho^* = 0.063$  \\[0.5ex] 
				&  $L_{\rm box} = 161.7,\hspace{0.4cm} N_{\rm ring} = 1000,\hspace{0.4cm} \rho/\rho^* = 0.74$  \\	[0.5ex] 
				&  $L_{\rm box} = 130.2,\hspace{0.4cm} N_{\rm ring} = 1000,\hspace{0.4cm} \rho/\rho^* = 1.42$  \\	[0.5ex] 
				&  $L_{\rm box} = 105.0,\hspace{0.4cm} N_{\rm ring} = 1000,\hspace{0.4cm} \rho/\rho^* = 2.71$  \\	[0.5ex] 
				&  $L_{\rm box} = 90.3,\hspace{0.4cm} N_{\rm ring} = 1000,\hspace{0.4cm} \rho/\rho^* = 4.26$  \\		[0.5ex] 
		\hline

	\end{tabular}
	\label{table:1s}
\end{table}

\normalsize

\noindent~Table T1. Simulation parameters of the solutions (all monomers and effective fluid) of linear and ring polymers with reversible bonds:
box size ($L_{\rm box}$), number of polymers ($N_{\rm lin}, N_{\rm ring}$), absolute ($\rho$) and reduced ($\rho/\rho^*$) concentration of monomers.

\newpage

\begin{table}[h!]
\small
	\begin{tabular}{| c | c |}	
		\hline
	        & Linear-Linear  \\[0.5ex]
		& \\[0.5ex]
		\hline
		all monomers (AM) &  $L_{\rm box} = 175,\hspace{0.3cm} N_{\rm lin,1} = 54,\hspace{0.3cm} N_{\rm lin,2} = 54,\hspace{0.3cm}  \rho = 0.004,\hspace{0.3cm} \rho/\rho^* = 0.16$  \\[0.5ex] 
			&  $L_{\rm box} = 77,\hspace{0.3cm} N_{\rm lin,1} = 54,\hspace{0.3cm} N_{\rm lin,2} = 54,\hspace{0.3cm}   \rho = 0.047,\hspace{0.3cm} \rho/\rho^* = 1.85$  \\	[0.5ex] 
			&  $L_{\rm box} = 64,\hspace{0.3cm} N_{\rm lin,1} = 54,\hspace{0.3cm} N_{\rm lin,2} = 54,\hspace{0.3cm}   \rho = 0.082,\hspace{0.3cm} \rho/\rho^* = 3.23$  \\	[0.5ex] 
			&  $L_{\rm box} = 55,\hspace{0.3cm} N_{\rm lin,1} = 54,\hspace{0.3cm} N_{\rm lin,2} = 54,\hspace{0.3cm}   \rho = 0.129,\hspace{0.3cm} \rho/\rho^* = 5.08$   \\	[0.5ex] 
			&  $L_{\rm box} = 50,\hspace{0.3cm} N_{\rm lin,1} = 54,\hspace{0.3cm} N_{\rm lin,2} = 54,\hspace{0.3cm}   \rho = 0.173,\hspace{0.3cm} \rho/\rho^* = 6.77$  \\	[0.5ex] 
		\hline
			effective fluid (EF) &  $L_{\rm box} = 367.5,\hspace{0.3cm} N_{\rm lin,1} = 500,\hspace{0.3cm} N_{\rm lin,2} = 500,\hspace{0.3cm} \rho/\rho^* = 0.16$  \\[0.5ex] 
			&  $L_{\rm box} = 161.7,\hspace{0.3cm} N_{\rm lin,1} = 500,\hspace{0.3cm} N_{\rm lin,2} = 500,\hspace{0.3cm} \rho/\rho^* = 1.85$  \\	[0.5ex] 
			&  $L_{\rm box} = 134.4,\hspace{0.3cm} N_{\rm lin,1} = 500,\hspace{0.3cm} N_{\rm lin,2} = 500,\hspace{0.3cm} \rho/\rho^* = 3.23 $  \\	[0.5ex] 
			&  $L_{\rm box} = 115.5,\hspace{0.3cm} N_{\rm lin,1} = 500,\hspace{0.3cm} N_{\rm lin,2} = 500,\hspace{0.3cm} \rho/\rho^* = 5.08 $  \\	[0.5ex] 
			&  $L_{\rm box} = 105.0,\hspace{0.3cm} N_{\rm lin,1} = 500,\hspace{0.3cm} N_{\rm lin,2} = 500,\hspace{0.3cm} \rho/\rho^* = 6.77 $  \\		[0.5ex]  
		\hline
	        & Linear-Ring  \\[0.5ex]
		& \\[0.5ex]
		\hline
				all monomers (AM) &  $L_{\rm box} = 175,\hspace{0.3cm} N_{\rm lin} = 54,\hspace{0.3cm} N_{\rm ring} = 54,\hspace{0.3cm}  \rho = 0.004,\hspace{0.3cm} \rho/\rho^* = 0.11 $  \\[0.5ex] 
				&  $L_{\rm box} = 77,\hspace{0.3cm} N_{\rm lin} = 54,\hspace{0.3cm} N_{\rm ring} = 54,\hspace{0.3cm}  \rho = 0.047,\hspace{0.3cm} \rho/\rho^* = 1.30$  \\	[0.5ex] 
				&  $L_{\rm box} = 62,\hspace{0.3cm} N_{\rm lin} = 54,\hspace{0.3cm} N_{\rm ring} = 54,\hspace{0.3cm}  \rho = 0.090,\hspace{0.3cm} \rho/\rho^* = 2.49$  \\	[0.5ex] 
				&  $L_{\rm box} = 50,\hspace{0.3cm} N_{\rm lin} = 54,\hspace{0.3cm} N_{\rm ring} = 54,\hspace{0.3cm}  \rho = 0.173,\hspace{0.3cm} \rho/\rho^* = 4.74$  \\	[0.5ex] 
				&  $L_{\rm box} = 43,\hspace{0.3cm} N_{\rm lin} = 54,\hspace{0.3cm} N_{\rm ring} = 54,\hspace{0.3cm}  \rho = 0.272,\hspace{0.3cm} \rho/\rho^* = 7.45$  \\		[0.5ex] 
				\hline
				efective fluid (EF) &  $L_{\rm box} = 367.5,\hspace{0.3cm} N_{\rm lin} = 500,\hspace{0.3cm} N_{\rm ring} = 500,\hspace{0.3cm} \rho/\rho^* = 0.11$  \\[0.5ex] 
				&  $L_{\rm box} = 161.7,\hspace{0.3cm} N_{\rm lin} = 500,\hspace{0.3cm} N_{\rm ring} = 500,\hspace{0.3cm} \rho/\rho^* = 1.30$  \\	[0.5ex] 
				&  $L_{\rm box} = 130.2,\hspace{0.3cm} N_{\rm lin} = 500,\hspace{0.3cm} N_{\rm ring} = 500,\hspace{0.3cm} \rho/\rho^* = 2.49$  \\	[0.5ex] 
				&  $L_{\rm box} = 105.0,\hspace{0.3cm} N_{\rm lin} = 500,\hspace{0.3cm} N_{\rm ring} = 500,\hspace{0.3cm} \rho/\rho^* = 4.74$  \\	[0.5ex] 
				&  $L_{\rm box} = 90.3,\hspace{0.3cm} N_{\rm lin} = 500,\hspace{0.3cm} N_{\rm ring} = 500,\hspace{0.3cm} \rho/\rho^* = 7.45$  \\		[0.5ex] 

			\hline
	\end{tabular}
	\label{table:2}
\end{table}

\normalsize

\noindent~Table T2. As Table T1 for the mixtures.

\newpage

\begin{figure}[htbp!]
	\centering
	\begin{minipage}[c]{.9\textwidth}
		\setlength{\captionmargin}{0pt}%
		\includegraphics[width=1.17\textwidth]{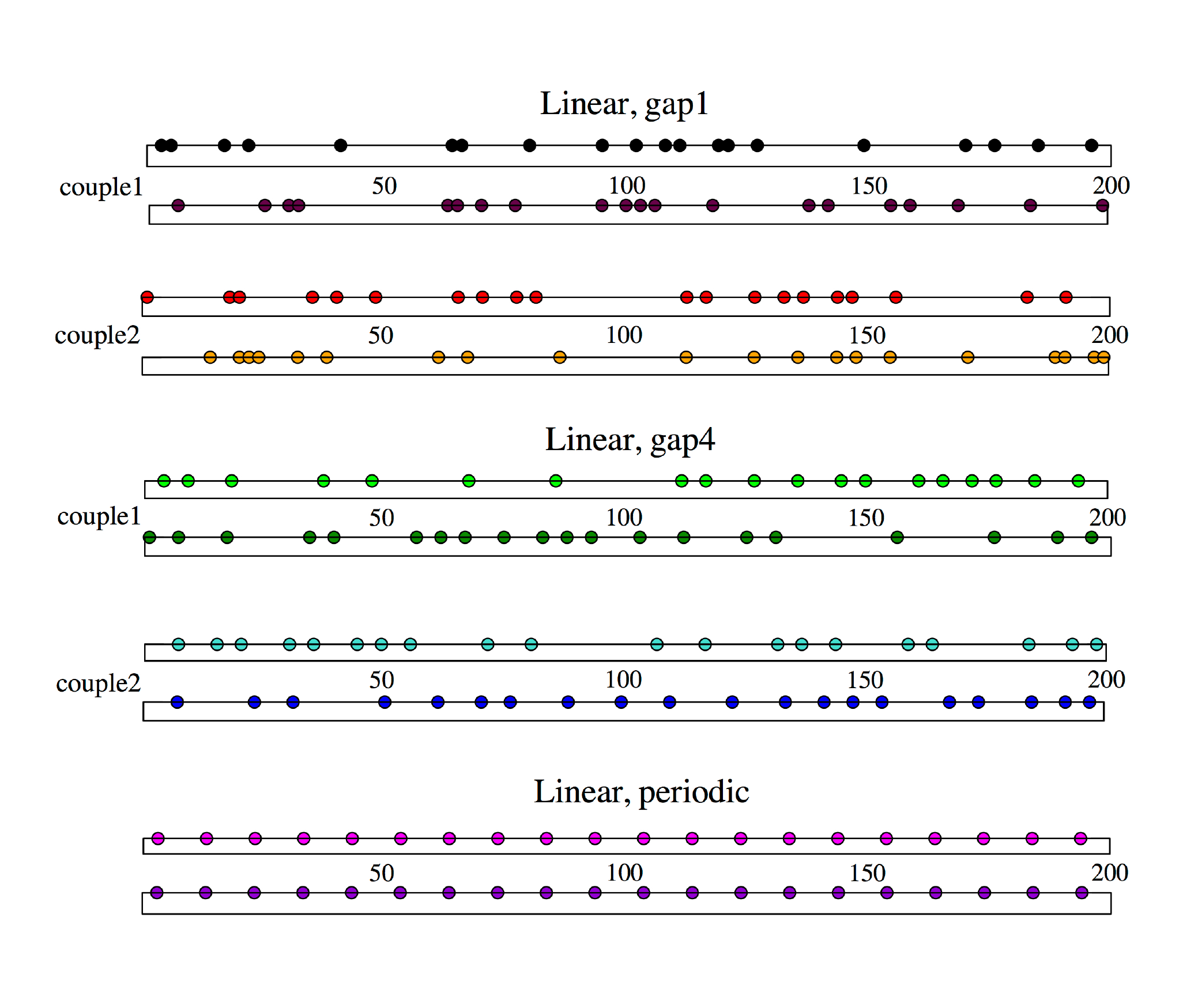}
	\end{minipage}%
\end{figure}
\noindent~Figure S1. Sequences of the reactive sites along the backbone for the linear chains with reversible bonds used to derive the effective potentials. Each row represents the two backbones of a pair of chains. The chain backbone consists of $N=200$ beads, 20 of them being reactive and able to form reversible bonds. Only the reactive beads are represented. The sequences are randomly generated with a minimum number of beads between consecutive reactive sites (1 and 4 for gap1 and gap4 cases, respectively). Two couples (1 and 2) are generated for both the gap1 and gap4 cases. A couple of chains with a periodic sequence of reactive sites is also studied (bottom row).

\newpage

\begin{figure}[htbp!]
	\centering
	\begin{minipage}[c]{.9\textwidth}
		\setlength{\captionmargin}{0pt}%
		\includegraphics[width=1.17\textwidth]{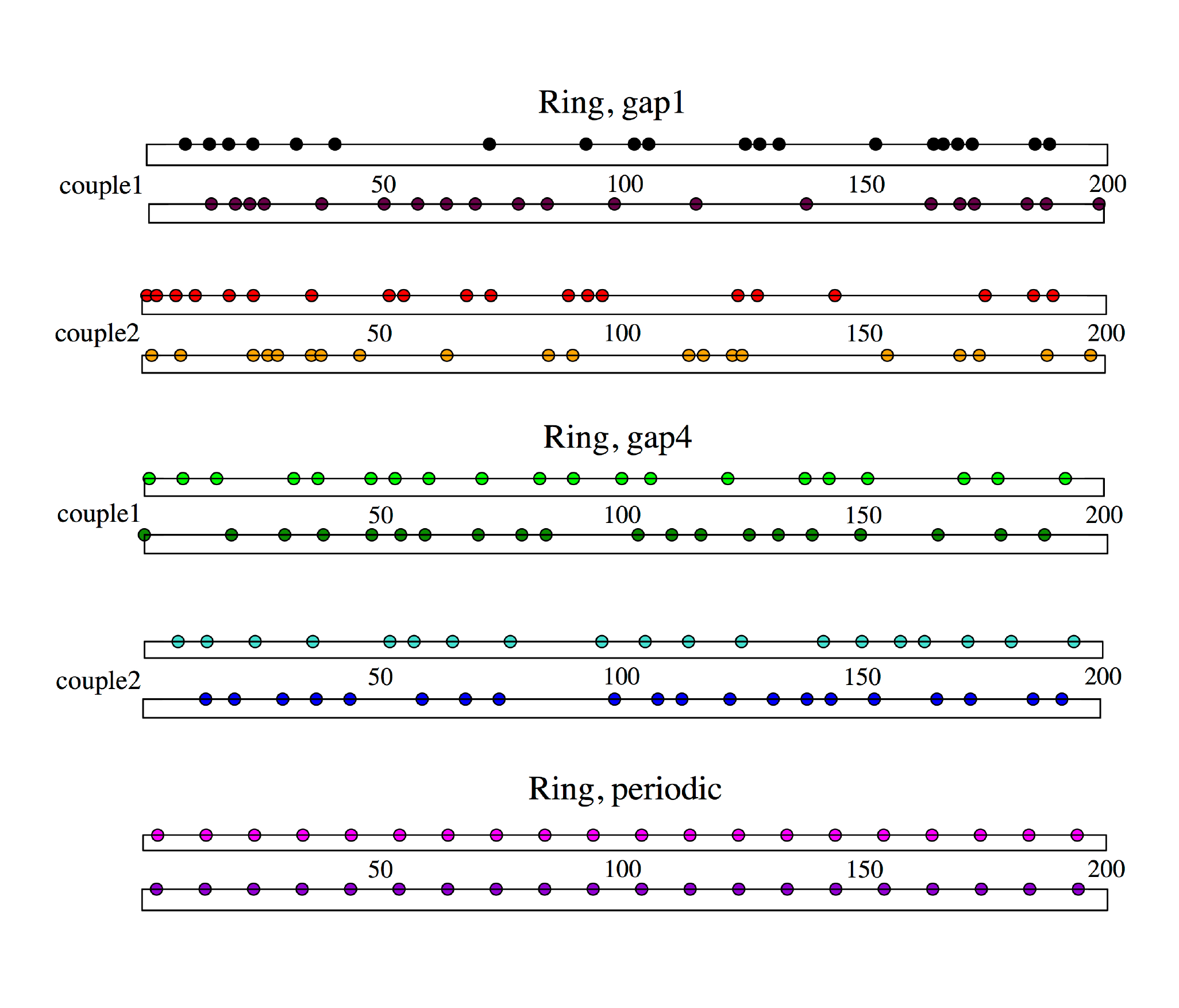}
	\end{minipage}%
\end{figure}
\noindent~Figure S2. As Figure S1 for the rings with reversible bonds. Although a linear representation 
of the backbone is used, it should be noted that both ends are mutually and permanently connected because of the ring geometry.

\newpage 

\begin{figure}[htbp!]
	\begin{minipage}[c]{.5\textwidth}
		\setlength{\captionmargin}{0pt}%
		\includegraphics[width=1.17\textwidth]{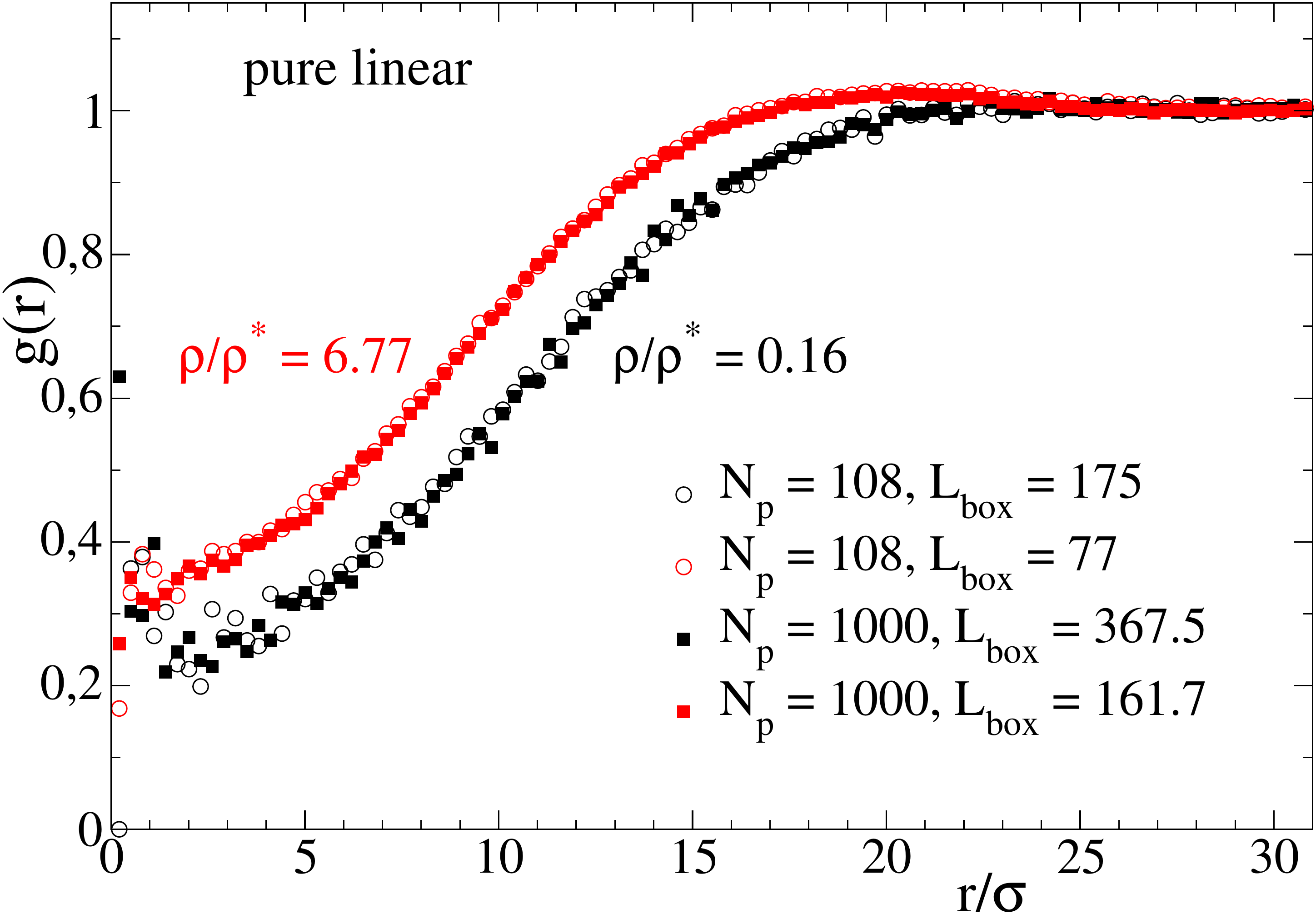}
	\end{minipage}%
\end{figure}
\noindent~Figure S3. Radial distribution functions of the effective fluid for the system 
of linear chains with reversible bonds. Data are shown at the lowest and highest investigated concentrations, and in both cases for $N_{\rm p} = 108$ and 1000 effective particles (with the respective rescaling of the box size to produce 
the same concentration, see legend). No significant size effects are observed.

\newpage

\begin{figure}[htbp!]
	\begin{minipage}[c]{.8\textwidth}
		\setlength{\captionmargin}{0pt}%
		\includegraphics[width=1.17\textwidth]{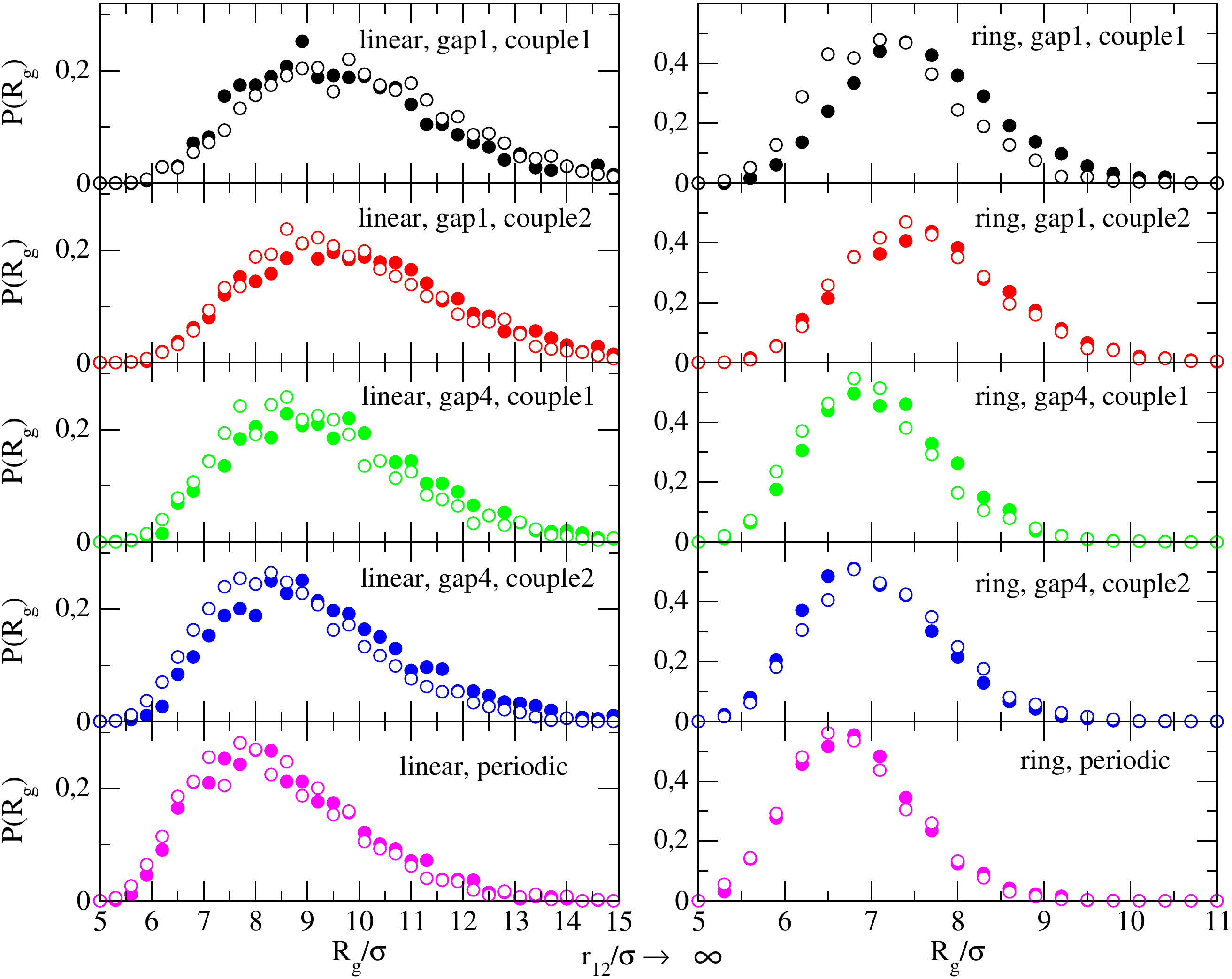}
	\end{minipage}%
\end{figure}
\noindent~Figure S4. Distribution of the radius of gyration for pairs of polymers at a very large distance ($25\sigma$) where there are no mutual contacts and the effective potential is zero.

\newpage

\begin{figure}[htbp!]
	\begin{minipage}[c]{.8\textwidth}
		\setlength{\captionmargin}{0pt}%
		\includegraphics[width=1.17\textwidth]{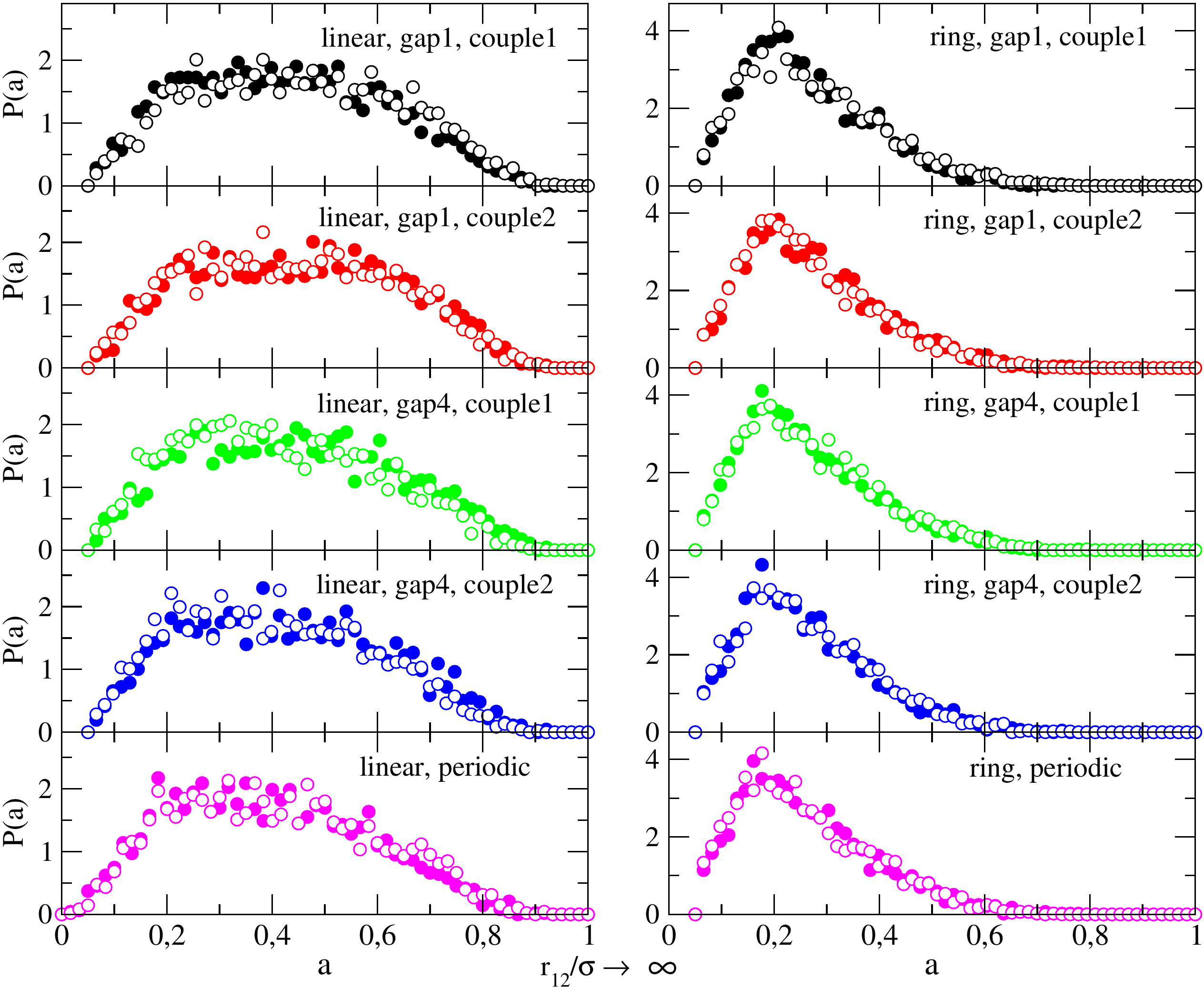}
	\end{minipage}%
\end{figure}
\noindent~Figure S5. Distribution of the asphericities for pairs of polymers at a very large distance ($25\sigma$) where there are no mutual contacts and the effective potential is zero.

\newpage

\begin{figure}[htbp!]
	\begin{minipage}[c]{.8\textwidth}
		\setlength{\captionmargin}{0pt}%
		\includegraphics[width=1.17\textwidth]{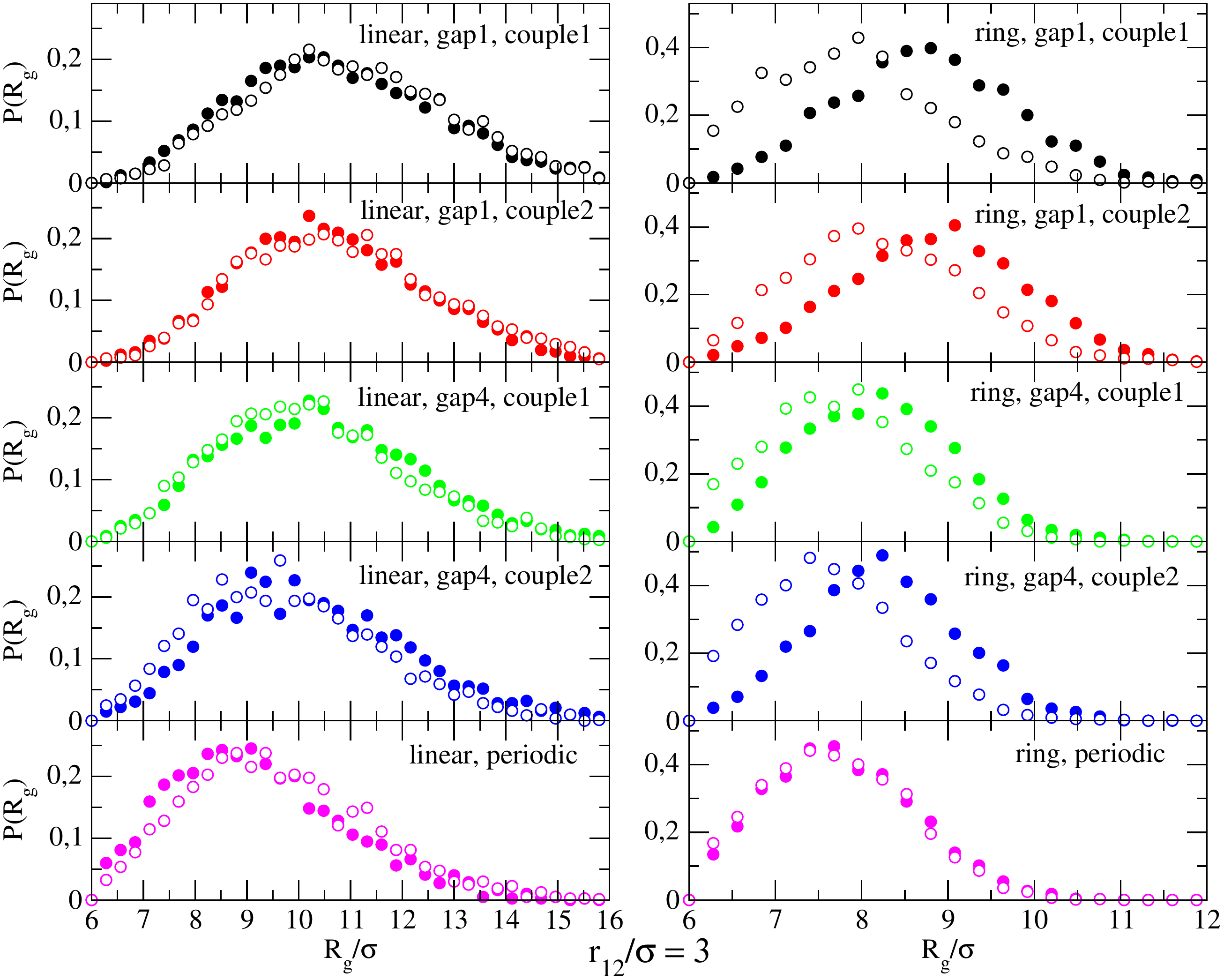}
	\end{minipage}%
\end{figure}
\noindent~Figure S6. Distribution of the radius of gyration for pairs of polymers at a distance $3\sigma$ where the force $\textbf{F}_{12}$ acting between them has its maximum.

\newpage

\begin{figure}[htbp!]
	\begin{minipage}[c]{.8\textwidth}
		\setlength{\captionmargin}{0pt}%
		\includegraphics[width=1.17\textwidth]{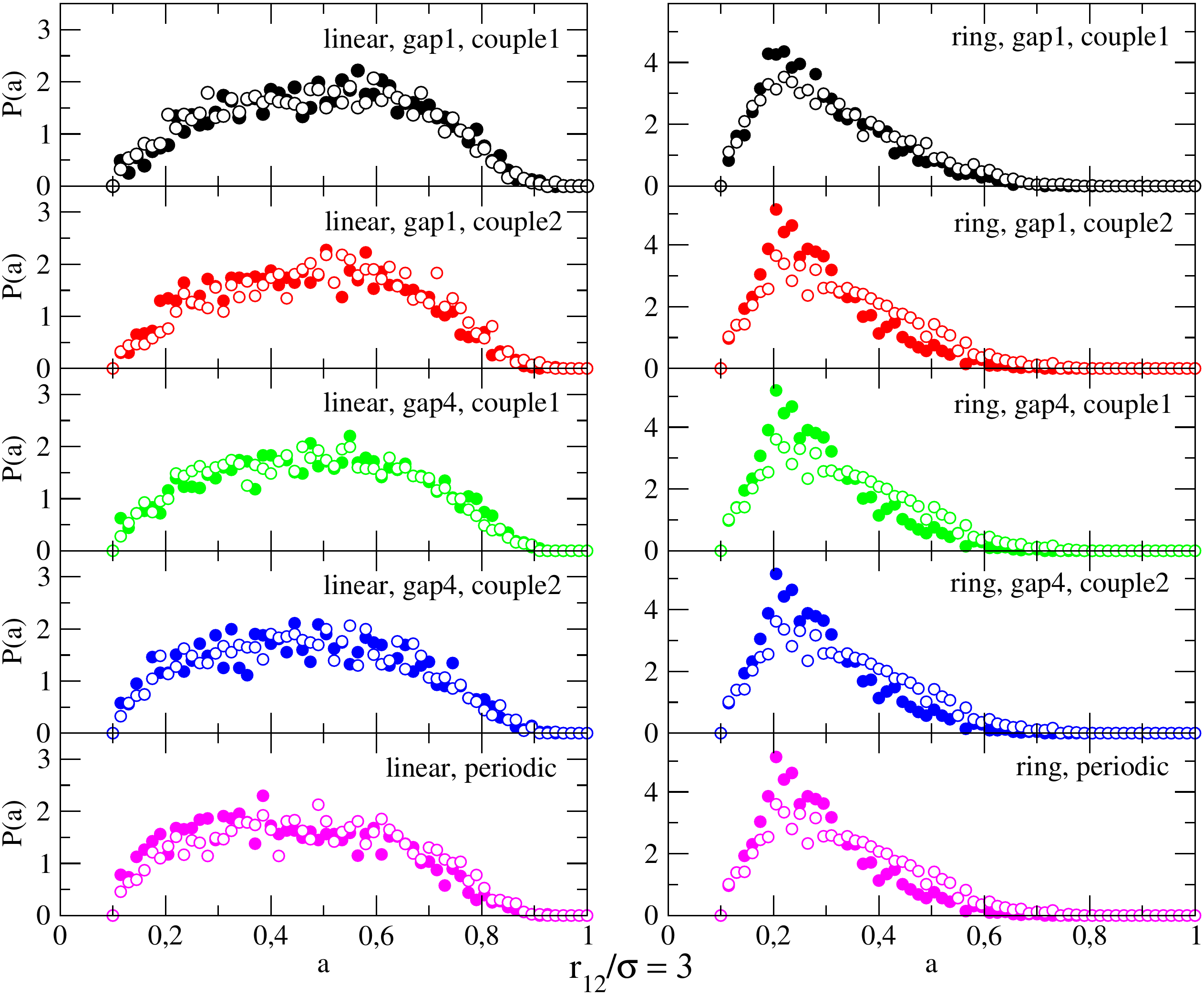}
	\end{minipage}%
\end{figure}
\noindent~Figure S7. Distribution of the asphericity for pairs of polymers at a distance $3\sigma$ where the force $\textbf{F}_{12}$ acting between them has its maximum.

\newpage

\begin{figure}[htbp!]
	\centering
	\begin{minipage}[c]{.8\textwidth}
		\setlength{\captionmargin}{0pt}%
		\includegraphics[width=1.2\textwidth]{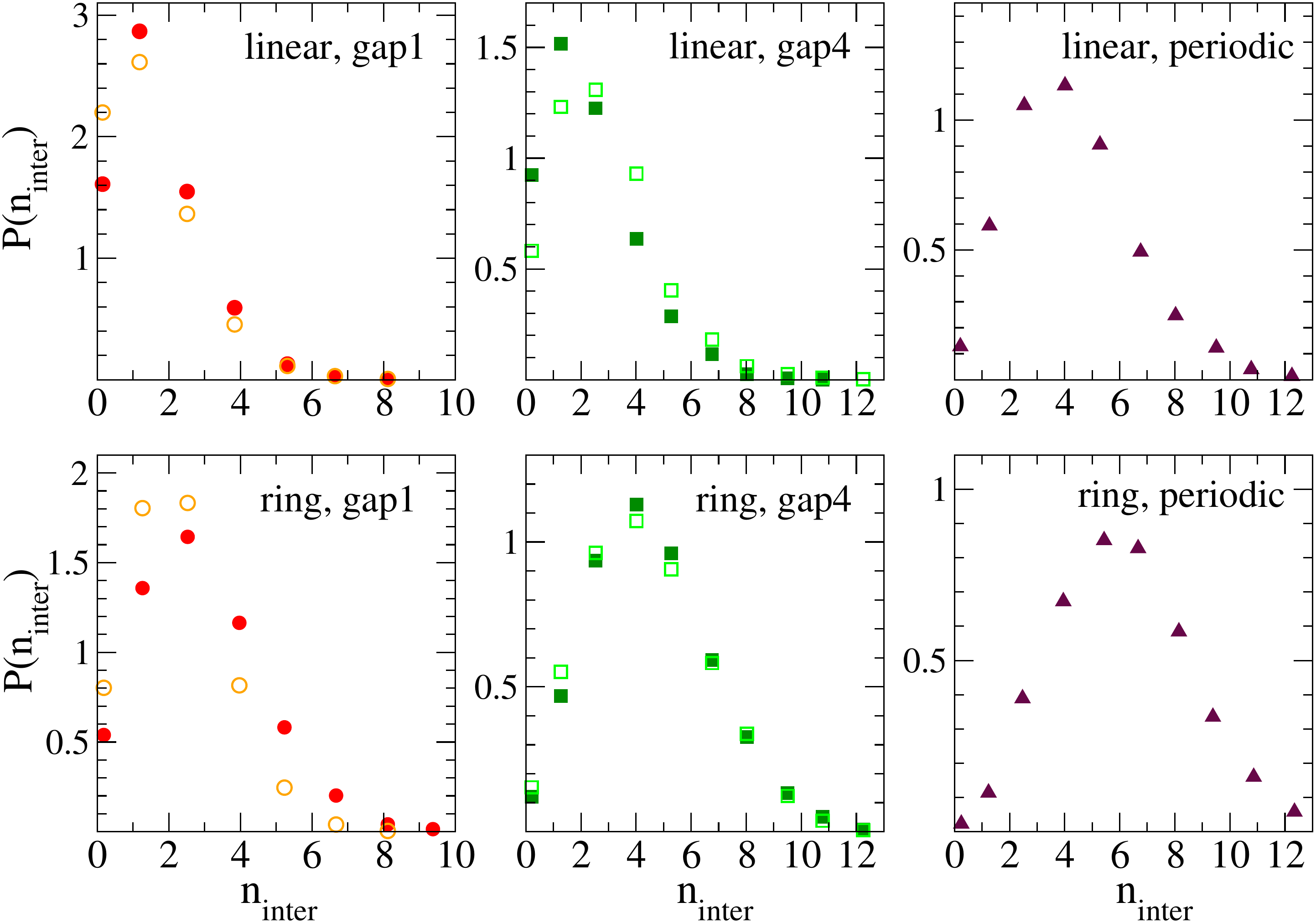}
	\end{minipage}%
\end{figure}
\noindent~Figure S8. Distribution of instantaneous values of the number of intermolecular bonds for two interacting polymers
(linear-linear and ring-ring) with reversible bonds and different sequences of reactive sites, at a mutual distance $r=3\sigma$ corresponding to the maximum of the effective force.

\newpage

\begin{figure}[htbp!]
	\centering
	\begin{minipage}[c]{.5\textwidth}
		\setlength{\captionmargin}{0pt}%
		\includegraphics[width=1.2\textwidth]{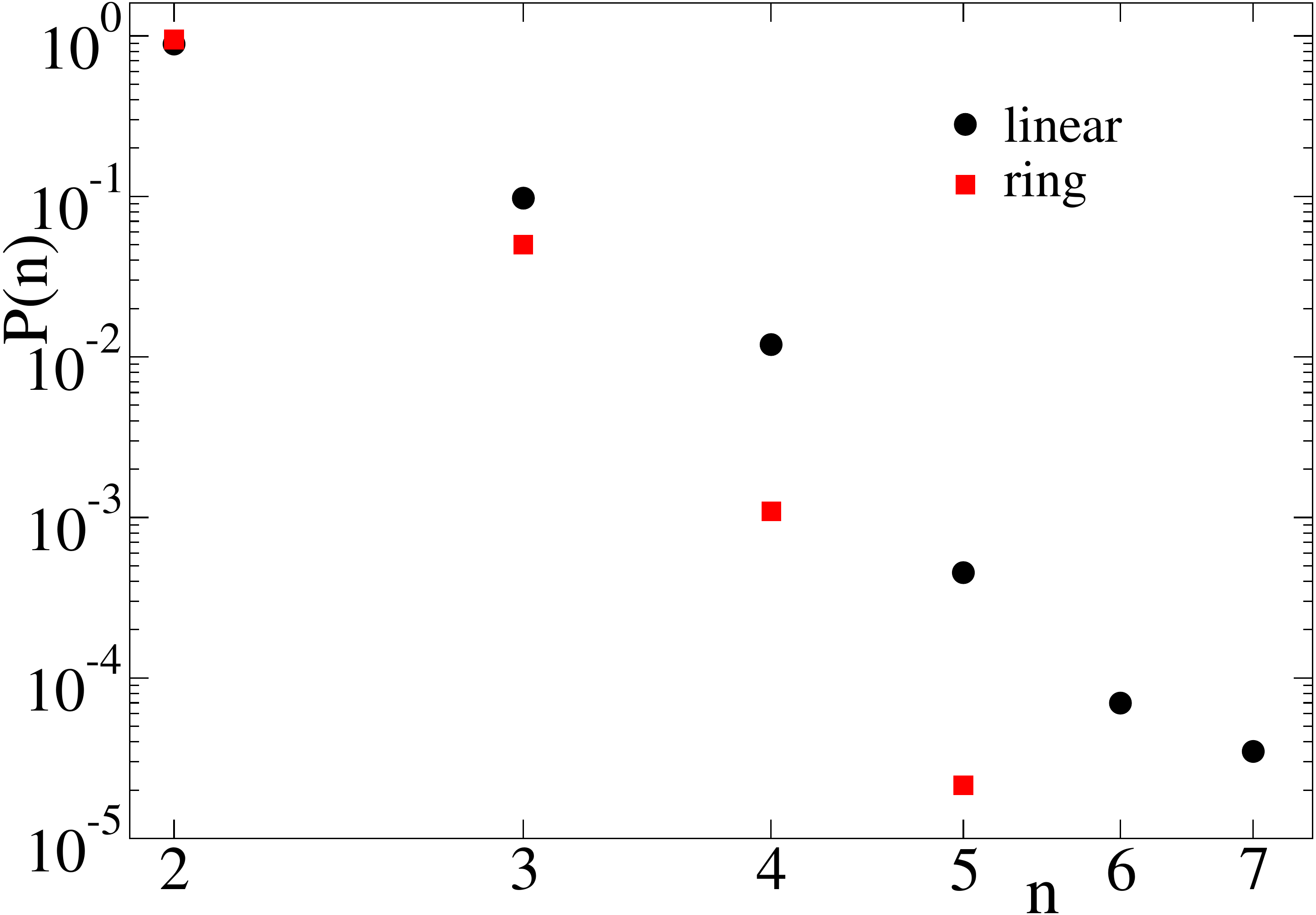}
	\end{minipage}%
\end{figure}
\noindent~Figure S9. Cluster size distribution in the pure solutions of linear chains and rings with reversible bonds, at the lowest investigated concentration. The cluster size is the number of polymers in the cluster. Two polymers belong to the same cluster if they are mutually linked by at least one intermolecular bond.

\end{document}